\documentclass[11pt,aps,prd,twocolumn,nofootinbib,reprint,longbibliography, superscriptaddress]{revtex4-2}
\pdfoutput=1
\usepackage{amsmath, amssymb, amsfonts, mathrsfs, slashed, bm, graphicx, xcolor, multirow}
\usepackage[colorlinks=true]{hyperref} 
\newcommand{\cmmnt}[1]{\ignorespaces}

\begin{document}
\pagestyle{plain}

\title{EFT analysis of leptophilic dark matter at future electron-positron colliders in the mono-photon and mono-$Z$ channels}

\author{Saumyen Kundu}
\email{p20170022@goa.bits-pilani.ac.in}
\affiliation{Department of Physics, Birla Institute of Technology and Science Pilani, K K Birla Goa Campus, NH-17B, Zuarinagar, Goa 403726, India}

\author{Atanu Guha}
\email{atanu@cnu.ac.kr}
\affiliation{Department of Physics, Chungnam National University, Daejeon 34134, Republic of Korea}

\author{Prasanta Kumar Das}
\email{pdas@goa.bits-pilani.ac.in}
\affiliation{Department of Physics, Birla Institute of Technology and Science Pilani, K K Birla Goa Campus, NH-17B, Zuarinagar, Goa 403726, India}

\author{P. S. Bhupal Dev}
\email{bdev@wustl.edu}
\affiliation{Department of Physics and McDonnell Center for the Space Sciences, Washington University, St.\,Louis, Missouri 63130, USA}

\begin{abstract}
We consider the possibility that dark matter (DM) only interacts with the Standard Model leptons, but not quarks at tree level, and analyze the future lepton collider prospects of such leptophilic DM in the monophoton and mono-$Z$ (both leptonic and hadronic) channels. Adopting a model-independent effective field theory framework, we consider all possible dimension-six operators of scalar-pseudoscalar (SP), vector-axial vector (VA), and tensor-axial tensor (TAT) types for a fermionic DM and derive the collider sensitivities on the effective cutoff scale $\Lambda$ as a function of the DM mass. As a concrete example, we take the beam configurations of the International Linear Collider with $\sqrt s=1$ TeV and $8$ ab$^{-1}$ integrated luminosity, including the effect of beam polarization, and show that it can probe leptophilic DM at $3\sigma$ level up to $\Lambda$ values of $6.6$, $8.8$, and $7.1$~TeV for the SP-, VA- and TAT-type operators, respectively. This is largely complementary to the direct and indirect searches for leptophilic DM and can potentially provide the best-ever sensitivity in the low-mass DM regime.  
\end{abstract}
\maketitle

\section{Introduction}
\label{sec:intro}
There is overwhelming evidence for the existence of dark matter (DM) in our Universe from various astrophysical and cosmological observations, such as galactic rotation curves, velocities of stars in dwarf galaxies, velocities of galaxies in clusters, hot gas in galaxy clusters, collisions of galaxy clusters, gravitational lensing, cosmic microwave background (CMB) and large-scale structure measurements (see e.g.~Refs.~\cite{Bertone:2004pz, Green:2021jrr} and references therein). However, apart from its gravitational interaction and the fact that it constitutes 26.8\% of the total energy budget of the Universe~\cite{Planck:2018vyg}, very little is known about the true nature and properties of DM. On the theory front, there exists a plethora of particle DM candidates~\cite{Bertone:2018krk} and identifying the correct one is a pressing issue in beyond the Standard Model (BSM) physics. On the experimental front, a number of ongoing and planned efforts in direct~\cite{Billard2021uyg}, indirect~\cite{Gaskins:2016cha} and collider~\cite{Boveia:2018yeb} searches for DM are poised to shed more light on the DM mystery~\cite{Battaglieri:2017aum}. 

Most of the current DM searches are traditionally motivated by the so-called weakly interacting massive particle (WIMP) miracle (see, e.g., Ref.~\cite{Feng:2010gw}) -- the simple observation that an electrically neutral, cosmologically stable WIMP $\chi$ with mass around the electroweak scale and couplings of the order of weak coupling strength to the SM sector can successfully reproduce the observed DM relic density of $\Omega_\chi h^2=0.120\pm 0.001$~\cite{Planck:2018vyg}. The WIMP DM mass can actually be in a wide range varying from $\mathcal{O}(1)$ MeV to $\mathcal{O}(100)$ TeV, although the possibility of an electroweak scale mass is much more appealing as a host of well-motivated BSM scenarios, from the supersymmetric extensions to the extra dimension models, can easily accommodate such a WIMP DM. The robustness of the WIMP paradigm comes from the fact that the interactions of the WIMP with the SM keep it in thermal equilibrium at high temperatures in the early Universe. Later, as the Universe expands and cools, the rate of interaction of WIMPs with the SM sector falls below the Hubble expansion rate, and as a result, the WIMP DM decouples (``freezes out") from the thermal plasma and explains the observed relic density in the present Universe.  

However, the WIMP paradigm is under significant strain, as the stringent exclusion limits from DM direct detection experiments, such as LUX~\cite{LUX:2016ggv}, XENON1T~\cite{XENON:2018voc} and PANDAX-4T~\cite{PandaX-4T:2021bab} have pushed the WIMP parameter space down to an awkward corner fast approaching the neutrino floor~\cite{Billard:2013qya, OHare:2021utq}, below which it will be extremely challenging to disentangle the DM signal from the coherent neutrino background. Therefore, it is important to explore other DM ideas that could potentially explain the absence of a signal in these experiments, while having promising prospects at other experiments. It is interesting to note that a majority of the existing experimental constraints  crucially rely on the WIMP interactions with nucleons and therefore can be largely weakened if the WIMP predominantly interacts with the SM leptons, but not quarks at tree level. Such ``leptophilic" DM could  indeed arise naturally in many BSM scenarios~\cite{Krauss:2002px, Baltz:2002we, Ma:2006km, Hambye:2006zn, Bernabei:2007gr, Cirelli:2008pk, Chen:2008dh, Bi:2009md, Cao:2009yy, Goh:2009wg, Bi:2009uj, Ibarra:2009bm, Davoudiasl:2009dg,Dedes:2009bk, Kopp:2009et, Cohen:2009fz, Chun:2009zx, Chao:2010mp, Haba:2010ag,Ko:2010at, Carone:2011iw, Schmidt:2012yg, Das:2013jca, Dev:2013hka, Chang:2014tea, Agrawal:2014ufa, Bell:2014tta, Freitas:2014jla,Cao:2014cda, Boucenna:2015tra,Lu:2016ups, Duan:2017pkq,Duan:2017qwj,Chao:2017emq,Li:2017tmd,Ghorbani:2017cey,Sui:2017qra,Han:2017ars,  Madge:2018gfl,Junius:2019dci, YaserAyazi:2019psw, Ghosh:2020fdc,  Chakraborti:2020zxt, Horigome:2021qof, Garani:2021ysl}, some of which could even explain various experimental anomalies, such as the muon anomalous magnetic moment
~\cite{Bennett:2006fi, Abi:2021ojo}, DAMA/LIBRA annual modulation~\cite{Bernabei:2020mon}, 
anomalous cosmic ray positron excess as observed in ATIC~\cite{Chang:2008aa}, PAMELA~\cite{Adriani:2008zr, Adriani:2013vxg}, Fermi-LAT~\cite{Fermi-LAT:2011baq, Abdollahi:2017nat}, AMS-02~\cite{Accardo:2014lma, Aguilar:2014mma, AMS:2021nhj}, 
 DAMPE~\cite{DAMPE:2017fbg} and CALET~\cite{Adriani:2018ktz},  
the Galactic Center gamma-ray excess~\cite{TheFermi-LAT:2015kwa}, 511 keV gamma-ray line~\cite{Siegert:2015knp}, IceCube ultrahigh-energy neutrino excess~\cite{IceCube:2020wum}, and XENON1T electron excess~\cite{XENON:2020rca}. Dedicated searches for leptophilic DM in direct detection experiments~\cite{XENON100:2015tol, XENON:2019gfn, LZ:2021xov}, beam dump experiments~\cite{Chen:2018vkr, Marsicano:2018vin}, and gravitational wave detectors~\cite{Madge:2018gfl}, as well as using terrestrial~\cite{Chauhan:2016joa} and celestial objects like Sun~\cite{Garani:2017jcj, Liang:2018cjn}, supernovae~\cite{Guha:2018mli} and neutron stars~\cite{Bell:2019pyc, Garani:2019fpa, Joglekar:2019vzy, Joglekar:2020liw, Bell:2020lmm} have also been discussed. 

 In this paper, we focus on the leptophilic DM searches at colliders, which are complementary to the direct and indirect detection searches. In order to keep our discussion as general as possible, we will adopt an effective field theory (EFT) approach, which has been widely used in the context of collider searches for DM following the early works of Refs.~\cite{Kopp:2009et, Beltran:2010ww,Goodman:2010yf,Bai:2010hh, Goodman:2010ku, Fox:2011fx,Fox:2011pm, Rajaraman:2011wf,Chae:2012bq} (see Ref.~\cite{Kahlhoefer:2017dnp} for a review).  The same interactions responsible for WIMP DM pair annihilation in the early Universe leading to their thermal freeze-out guarantee their direct production at colliders, as long as kinematically allowed. This will give a characteristic mono-$X$ signature, where the large missing transverse momentum carried away by the DM pair is balanced by a visible sector particle $X$ (which can be either a photon, jet, $W$, $Z$, or Higgs, depending on the model) emitted from an initial, intermediate or final state (see Refs.~\cite{Kahlhoefer:2017dnp, Penning:2017tmb} for reviews). Specifically, the monojet signature has become emblematic for LHC DM searches~\cite{CMS:2014jvv, ATLAS:2015qlt, CMS:2017zts, ATLAS:2021kxv}. However, for a leptophilic DM with loop-suppressed interactions to the SM quarks, hadron colliders like the LHC are not expected to provide a better model-independent limit than the existing constraints from indirect searches, such as from AMS-02~\cite{Cavasonza:2016qem, John:2021ugy}, at least within the EFT framework with contact interactions. 
 
 On the other hand, lepton colliders provide an ideal testing ground for the direct production of leptophilic DM and its subsequent detection via either monophoton~\cite{DELPHI:2003dlq, Birkedal:2004xn, Fox:2008kb,Konar:2009ae, Fox:2011fx,Bartels:2012ex, Dreiner:2012xm,Chae:2012bq, Liu:2019ogn,Habermehl:2020njb, Kalinowski:2021tyr, Barman:2021hhg} or mono-$Z$~\cite{Wan:2014rhl, Yu:2014ula, Dutta:2017ljq, Grzadkowski:2020frj} signatures. In this paper, we go beyond the existing literature and perform a comprehensive and comparative study of both monophoton and mono-$Z$ signatures of leptophilic DM at future $e^+e^-$ colliders in a  model-independent, EFT approach. Our analysis is generically applicable to all future $e^+e^-$ colliders, such as  the International Linear Collider (ILC)~\cite{Bambade:2019fyw}, Compact Linear Collider~\cite{CLIC:2016zwp}, Circular Electron Positron Collider~\cite{CEPCStudyGroup:2018ghi} and the electron-positron Future Circular Colliders~\cite{FCC:2018evy}, but for concreteness, we have taken the $\sqrt s=1$ TeV ILC as our case study for numerical simulations. We also assume the DM to be fermionic and limit ourselves to the dimension-six operators, but taking into consideration all possible dimension-six operators of scalar-pseudoscalar (SP), vector-axial vector (VA) and tensor-axialtensor (TAT)-type as applicable for the most general DM-electron coupling. 
 Within the minimal EFT approach, the only relevant degrees of freedom in our analysis are the DM mass and an effective cutoff scale $\Lambda$ which determines the strength of the four-Fermi operators. This enables us to derive model-independent ILC sensitivities on leptophilic DM in the $(m_\chi, \Lambda)$ plane in both monophoton and mono-$Z$ (leptonic and hadronic) channels, after taking into account all relevant backgrounds and systematic uncertainties.  We consider both unpolarized and polarized beam options~\cite{Bambade:2019fyw, Barklow:2015tja}, and find that with the proper choice of polarizations for the $e^-$ and $e^+$ beams (which depends on the operator type), the DM sensitivities could be significantly enhanced. We also find that the hadronic mono-$Z$ channel provides the best sensitivities for the SP and TAT-type operators, while the monophoton channel gives better sensitivity for the VA-type operators. For example, for a $1$ GeV DM mass, the monophoton channel can probe $\Lambda$ values at $3\sigma$ confidence up to $6.6$ TeV, $8.8$ TeV and $7.1$ TeV for the SP, VA and TAT-type operators respectively at $\sqrt s=1$ TeV ILC with 8 ${\rm ab}^{-1}$ integrated luminosity, whereas the mono-$Z$ hadronic (leptonic) channel can probe $\Lambda$ values up to $5.5\; (4.3)$ TeV, $4.2\; (2.7)$ TeV and $6.7\; (5.2)$ TeV respectively. We compare our results with the existing literature wherever applicable and also with the current constraints from direct and indirect detection, as well as from relic density considerations. We find that the $\sqrt s=1$ TeV ILC sensitivities could surpass the existing constraints for  light DM with masses below about 300 GeV, above which the collider reach is weakened mainly due to the kinematic suppression.   
 
 The rest of the paper is organized as follows. In \autoref{sec:2}, we briefly describe the EFT approach and list all possible dimension-six operators for the DM-electron coupling in our leptophilic DM scenario. 
In \autoref{sec:3}, we present our cut-based analysis for the monophoton+$\slashed{E}_T$ signal and background, both with and without beam polarization. In \autoref{sec:4}, we repeat the cut-based analysis for the mono-$Z$+$\slashed{E}_T$ channel, with the leptonic case of $Z\to \ell^+ \ell^-$ (with $\ell = e, \mu$) in \autoref{sec:4.1} and the hadronic case of $Z\to {\rm jets}$ in \autoref{sec:4.2}. Our conclusions are given in \autoref{sec:5}. 

\section{Effective operators}\label{sec:2}

\begin{figure}[!t]
\includegraphics[width=0.35\textwidth]{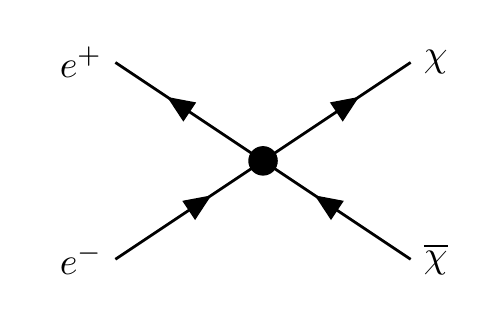}
\caption{Feynman diagram for the effective four-Fermi interactions between the DM and electrons induced by the dimension-six operator~\eqref{eq:EFT}.}
\label{fig:EFT}
\end{figure}

Our goal in this paper is to perform a model-independent collider analysis for the leptophilic DM scenario in an EFT approach, with the primary assumptions that (i) the DM particle $\chi$ couples directly only to the SM leptons but not to the quarks (hence leptophilic), and (ii) the energy scale of the associated new physics is large compared to the collider energies under consideration, thus allowing us to integrate out the heavy mediators and parametrize the DM-SM interactions using effective higher-dimensional operators.    
For concreteness, we assume that the DM particles are Dirac fermions, and therefore, the leading order DM-SM interactions are the dimension-six four-Fermi interactions shown in \autoref{fig:EFT}, with the most general effective Lagrangian given by~\cite{Kopp:2009et} 
\begin{align}
    \mathcal{L}_{\rm eff}  =  \frac{1}{\Lambda^2}\sum_j\left(\overline{\chi}\Gamma^{j}_{\chi}\chi\right)\left(\overline{\ell}\Gamma^{j}_{\ell}\ell\right) \, , 
    \label{eq:EFT}
\end{align}
 where $\Lambda$ is the cutoff scale for the EFT description and the index $j$ corresponds to different Lorentz structures, as shown below. Since our main focus is on $e^+e^-$ colliders, we will just set $\ell=e$ in Eq.~\eqref{eq:EFT} and assume this to be the only leading-order coupling, but our discussion below could be easily extended to other cases, e.g. future muon colliders~\cite{Delahaye:2019omf} by setting $\ell=\mu$. 
 
 A complete set of Lorentz-invariant operators consists of scalar (S),  pseudo-scalar (P), vector (V), axial-vector (A), tensor (T) and axial-tensor (AT) currents. We classify them as follows:%
 \begin{widetext}
  \begin{align}
   &\text{Scalar-pseudoscalar (SP) type}: & &
   \Gamma_{\chi}  =  c^{\chi}_{S}+i c^{\chi}_{P} \gamma_5 \, ,  & & \Gamma_{e} = c^{e}_{S}+i c^{e}_{P} \gamma_5 \, , \nonumber \\
   &\text{Vector-axial vector (VA) type}: & &
   \Gamma_{\chi}^{\mu} =  \left( c^{\chi}_{V}+ c^{\chi}_{A} \gamma_5 \right) \gamma^{\mu} \, ,  & &
   \Gamma_{e\mu} = \left( c^{e}_{V}+ c^{e}_{A} \gamma_5 \right)\gamma_{\mu}  \, , \nonumber \\
   &\text{Tensor-axial Tensor (TAT) type}: & &
   \Gamma_{\chi}^{\mu \nu} = \left( c^{\chi}_{T}+i c^{\chi}_{AT} \gamma_5 \right) \sigma^{\mu \nu} \, , & & \Gamma_{e \mu \nu} = \sigma_{\mu \nu} \, ,
    \label{eq:operator}
   \end{align}
   \end{widetext}
   where $\sigma^{\mu \nu}=\frac{i}{2}[\gamma^\mu,\gamma^\nu] $ is the spin tensor and $c_j^{\chi,e}$ are dimensionless, real couplings. Note that for the TAT type, we did not write the AT term for electron separately, because the relation $\sigma^{\mu\nu}\gamma_5=\frac{i}{2}\epsilon^{\mu\nu\alpha\beta}\sigma_{\alpha\beta}$ implies that the ${\rm AT}\otimes{\rm AT}$ coupling is equivalent to ${\rm T}\otimes{\rm T}$, and similarly, ${\rm T}\otimes{\rm AT}={\rm AT}\otimes{\rm T}$. Also if we had considered Majorana DM, then vector- and TAT-type interactions are forbidden, i.e. $c_V^\chi=c_T^\chi=c_{AT}^\chi=0$. For simplicity, in Eq.~\eqref{eq:EFT} we have used a common cutoff scale $\Lambda$ for all Lorentz structures. Furthermore, in our subsequent numerical analysis, we will consider one type of operator at a time, by setting the corresponding couplings $c_j^{\chi,e}=1$ without loss of generality and all other couplings equal to zero, unless otherwise specified. For instance, setting $c_S^\chi=c_P^\chi=c_S^e=c_P^e=1$ and all other couplings equal to zero gives us the (S+P)-type operator, which we will simply refer to as the {\bf SP}-type in the following discussion. Similarly, we will denote the  $c_V^\chi=c_A^\chi=c_V^e=c_A^e=1$ case simply as the {\bf VA}-type, and  $c_T^\chi=c_{AT}^\chi=1$ as the {\bf TAT}-type for presenting our numerical results in the $(m_\chi,\Lambda)$ plane. For other choices of the absolute values of the couplings, our results for the sensitivity on $\Lambda$ can be easily scaled accordingly. 
   
   We do not discuss any specific realization of these effective operators since we are making a model-independent analysis. In the context of a given ultraviolet (UV)-complete theory or even a simplified model, the suppression or cutoff scale $\Lambda$ of the effective theory can be understood in terms of the mass $m_{\rm med}$ of a mediator (which couples to electrons as well as DM) as 
   \begin{align}
      \frac{1}{\Lambda^2} = \frac{g_e g_\chi}{m_{\rm med}^2} \, ,
      \label{eq:contact}
   \end{align}
   where $g_e$  and $g_\chi$ are the mediator couplings to electron and DM respectively, and the mediator is assumed to be heavier than the energy scale of interest. As a concrete example, let us consider the axial-vector-type operator, which can be obtained from a theory with a massive spin-1 particle $V^\mu$ with axial couplings to leptons and DM,
\begin{align}
    \mathcal{L} = \frac{1}{2}m^2_V V_\mu V^\mu + g_{Ae} \overline{e} \gamma_\mu \gamma^5 e V^\mu + g_{A\chi} \overline{\chi} \gamma_\mu \gamma^5 \chi V^\mu \, .
    \label{eq:spin1}
\end{align}
In the context of DM pair production at lepton colliders via the $e^+e^-\to \chi \overline{\chi}$ process, $V^\mu$ acts as the mediator of the interaction between the electrons and DM particles. For instance, if this is an $s$-channel process, then the corresponding matrix element will be   
\begin{align}
    \mathcal{M} & \sim \left(\frac{g_{Ae}~g_{A\chi}}{m^2_V - s} \right)\left(\overline{e} \gamma^\mu \gamma^5 e\right)\left(\overline{\chi} \gamma_\mu \gamma^5 \chi\right) \nonumber \\ 
& \simeq \left(\frac{g_{Ae}~g_{A\chi}}{m^2_V} \right) \left[1 + \frac{s}{m^2_V} +{\cal O}\left(\frac{s^2}{m_V^4}\right)\right]\left(\overline{e} \gamma^\mu \gamma^5 e\right)\left(\overline{\chi} \gamma_\mu \gamma^5 \chi\right) \, ,
\label{eq:M}
\end{align}
in the limit where the center-of-mass energy is smaller than the mediator mass, i.e. $s\ll m_V^2$. In this case, the leading term in Eq.~\eqref{eq:M} will be of the form $g_{Ae}g_{A\chi}/m^2_V$, which can be simply identified as $1/\Lambda^2$ in the contact interaction description of EFT [cf.~Eq.~\eqref{eq:contact}]. On the other hand, if $m^2_V \ll s$, i.e., in the light mediator scenario, Eq.~\eqref{eq:M} should be written as  
\begin{equation}
    \mathcal{M} \simeq -\left(\frac{g_{Ae}~g_{A\chi}}{s} \right) \left[1 + \frac{m^2_V}{s} +{\cal O}\left(\frac{m_V^4}{s^2}\right)\right]\left(\overline{e} \gamma^\mu \gamma^5 e\right)\left(\overline{\chi} \gamma_\mu \gamma^5 \chi\right) \, ,
\end{equation}%
causing the cross section to fall as $1/s$~\cite{Buchmueller:2013dya, DeSimone:2016fbz, Kalinowski:2021tyr}. For, $m_V^2\simeq s$, i.e.~for an on shell mediator, the cross section increases due to the Breit-Wigner resonance, but this resonant behavior is not realized in the EFT framework \cite{Buchmueller:2013dya,DeSimone:2016fbz}.

In this work we have assumed the heavy mediator scenario (i.e., $m_{\rm med}^2\gg s$). Then from Eq.~\eqref{eq:contact}, we have $\Lambda=m_{\rm med}/\sqrt{g_e g_\chi}>\sqrt s$, as long as the couplings are of ${\cal O}(1)$ or less. Hence, we will impose a theoretical limit of $\Lambda>\sqrt s$ for the EFT validity  with the couplings set to unity. 
 
For relatively larger DM mass, we must also have $\Lambda>2m_\chi$ in order to describe DM pair annihilation by the EFT. In fact, using the lower limit of $\Lambda=2m_\chi$ induces 100\% error in the EFT prediction for $s$-channel UV completions. Therefore, we will use $\Lambda>{\rm max}\{\sqrt s, 3m_\chi\}$ as a conservative lower bound~\cite{Matsumoto:2016hbs} to ensure the validity of our EFT approach.

As opposed to the lepton colliders, the validity of the EFT approach for mono-$X$ searches at the hadron colliders is problematic~\cite{Busoni:2013lha}. Instead, simplified models~\cite{Alwall:2008ag, Abdallah:2015ter} are more appropriate for the interpretation of the LHC results. Therefore, it becomes difficult to directly compare the results from lepton and hadron colliders. 

As for the specific model realizations of the other operators listed in Eq.~\eqref{eq:operator}, the vector-type operator can be easily realized by a  similar spin-1 mediator as in Eq.~\eqref{eq:spin1}, with vector instead of axial-vector couplings. Similarly, the SP-type operators can be realized by a scalar/pseudoscalar mediator, whereas the TAT-type operators can be realized by a spin-2 mediator. A detailed discussion of these model realizations is beyond the scope of the current work.  

\section{monophoton channel} \label{sec:3}

\begin{figure}[!t]
\includegraphics[width=0.35\textwidth]{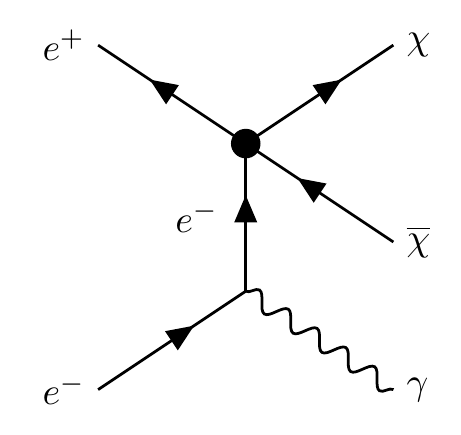}
\caption{Feynman diagram for the monophoton signal at an $e^+e^-$ collider. The photon can be radiated off either the electron or the positron leg.}
\label{fig:mono-g_Feyn}
\end{figure}
Since DM behaves as missing energy in the collider environment, we need a visible particle along with the DM pair to identify the DM production process. In the context of leptophilic DM at lepton colliders, monophoton and mono-$Z$ are the most promising channels, where the extra $\gamma/Z$ comes from initial-state radiation (ISR) from either the electron or positron leg. In this section, we analyze the monophoton signal and backgrounds in detail and present our numerical sensitivity results for $\sqrt s=1$ TeV ILC with 8 ab$^{-1}$ integrated luminosity. In the next section, we will repeat the same analysis for the mono-$Z$ case. 

For the monophoton signal  $e^+e^-\to\chi\overline{\chi}\gamma$ (as shown in \autoref{fig:mono-g_Feyn}), the $\chi$'s will contribute to the missing transverse energy at the detector. The dominant irreducible SM background to this process comes from neutrino pair production with an associated ISR photon, i.e. $e^+e^-\to\nu\overline{\nu}\gamma$. Since neutrinos are practically indistinguishable from WIMPs on an event-by-event basis, the majority of this background survives the event selection cuts. However, as we will show later, this background is highly polarization dependent and therefore can be significantly reduced by the proper choice of polarized beams, without affecting the signal much. 

Apart from the neutrino background, any SM process with a single photon in the final state can contribute to the total background if all other visible particles escape detection. The SM processes containing either jets or charged particles are relatively easy to distinguish from a DM event, so their contribution to the total background is negligible~\cite{Bartels:2012ex}. The only exception is the ``Bhabha scattering" process associated with an extra photon (either from initial or final-state radiation), i.e., $e^+e^-\to e^+e^-\gamma$, which has a large cross section, is polarization independent, and can significantly contribute to the total background whenever the final-state electrons and positrons go undetected, e.g. along beam pipes. In our following analysis, we consider both neutrino and radiative Bhabha backgrounds.  

\subsection{Cross sections}\label{sec:3.1} 
The cross sections for the monophoton signal $e^+e^-\to \chi\overline{\chi}\gamma$  at $\sqrt s=1$ TeV ILC are estimated using \texttt{CalcHEP}~\cite{Belyaev:2012qa} with proper implementation of ISR and beamstrahlung effects, which significantly affect the width and position of the neutrino $Z$ resonance. For this purpose, the EFT Lagrangian~\eqref{eq:EFT} is implemented in \texttt{FeynRules}~\cite{Alloul:2013bka} to generate the model library files suitable for \texttt{CalcHEP}. To avoid collinear and infrared divergences, we limit the phase space in the event generation with the following cuts on the outgoing photon: 
\begin{equation} \label{eq:CutsPh}
 p_T^{\gamma} > 2\text{ GeV}, \;\;E_{\gamma} > 1\text{ GeV},\;\; |\cos\theta_\gamma|\le 0.9975 \, .
\end{equation}
\begin{figure*}[ht]
\includegraphics[width=0.48\linewidth]{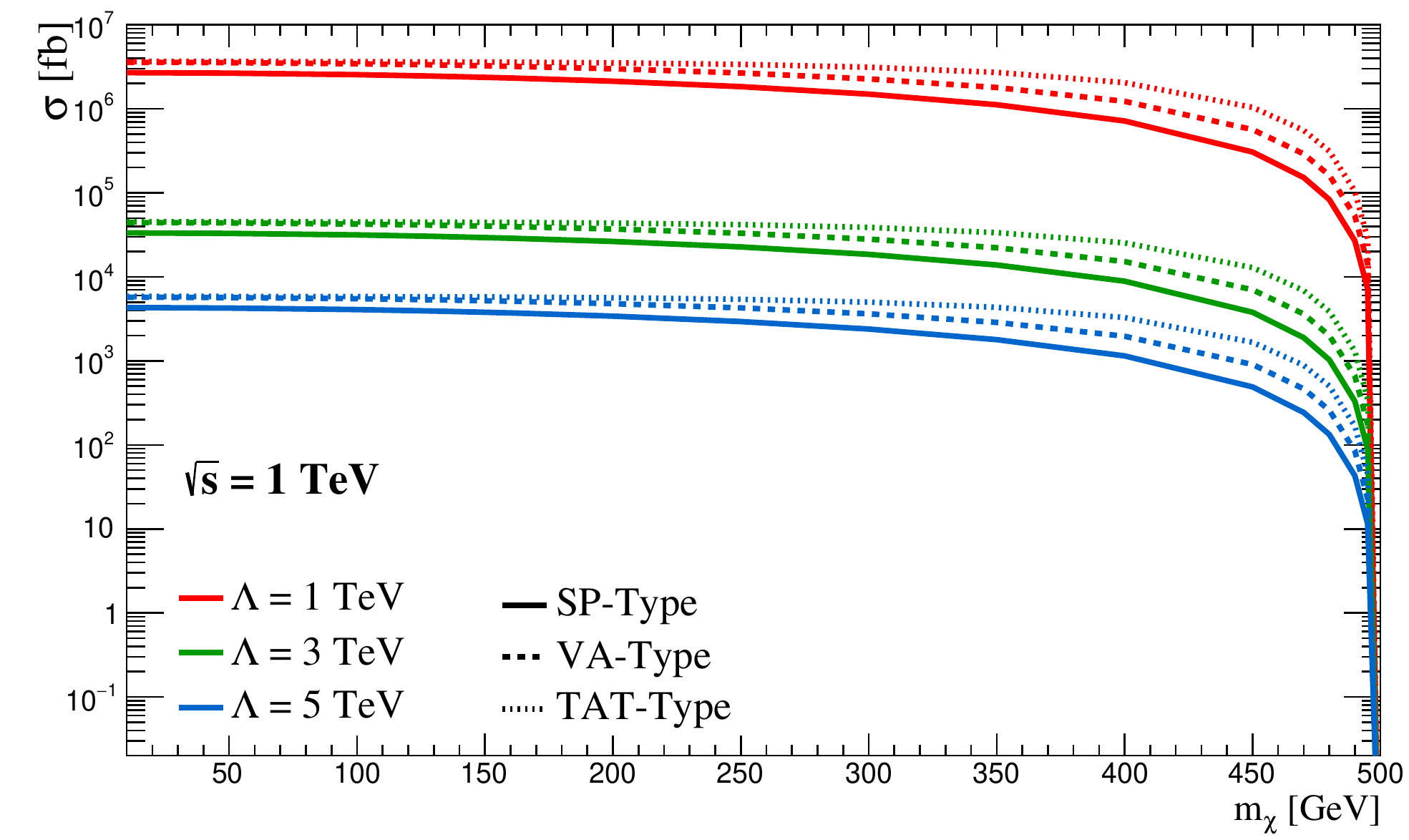}
\includegraphics[width=0.48\linewidth]{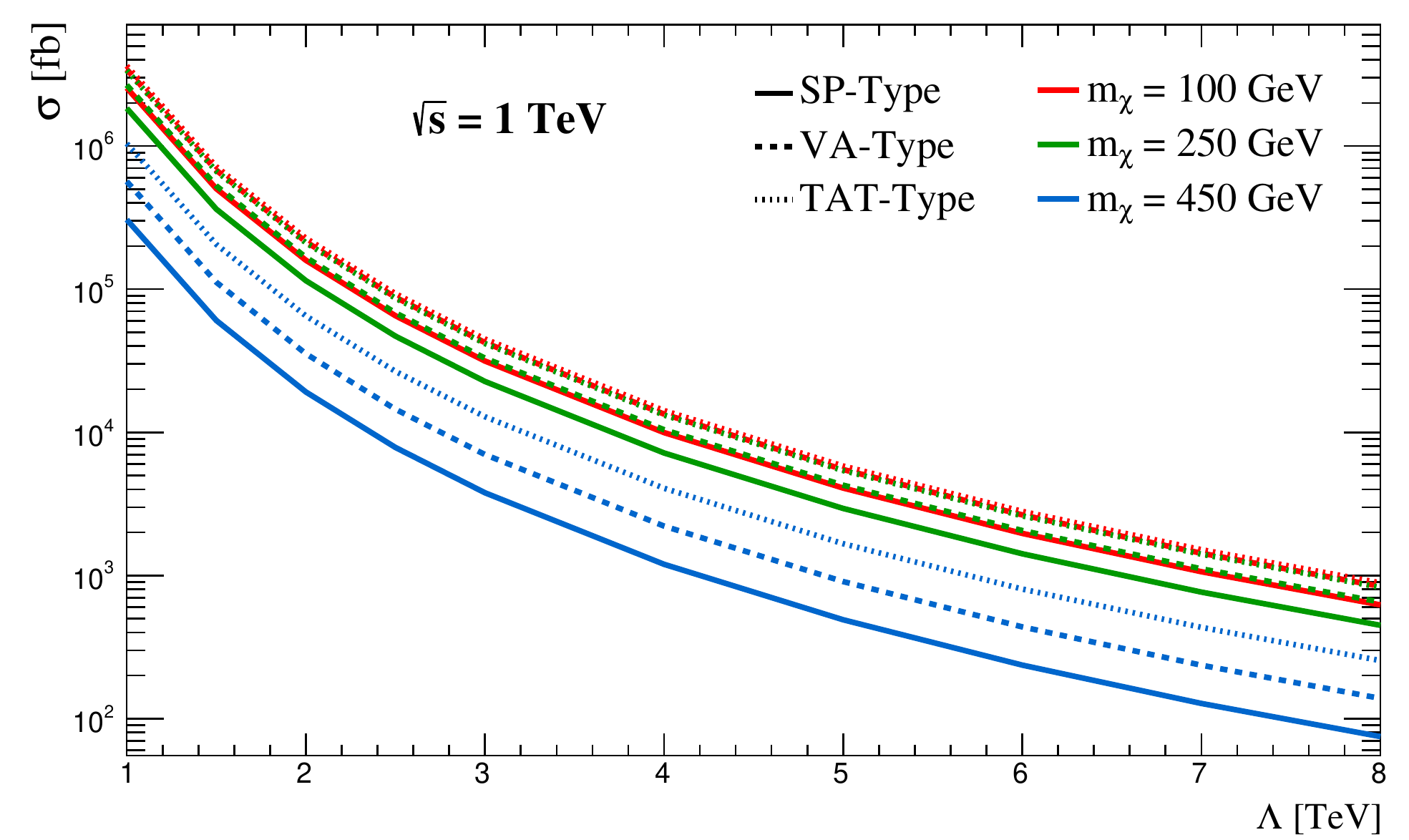}
\caption{Variation of monophoton signal cross section with the DM mass (left) and the cutoff scale (right) at $\sqrt s=1$ TeV ILC. The solid, dashed and dotted lines are for the SP-, VA- and TAT-type operators respectively. In the left panel, the red, green and blue curves correspond to different values of the cutoff scale $\Lambda=1$, 3, and 5 TeV, respectively, whereas in the right panel, they correspond to different values of the DM mass $m_\chi=100$, 250, and 450 GeV respectively.}
\label{fig:gCS}
\end{figure*}
The \emph{radiative} neutrino-pair and the \emph{radiative} Bhabha background events are generated using \texttt{WHIZARD}~\cite{Kilian:2007gr} (to better handle the singularities)  with the same set of cuts as in Eq.~\eqref{eq:CutsPh} to the matrix element (ME) photon (i.e., excluding the ISR and  beamstrahlung photons). Also, some additional cuts are implemented on the invariant masses for the Bhabha process to take care of the soft and collinear divergences: 
\begin{align}
&M_{e^{\pm}_{\rm in}, e^{\pm}_{\rm out}}< -2 \text{ GeV},~M_{e^{\pm}_{\rm out}, e^{\pm}_{\rm out}}> 2 \text{ GeV},\nonumber\\
&M_{e^{\pm},\gamma_{i}}>2\text{ GeV},~M_{e^{\pm},\gamma}> 4\text{ GeV} \, ,
\end{align}
where $\gamma_{i}$ are the ME photons and $\gamma$ is the signal photon. \par
To properly merge the ISR and ME photons in the events we also incorporated the ME-ISR merging technology as discussed in Ref.~\cite{Kalinowski:2020lhp}. This takes into account the higher-order corrections (which contribute to about 20\% of cross section) while avoiding double counting of multiple photon emission. We considered the values of the independent variables $q_{-}$ and $q_{+}$ to be $1$ GeV~\cite{Kalinowski:2020lhp} where $q_{\pm}$ are defined as 
\begin{eqnarray}
     q_{-}&= \sqrt{2E_{\rm cm}E_{\gamma}}\cdot \sin{\theta_{\gamma}/2} \, , \\
     q_{+}&= \sqrt{2E_{\rm cm}E_{\gamma}}\cdot \cos{\theta_{\gamma}/2}  \, ,
\end{eqnarray}
where $E_{\rm cm}\equiv\sqrt{s}$, and $E_{\gamma}$ and $\theta_{\gamma}$ are the photon energy and polar angle respectively. We rejected all events with $q_{\pm}<1$ GeV  or $E_{\gamma}<1$ GeV for any ME photons, and with $q_{\pm}>1$ GeV and $E_{\gamma}>1$ GeV for any ISR photons. After generating the signal and background events, we performed a fast detector simulation using \texttt{Delphes3}~\cite{deFavereau:2013fsa} with the \texttt{ILCgen} configuration card.

The variations of the unpolarized signal cross section as a function of the DM mass and the cutoff scale are shown in \autoref{fig:gCS}---left and right panels, respectively--- for all three operator types, namely, SP (solid), VA (dashed) and TAT (dotted) types. We find that the cross section is the smallest (largest) for the SP (TAT)-type operator at any given DM mass. In the left panel, the sudden drop in the cross section as $m_\chi$ approaches $\sqrt s/2$ is due to phase-space suppression. Otherwise, for smaller DM masses, the cross section for a given operator type and a given cutoff scale is almost independent of the DM mass. In the right panel, we see that for a given DM mass the cross section drops as $\Lambda^{-4}$, as expected.      
As for the background, we find that the neutrino background cross section at $\sqrt s=1$ TeV is 6.63$\times10^3$ fb, while the radiative Bhabha background is 1.07$\times10^5$ fb before ME-ISR merging. After ME-ISR merging these cross sections reduce to 5.08$\times10^3$ fb and 7.68$\times10^4$ fb, respectively for neutrino-pair and Bhabha background. On the other hand, the DM signal cross section is found to be much smaller, as shown in \autoref{table:PolG} for a benchmark DM mass of $m_\chi=100$ GeV and the cutoff scale $\Lambda=3$ TeV. 
\begin{table*}[!t]%
\caption{\label{table:PolG}%
Comparison of the SM backgrounds and signal cross sections for the monophoton channel for the unpolarized beam, as well as for different choices of beam polarization at $\sqrt{s} = 1~\rm{TeV}$. For the signal, we have chosen benchmark values of $m_\chi = 100~\rm{GeV}$ and $\Lambda = 3~\rm{TeV}$. The numbers in bold highlight the optimal polarization choice for a given operator type.}
\begin{ruledtabular}
\begin{tabular}{ccccccc} 
 &  &  &  \multicolumn{4}{c}{Polarized cross section (fb)}   \\
  \cline{4-7}
 Process type & Unpolarized cross section (fb) & Polarization $P(e^{-},e^{+})$ & $(+,+)$ & $(+,-)$ & $(-,+)$ & $(-,-)$ \\
\colrule
           &  & $(80,0)$ & $1.15\times10^3$ & $1.15\times10^3$\phantom{1} & \phantom{1}$9.05\times10^3$\phantom{1} & \phantom{1}$9.05\times10^3$ \\
$\nu\overline{\nu}\gamma$ & $5.08\times10^3$ & $(80,20)$ & $1.31\times10^3$\phantom{1} & \phantom{1}$9.71\times10^2$\phantom{1} & \phantom{1}$1.08\times10^4$\phantom{1} & \phantom{1}$7.25\times10^3$ \\
           &  & $(80,30)$ & $1.40\times10^3$\phantom{1} & \phantom{1}$8.87\times10^2$\phantom{1} & \phantom{1}$1.18\times10^4$\phantom{1} & \phantom{1}$6.30\times10^3$ \\

\colrule

           &  & $(80,0)$ & $3.02\times10^4$\phantom{1} & \phantom{1}$3.02\times10^4$\phantom{1} & \phantom{1}$2.99\times10^4$\phantom{1} & \phantom{1}$2.99\times10^4$ \\           
$e^- e^+ \gamma$ & $2.97\times10^4$ & $(80,20)$ & $3.13\times10^4$\phantom{1} & \phantom{1}$2.99\times10^4$\phantom{1} & \phantom{1}$3.08\times10^4$\phantom{1} & \phantom{1}$2.93\times10^4$ \\
           &  & $(80,30)$ & $3.02\times10^4$\phantom{1} & \phantom{1}$3.01\times10^4$\phantom{1} & \phantom{1}$3.16\times10^4$\phantom{1} & \phantom{1}$2.93\times10^4$ \\

\colrule

           &  & $(80,0)$ & 31.6 & 31.6 & 31.6 & 31.6 \\
SP & $31.6$ & $(80,20)$ & 36.6 & 26.5 & 26.5 & 36.6 \\
           &  & $(80,30)$ & \bf{39.1} & 24.0 & 24.0 & 39.1 \\

\colrule

           &  & $(80,0)$ & 76.5 & 76.5 & 8.5 & 8.5 \\
VA & $42.5$ & $(80,20)$ & 61.2 & 91.8 & 6.8 & 10.2 \\
           &  & $(80,30)$ & 53.5 & \bf{99.4} & 6.0 & 11.1 \\

\colrule

           &  & $(80,0)$ & 45.0 & 45.0 & 45.0 & 45.0 \\
TAT & $45.0$ & $(80,20)$ & 52.2 & 37.8 & 37.8 & 42.2 \\
           &  & $(80,30)$ & \bf{55.8} & 34.2 & 34.2 & 55.8 \\

\end{tabular}
\end{ruledtabular}
\end{table*}  
\subsection{Effect of polarization} \label{sec:3.2} 

One important advantage of lepton colliders is that the incoming beams can be polarized. This helps to reduce the neutrino background considerably, as shown in \autoref{table:PolG}. To utilize the full advantage of the beam polarization, we investigate the effect of different choices of polarization on the signal and background. At the ILC, the baseline design foresees at least 80\% electron-beam polarization at the interaction point, whereas the positron beam can be polarized up to 30\% for the undulator positron source (up to 60\% may be possible with the addition of a photon collimator)~\cite{Bambade:2019fyw}. 
For comparison, we show our results for three different nominal absolute values of polarization: $|P(e^-,e^+)|=(80,0)$,  (80,20) and (80,30). In each case, we can also have four different polarization configurations, namely, ${\rm sign}(P(e^-),P(e^+))=(+,+)$, $(+,-)$, $(-,+)$, and $(-,-)$, where $+$ and $-$ denote the right- and left-handed helicities respectively. According to the H20 running scenario, the standard sharing between the different beam helicity configurations is 40\% for $(+,-)$ and $(-,+)$ each and 10\% for $(+,+)$ and $(-,-)$ each~\cite{Barklow:2015tja}. With the luminosity upgrade, the target integrated luminosities per beam helicity configuration at $1$ TeV machine energy is 3200 fb$^{-1}$ for $(+,-)$ and $(-,+)$ each and 800 fb$^{-1}$ for $(+,+)$ and $(-,-)$ each, with the proposed total luminosity of 8000 fb$^{-1}$~\cite{Barklow:2015tja}. In our analysis, we have considered the H20 scenario as well as the optimal beam polarization case for better comparison at $8000$ fb$^{-1}$.

In \autoref{table:PolG}, we show the effect of different schemes of polarizations and helicity orientations on the monophoton signal and background cross sections. It is evident from the table that the radiative Bhabha background remains almost unchanged, as mentioned earlier. On the other hand, electron-beam polarization is very effective in reducing the neutrino background, as an 80\% \emph{right}-handed electron beam can reduce the neutrino background to $23\%$ of the unpolarized case, even without any polarization on the positron beam. The effect is further enhanced by a \emph{left}-handed positron beam. We see that for $20\%$ and $30\%$ left-handed positron beam polarization, the neutrino background is reduced to $26\%$ and $18\%$ of its unpolarized value, respectively. 

The signals are also affected to some extent by beam polarization and the optimal helicity configuration depends on the operator type. For SP- and TAT-type operators we see no effect of electron-beam polarization, but a 20\% (30\%) right-handed positron beam can enhance the signal by 16\% (24\%). The VA-type signal, on the other hand, prefers the $(+,-)$ helicity configuration --- the same choice for which the neutrino background is minimized. With the $(+80\%,-30\%)$ configuration, the VA-type signal is enhanced by a factor of 2.3, whereas the $(+80\%,+30\%)$ configuration enhances it by a modest 26\%. In \autoref{table:PolG}, it can be noted that the cross section of SP- and TAT-types are same for $(+,+)$ and $(-,-)$ helicity orientations, but it can be seen that the neutrino background enhances with $(-,-)$ helicity, whereas the Bhabha background remains same. This effectively renders the $(-80,-30)$ polarization choice to be nonoptimal for these two operator types. In \autoref{table:PolG}, the optimal polarization choices that give the largest signal-to-background ratio are shown in boldfaced fonts.
%
%
\subsection{Cut-based analysis}\label{sec:3.3}
Now we analyze various kinematic distributions and perform a cut-based analysis to optimize the signal-to-background ratio. This of course depends on the DM mass, so in \autoref{table:BPs&CutsGamma}, we list three benchmark points (BPs) with $m_\chi=100$, 250, and 350 GeV, respectively, and present the corresponding selection cuts optimized for each case. Here we fix $\Lambda=3$ TeV for illustration, but in the next subsection, we will vary both $m_\chi$ and $\Lambda$ to obtain the $3\sigma$ sensitivity limits. As for the choice of the DM mass values, since it was seen from \autoref{fig:gCS} that the signal cross sections are barely sensitive to the DM mass up to around $100$ GeV, our BP1 essentially captures the light DM scenario. Similarly, our BP3 is chosen moderately close to the kinematic limit of $\sqrt s/2$ (going too close to $\sqrt s/2$ will result in cross section values too low to give sizable event counts after all the selection cuts). BP2 is chosen for an intermediate DM mass in between BP1 and BP3. 
%
\begin{table*}[!t]
\caption{\label{table:BPs&CutsGamma}%
Monophoton selection cuts for different BPs across all operator types. Because of the dynamic nature of these cuts, the backgrounds will have to be separately analyzed for each BP.}
\begin{tabular}{@{}lccc@{}} \toprule
  & {\bf BP1} & {\bf BP2} & {\bf BP3} \\
  \colrule 
  \multirow{2}{4.5em}{\textbf{Definition}} & $m_{\chi} = 100\text{ GeV, }$ & $m_{\chi} = 250\text{ GeV, }$ & $m_{\chi} = 350\text{ GeV, }$ \\
  & $\Lambda = 3\text{ TeV}$ & $\Lambda = 3\text{ TeV}$ & $\Lambda = 3\text{ TeV}$ \\
  \colrule
  \multicolumn{4}{l}{\textbf{SP-type / VA-type / TAT-type}} \\
  \colrule
  Baseline selection & \multicolumn{3}{c}{$E_{\gamma} > 5\text{ GeV},\;\; |\eta_{\gamma}| < 2.8,$} \\
  Cut 1 & $E_{\gamma} < 450\text{ GeV}$\phantom{00} & \phantom{0}$E_{\gamma} < 350\text{ GeV}$\phantom{00} & \phantom{0}$E_{\gamma} < 250\text{ GeV}$ \\
  Cut 2 & \multicolumn{3}{c}{$|\cos\theta_{\gamma}| < 0.93$} \\
  Cut 3 & \multicolumn{3}{c}{Charged particle veto with $p_{T} > 5 $ GeV} \\
  Cut 4 &  \multicolumn{3}{c}{BeamCal veto} \\
\botrule
\end{tabular}
\end{table*}
\begin{figure*}[!t]
\centering 
\includegraphics[width=0.325\linewidth]{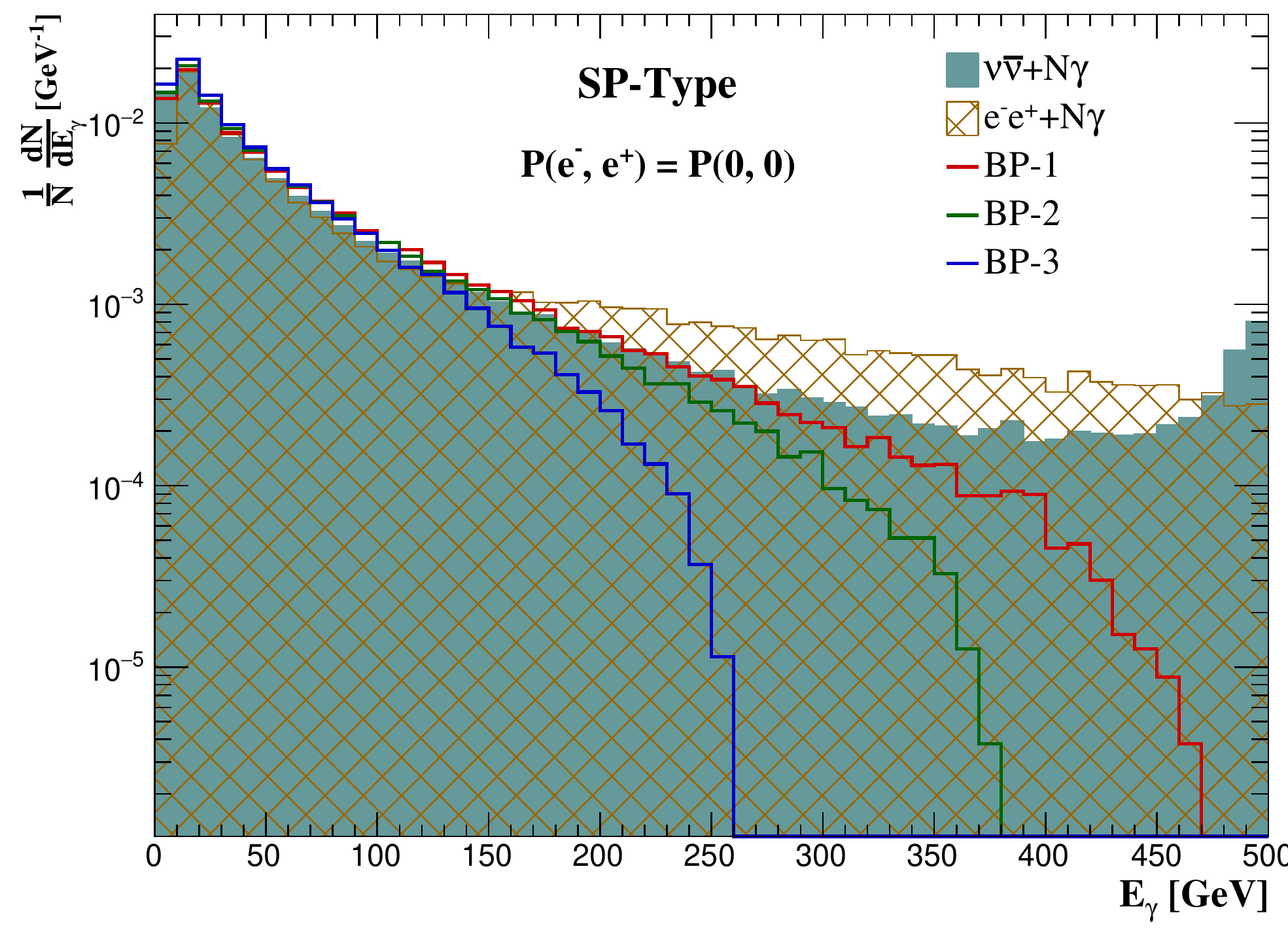}
\includegraphics[width=0.325\linewidth]{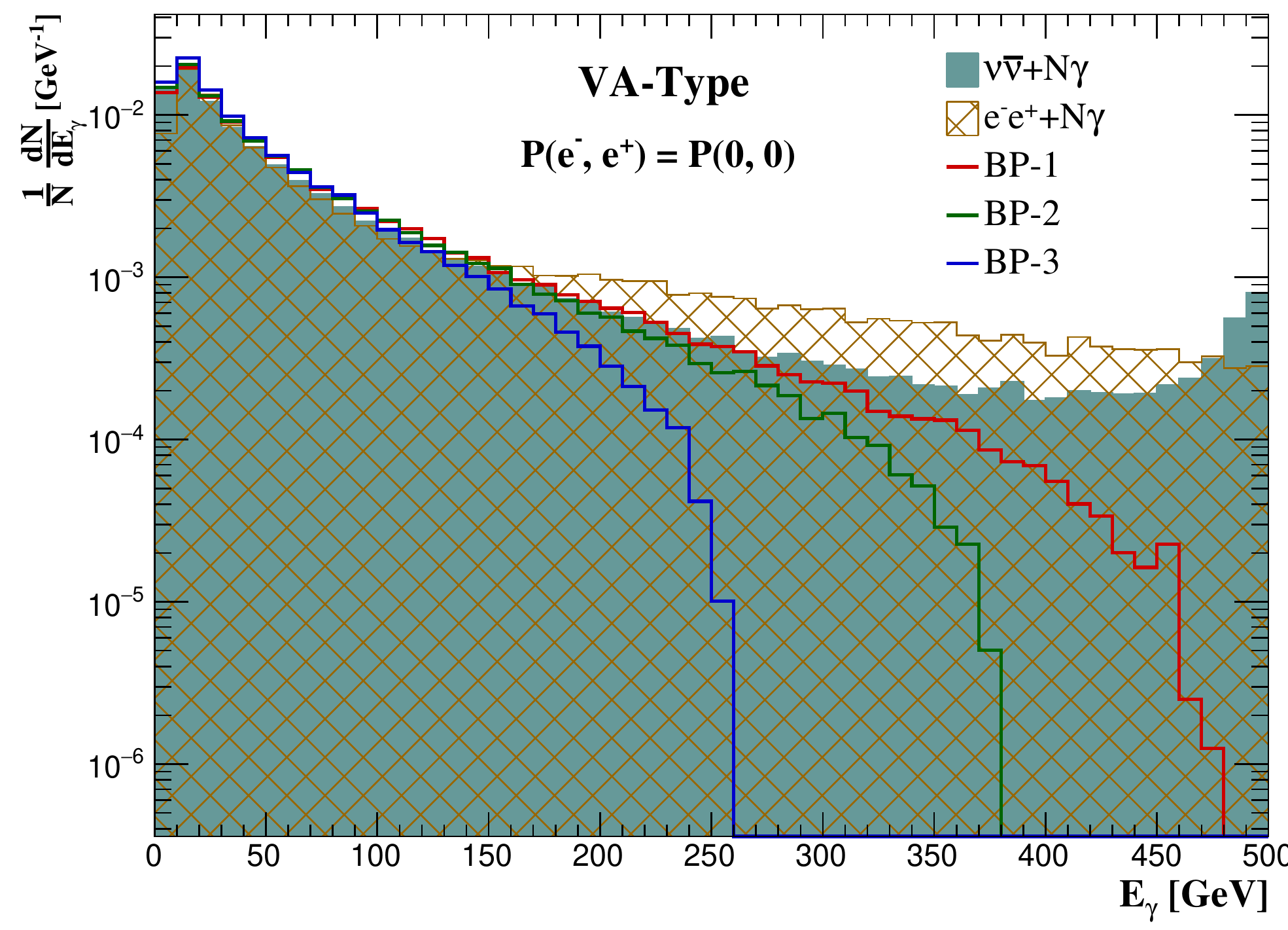}
\includegraphics[width=0.325\linewidth]{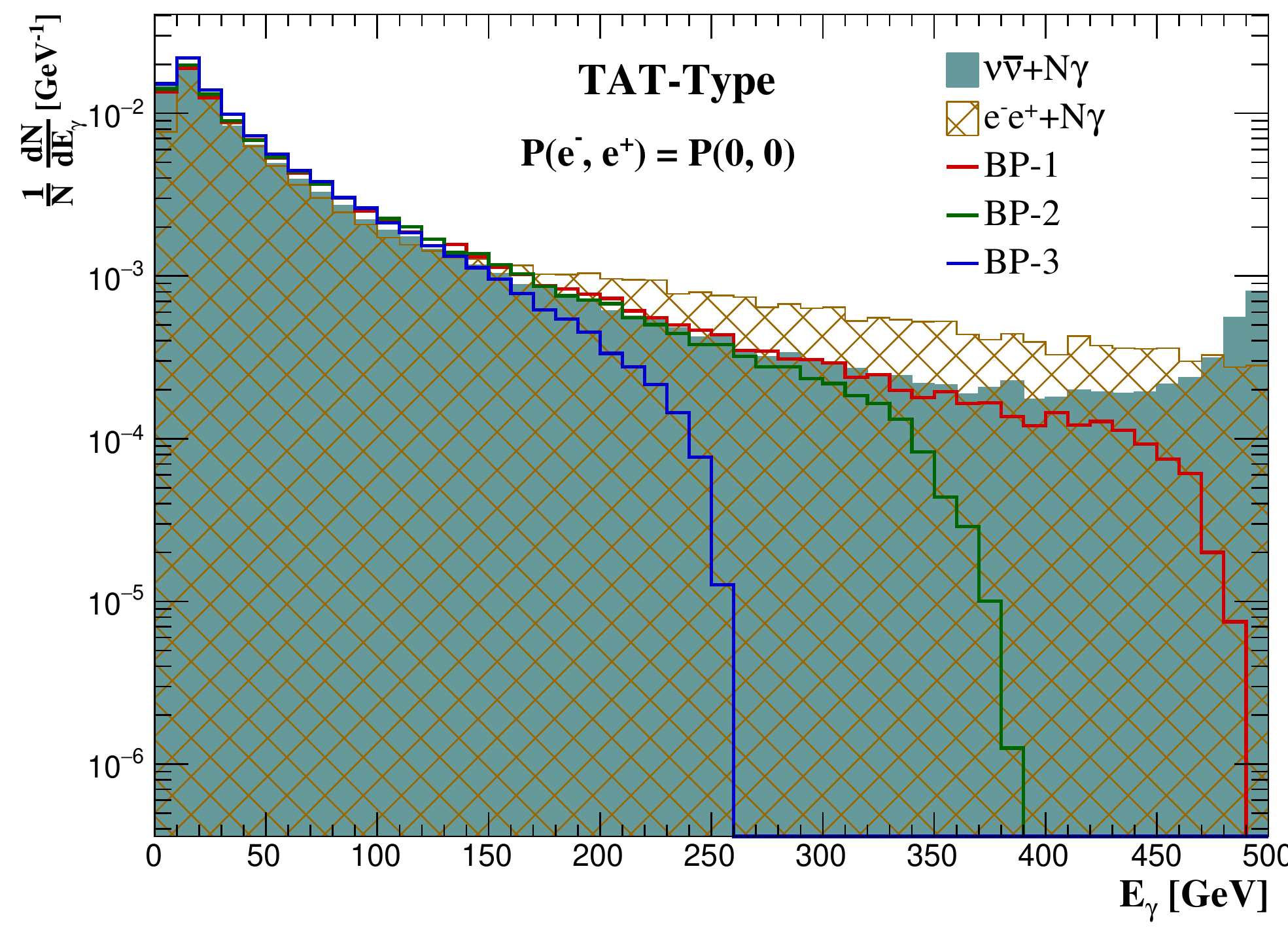}
\includegraphics[width=0.325\linewidth]{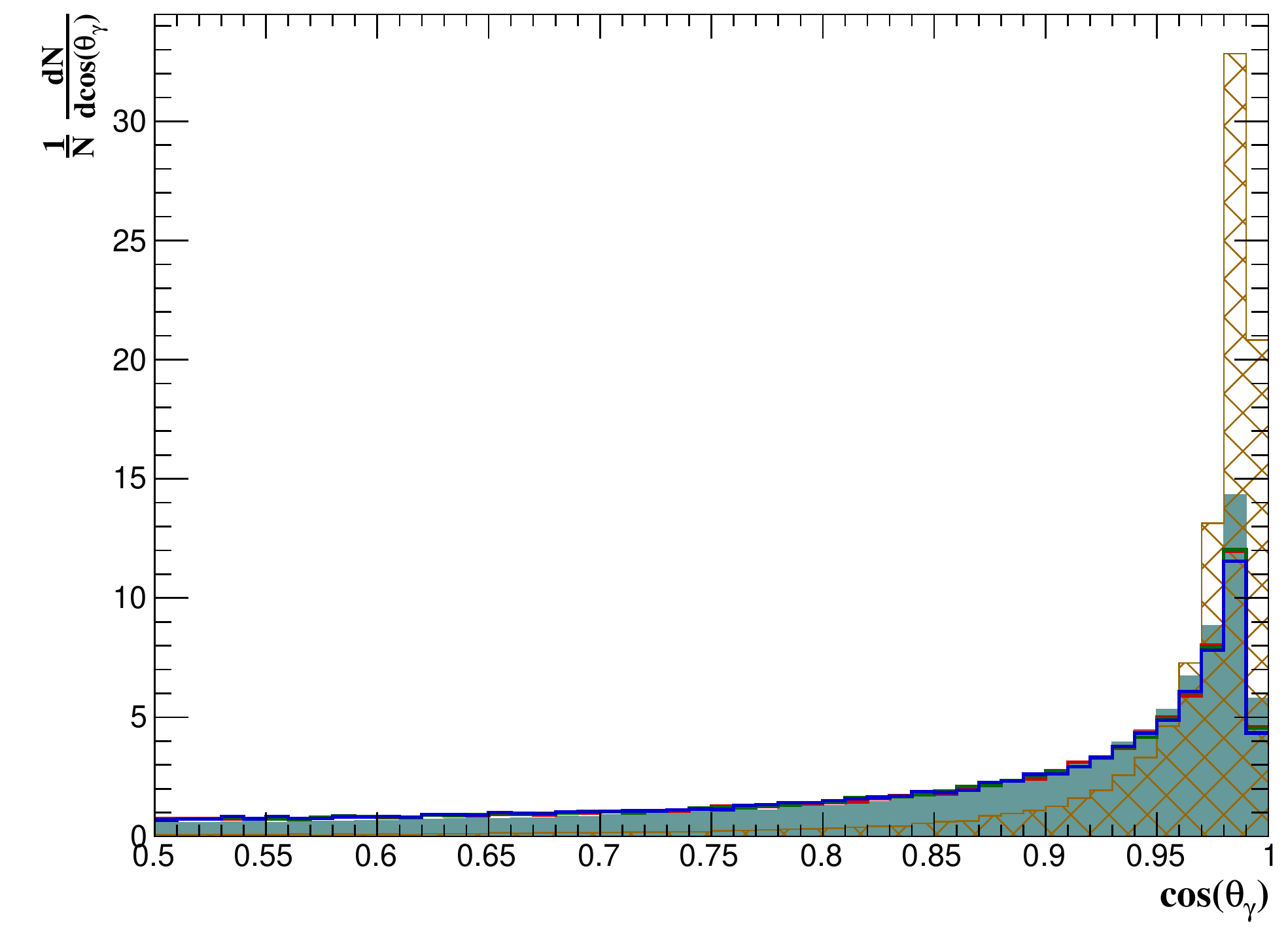}
\includegraphics[width=0.325\linewidth]{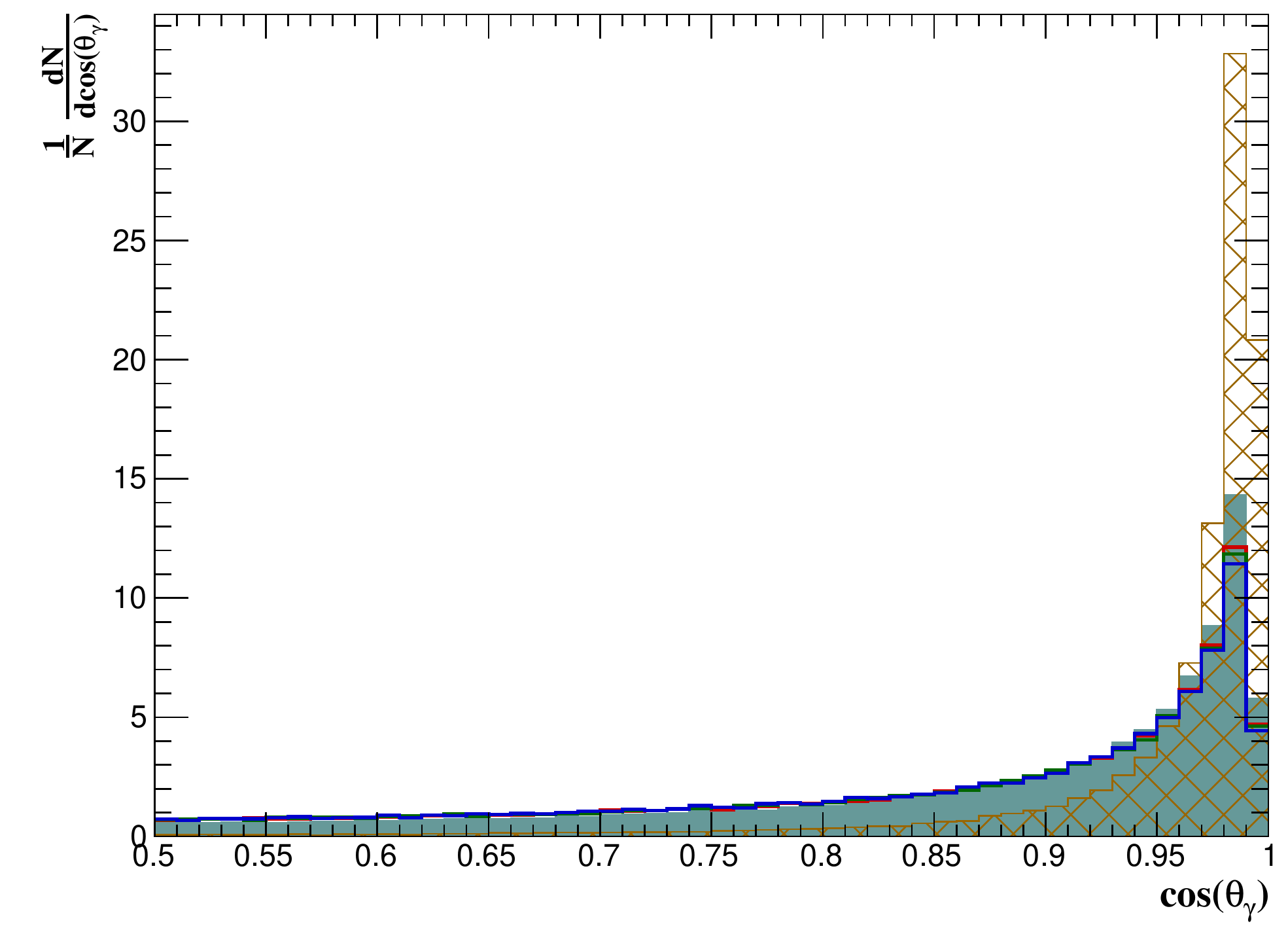}
\includegraphics[width=0.325\linewidth]{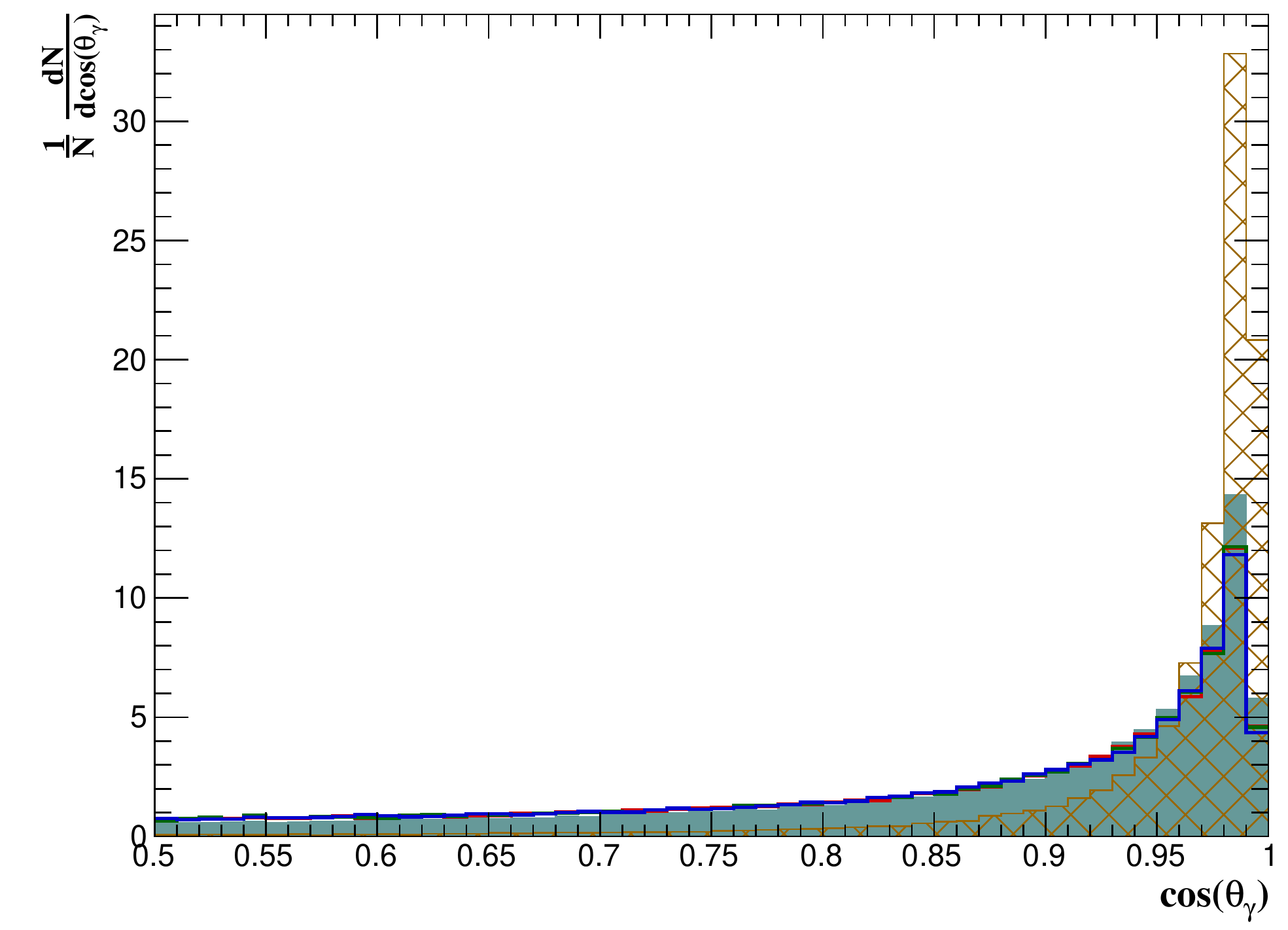}
\caption{Normalized differential distributions after baseline selection cuts for the set of five kinematic observables considered in \autoref{table:BPs&CutsGamma}, namely, $E_\gamma$, $\cos\theta_{\gamma}$ for the unpolarized case (the polarized case histograms are in the Appendix, see \autoref{figure:HistosPhpol}). In each panel, the shaded green histograms are for the neutrino background and the cyan hatched histograms are for the radiative Bhabha background. The unshaded red, green and blue histograms are for BP1, BP2, and BP3, respectively. The left, middle, and right panels correspond to the SP-, VA-, and TAT-type operators, respectively.}
\label{figure:HistosPh}
\end{figure*}

\begin{table*}[t]
\caption{\label{table:CutBGG}%
Cut-flow chart for different operators and BPs for the monophoton signal and backgrounds. The cut efficiencies are calculated with respect to the baseline selection (BS) cuts given in Eq.~\eqref{eq:baseline}, with the corresponding event numbers shown in the first row. Note that depending on the operator type we are interested in for the signal, the selection cuts are different (cf.~\autoref{table:BPs&CutsGamma}). Therefore, the corresponding backgrounds will also vary as shown here.}
  \begin{tabular}{l ccc ccc ccc}
\toprule
\multirow{3}{4.5em}{\bf Selection cuts} & \multicolumn{9}{c}{\bf Event Numbers ~(Cut efficiencies)}\\
  \cline{2-10}
  & \multicolumn{3}{c}{Neutrino pair} & \multicolumn{3}{c}{Radiative Bhabha} & 
  \multicolumn{3}{c}{Signal} \\
  \cline{2-10}
  & BP1 & BP2 & BP3 & BP1 & BP2 & BP3 & BP1 & BP2 & BP3 \\
  \colrule
  \multicolumn{10}{l}{\bf{SP-type:}}\\ 
  \colrule
  \multirow{2}{4.5em}{BS} & \multicolumn{3}{c}{$3.09\times10^7$} & \multicolumn{3}{c}{$6.70\times10^7$} & $2.01\times10^5$ & $1.45\times10^5$ & $8.74\times10^4$ \\
  & \multicolumn{3}{c}{$(100\%)$} & \multicolumn{3}{c}{$(100\%)$} & $(100\%)$ & $(100\%)$ & $(100\%)$ \\
  \multirow{2}{4.5em}{After Final cut} 
   & $1.76\times10^7$ & $1.73\times10^7$ & $1.70\times10^7$ & $6.08\times10^5$ & $6.08\times10^5$ & $6.08\times10^5$ & $1.24\times10^5$ & $ 8.97\times10^4 $ & $ 5.50\times10^4 $ \\
   & $(56.75\%)$ & $(56.04\%)$ & $(54.95\%)$ & $(0.91\%)$ & $(0.91\%)$ & $(0.91\%)$ & $(61.68\%)$ & $(61.97\%)$ & $(62.97\%)$ \\
  \colrule
  \multicolumn{10}{l}{\bf{VA-type:}}\\ 
  \colrule
  \multirow{2}{4.5em}{BS} & \multicolumn{3}{c}{$3.09\times10^7$} & \multicolumn{3}{c}{$6.70\times10^7$} & $2.71\times10^5$ & $2.10\times10^5$ & $1.41\times10^5$ \\
  & \multicolumn{3}{c}{$(100\%)$} & \multicolumn{3}{c}{$(100\%)$} & $(100\%)$ & $(100\%)$ & $(100\%)$ \\

  \multirow{2}{4.5em}{After Final cut} 
   & $1.76\times10^7$ & $1.73\times10^7$ & $1.70\times10^7$ & $6.08\times10^5$ & $6.08\times10^5$ & $6.08\times10^5$ & $ 1.66\times10^5 $ & $ 1.30\times10^5 $ & $ 8.84\times10^4 $ \\
   & $(56.75\%)$ & $(56.04\%)$ & $(54.95\%)$ & $(0.91\%)$ & $(0.91\%)$ & $(0.91\%)$ & $(61.07\%)$ & $(61.97\%)$ & $(62.78\%)$ \\
  \colrule
  \multicolumn{10}{l}{\bf{TAT-type:}}\\ 
  \colrule
  \multirow{2}{4.5em}{BS} & \multicolumn{3}{c}{$3.09\times10^7$} & \multicolumn{3}{c}{$6.70\times10^7$} & $2.88\times10^5$ & $2.67\times10^5$ & $2.13\times10^5$ \\
  & \multicolumn{3}{c}{$(100\%)$} & \multicolumn{3}{c}{$(100\%)$} & $(100\%)$ & $(100\%)$ & $(100\%)$ \\

  \multirow{2}{4.5em}{After Final cut} 
   & $1.76\times10^7$ & $1.73\times10^7$ & $1.70\times10^7$ & $6.08\times10^5$ & $6.08\times10^5$ & $6.08\times10^5$ & $ 1.78\times10^5 $ & $ 1.64\times10^5 $ & $ 1.34\times10^5 $ \\
   & $(56.75\%)$ & $(56.04\%)$ & $(54.95\%)$ & $(0.91\%)$ & $(0.91\%)$ & $(0.91\%)$ & $(61.77\%)$ & $(61.67\%)$ & $(62.64\%)$ \\
\botrule
\end{tabular}
\end{table*}

Baseline selection cuts:
We define our monophoton signals by those events that pass through the baseline selection criteria as defined below, in addition to the cuts given in Eq.~\eqref{eq:CutsPh},

\begin{equation}
E_{\gamma} > 5\text{ GeV},\;\; |\eta_{\gamma}| < 2.8
\label{eq:baseline}
\end{equation}

where the photon with highest transverse momentum in an event is considered as the signal photon. 

After implementing these baseline selection criteria, the signal and background events (selection efficiencies) are given in the first rows of \autoref{table:CutBGG}. We find that the signal and the neutrino background are reduced to about $80\%$ and $76\%$ of their original values, respectively as in \autoref{table:PolG}. Similarly, the actual Bhabha-induced background relevant for our signal is found to be only about 34\% of its original value quoted in \autoref{table:PolG} after the baseline selection cuts. To further enhance our signal-to-background ratio, we then examine the signal versus background distributions of some relevant kinematic variables and devise further cuts. The normalized distributions are given in \autoref{figure:HistosPh} for the unpolarized case. The corresponding distributions for the optimally polarized case are shown in \autoref{figure:HistosPhpol} in the Appendix.  
From \autoref{figure:HistosPh}, we see some variations of the signal distributions between the three BPs. Therefore, the specialized selection criteria discussed below are dynamic with respect to different BPs, as also summarized in \autoref{table:BPs&CutsGamma}.

Cut 1: 
 From the $E_\gamma$ distribution in \autoref{figure:HistosPh} (top row), we see that although the neutrino background is mostly signallike, it has a large number of events around $E_\gamma\approx496$ GeV, due to the radiative return of the $Z$ resonance which is determined by $(s-M_Z^2)/2\sqrt{s}$. On the other hand, the $E_\gamma$ distribution for the signal drops at a lower value, depending on the DM mass, due to phase-space suppression. Therefore, we impose a dynamic maximum $E_\gamma$ cut (cf.~\autoref{table:BPs&CutsGamma}) to account for both the effects. %
Cut 2: 
Detection of collinear photons is not possible. The limitation of the detector reflects in the baseline pseudorapidity cut $|\eta_\gamma| < 2.8$ or $7^\circ<\theta <173^\circ$. However, for the purpose of background suppression, a stricter cut in the form of $\cos{\theta_\gamma}$ was applied, based on the event distribution in \autoref{figure:HistosPh}. This cut is particularly effective in reducing the Bhabha background significantly to about 16\%-17\% of its yield after baseline selection, whereas the neutrino-pair background and signal drop to about 56\%-58\% and 65\%-66\% respectively. 

Cut 3: 
Next we apply a veto condition on the charged particles that are reconstructed as electrons as well as muon and jets. Here, we selected the jets and muons to have $p_T>5$ GeV and $|\eta|<2.8$, whereas for the electrons we required $p_T>2$ GeV. This kind of selection criteria was also used in Ref.~\cite{Habermehl:2018yul} which substantially reduces the Bhabha background. We note that after this cut the Bhabha event yield drops from around $17\%$ (after cut 2) to approximately $7\%$.

Cut 4: 
Finally, we apply a veto condition to reject all the events that have energy deposits in the electromagnetic calorimeter in the very forward direction of the beamline (BeamCal)~\cite{Abramowicz:2010bg}. This calorimeter is useful in detecting and tagging the highly energetic electrons from the Bhabha events that otherwise escape the central tracker, thereby minimizing their contribution to the background. As can be seen in \autoref{table:CutBGG}, the Bhabha background is now reduced to $\sim2\%$, whereas neutrino-pair background and signal events are left at $\sim55\%$ and $\sim60\%$ of their respective numbers after baseline selection. 

The above-mentioned cuts are optimized for the unpolarized beam case, and so in the cut-flow table we only show the numbers for the unpolarized case. A similar table for polarized case can be found in the Appendix in \autoref{table:CutBGGpol}, where we see that the final selection efficiencies are similar. However, as shown in the \autoref{table:PolG}, due to enhanced cross sections of the signals and reduced production of the neutrino-pair background (while Bhabha remains the same) in the respective optimal polarization scenarios of the operators, it translates into significantly better signal significance than the unpolarized case as discussed in the next section. 

\subsection{Signal significance} \label{sec:3.4}
\begin{table*}[htb]
\caption{Signal significance in the monophoton channel at $\sqrt{s}=1$ TeV with $\mathcal{L}_{\rm int} = 8~{\rm ab}^{-1}$ integrated luminosity. These are the optimized numbers after implementing all the selection cuts. The values in the parentheses denote the significances with background systematic  uncertainty.}
\begin{ruledtabular}
\begin{tabular}{lccc ccc ccc} 
  \multirow{3}{3.5em}{} & \multicolumn{9}{c}{Signal significance for ${\cal L}_{\rm int}=8\,{\rm ab}^{-1}$}\\
   \cline{2-10}
   & \multicolumn{3}{c}{Unpolarized beams} & \multicolumn{3}{c}{H20 scenario} & \multicolumn{3}{c}{Optimally polarized beams}\\
   \cline{2-10}
  Operator type & BP1 & BP2 & BP3 & BP1 & BP2 & BP3 & BP1 & BP2 & BP3 \\
  \colrule
  SP & $29.0\;(3.5)$ & $21.1\;(2.5)$ & $13.1\;(1.6)$ 
  & $40.1\;(15.5)$ & $29.7\;(11.5)$ & $18.5\;(7.1)$  
  & $64.6\;(12.6)$ & $47.8\;(9.3)$ & $29.8\;(5.8)$ \\
  VA & $38.7\;(4.6)$ & $30.6\;(3.7)$ & $21.0\;(2.5)$ 
  & $129.3\;(45.6)$ & $104.3\;(36.5)$ & $72.6\;(25.2)$   
  & $199.3\;(44.7)$ &$161.1\;(35.8)$ & $112.3\;(24.6)$ \\
  TAT & $41.5\;(5.0)$ & $38.6\;(4.6)$ & $31.7\;(3.8)$ 
  & $57.6\;(22.4)$ & $54.2\;(21.0)$ & $44.5\;(17.3)$   
  & $92.6\;(18.2)$ & $86.8\;(17.1)$ & $71.5\;(14.1)$ \\
   
 \end{tabular}
\end{ruledtabular}
\label{table:SigG}%
\end{table*}

After implementing all the cuts mentioned above, we calculate the final signal significance for our benchmark scenarios using the definition
\begin{equation}
    {\rm Sig} = \frac{S}{\sqrt{S+ \sum_i \left(B_i+\epsilon_i^2 B_i^2\right)}} \, ,
     \label{eq:significance}
\end{equation}
where $S$ and $B_i$ are the number of signal and ($i$th channel) background  events, respectively, for a given integrated luminosity, and  $\epsilon_i$ is the corresponding systematic uncertainty in the background estimation. Our results are given in \autoref{table:SigG} for the three BPs. 
We show the numbers for an ideal case with zero systematics and also for a more conservative case with 0.2\% and 1\% systematics for the dominant and subdominant backgrounds, respectively (in parentheses). 
The results are significantly weakened in the latter case because of the relatively large background compared to the signal.
For the mixed polarization case in H20 scenario, we calculated the significances for individual polarization cases and assumed them to be uncorrelated, calculated the final signal significance as 
\begin{equation}
    {\rm Sig_{\,H20}} = \sqrt{{\rm Sig}_{(+,+)}^2+{\rm Sig}_{(+,-)}^2+{\rm Sig}_{(-,+)}^2+{\rm Sig}_{(-,-)}^2}
    \label{eq:simp}
\end{equation}

\par
The possible sources of systematic uncertainties at the ILC are \cite{Habermehl:2018yul,Habermehl:2020njb, Kalinowski:2021tyr} (i) event  estimation, (ii) integrated luminosity, (iii) beam polarization, and (iv) shape of the luminosity spectrum. Here we did a simplified analysis and considered only uncertainties in background estimation in a cut-and-count  method. We considered here an uncertainty of $0.2\%$ for the neutrino-pair background and $1\%$ for the radiative Bhabha background \cite{Habermehl:2018yul,Blaising:2021vhh,Kalinowski:2020lhp}. We admit that for a more realistic uncertainty analysis one should consider all the sources of uncertainty based on the event distributions and their possible correlations. Also it is possible to constrain systematic uncertainties with the data alone, i.e., by looking at the $Z$-return peak~\cite{Habermehl:2018yul}. Since, in practice, uncertainty in background normalization is $100\%$ correlated between different polarizations, its impact on the signal sensitivity will be reduced compared to our predicted sensitivities. However, such a detailed analysis of systematic uncertainties is beyond the scope of our work. When the actual experiment is performed, we expect the realistic sensitivities to lie somewhere between the two cases presented here, i.e., the ideal situation with no systematics and the conservative case with \cmmnt{1\%} our choice of systematics.  

\par
From \autoref{table:SigG}, we see that the significance enhances as we go to lower DM mass regions, as expected because of kinematic reasons. Operatorwise we see that TAT- and VA-type operators perform better than the SP-type. We also find substantial increase in significance on application of optimal beam polarization. We note that the optimal beam polarization gives the best significance when there are no systematics, but the mixed polarization case of the H20 running scenario gives better significance when uncertainty in background estimation is taken into account.

Using a machine-learning-based multivariate (MV) analysis might improve the signal significance over that obtained here using a simple cut-based analysis, but a detailed MV analysis is beyond the scope of the current work and will be reported in a follow-up study.  

\begin{figure*}[!t]
\centering 
\includegraphics[width=0.51\linewidth]{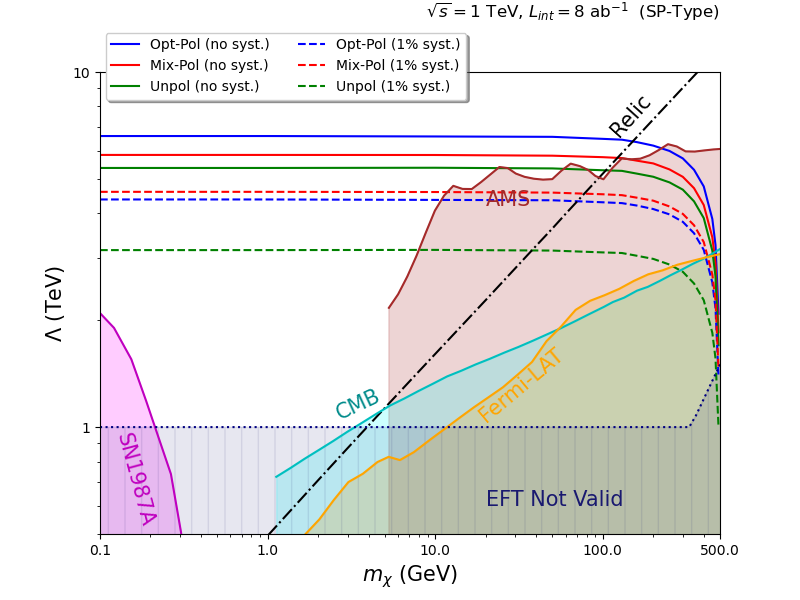}
\hspace{-0.7cm}
\includegraphics[width=0.51\linewidth]{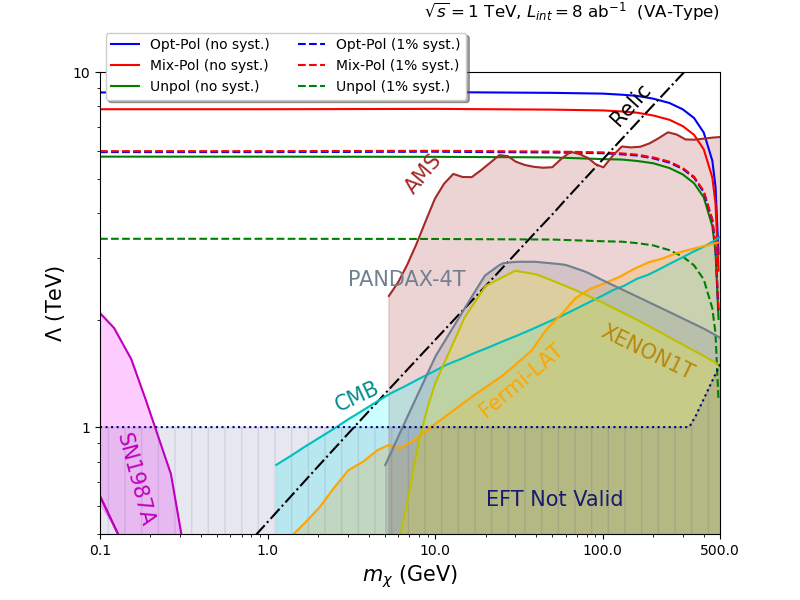}
\includegraphics[width=0.51\linewidth]{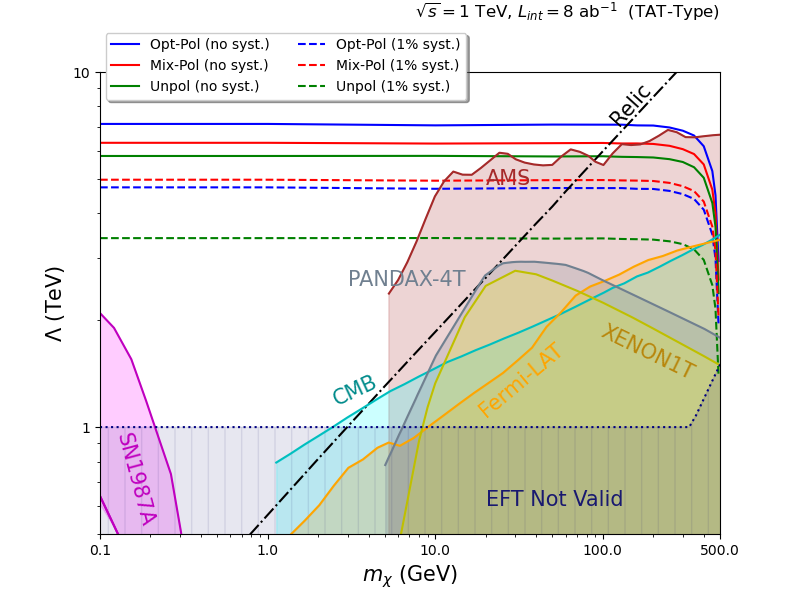}
\caption{$3\sigma$ sensitivity contours in the monophoton channel for the SP-type (top left panel), VA-type (top right panel) and TAT-type (bottom panel) operators with unpolarized (Unpol, green lines), mixed polarized (Mix-Pol, red lines) and optimally polarized (Opt-pol, blue lines) $e^+e^-$ beams at $\sqrt s=1$ TeV center-of-mass energy and with  ${\cal L}_{\rm int}=8$ ab$^{-1}$ integrated luminosity. The solid (dashed) contours are assuming zero (1\%) background systematics. The various shaded regions are excluded by direct detection (XENON1T, PANDAX-4T), indirect detection (Fermi-LAT, AMS), astrophysics (SN1987A) and cosmology (CMB) constraints. In the shaded region below $\Lambda={\rm max}\{\sqrt{s},3m_\chi\}$, our EFT framework is not valid. Along the dot-dashed line, the observed DM relic density is reproduced for a thermal WIMP assuming only DM-electron effective coupling. }
\label{figure:ContourG}
\end{figure*}

Going beyond the three BPs, we now vary the DM mass and calculate the signal significance following the same cut-based analysis procedure outlined above. Our results for the $3\sigma$ sensitivity contours in the $(m_\chi,\Lambda)$ plane are shown in \autoref{figure:ContourG} for all the operator types. The green, blue and red contours are for the unpolarized, optimal and mixed polarization cases, respectively, and the solid (dashed) contours are assuming zero (1\%) background systematics. The shaded regions are excluded by various constraints, as follows: first of all, for $\Lambda<{\rm max}\{\sqrt s,3m_\chi\}$, our EFT framework is not valid (cf.~\autoref{sec:2}). This is shown by the navy-blue-shaded regions in \autoref{figure:ContourG}. For $\sqrt s=1$ TeV as considered here, this EFT validity limit supersedes the previous limit from the Large Electron-Positron collider~\cite{Fox:2011fx}. 

The same effective operator given in Eq.~\eqref{eq:EFT} also gives rise to DM scattering with electrons $\chi e^-\to \chi e^-$. The exact analytic expressions for these cross sections in our EFT framework can be found in Appendix C of Ref.~\cite{Guha:2018mli} for all the operator types. Up to velocity or electron mass suppression, the typical size of the cross section is given by 
\begin{align}
    \sigma^0_{\chi e} \simeq \frac{m_e^2}{\pi \Lambda^4} \approx 3.2\times 10^{-47}{\rm cm}^2\left(\frac{1~{\rm TeV}}{\Lambda}\right)^4 \, .
    \label{eq:direct-e}
\end{align}
Comparing this with the experimental upper limits on $\sigma_{\chi e}$ from dedicated direct detection experiments~\cite{XENON100:2015tol, XENON:2019gfn}, we can derive a {\it lower} limit on the cutoff scale $\Lambda$ as a function of the DM mass $m_\chi$. However, the current best limit on $\sigma_{\chi e}$ from XENON1T is at the level of $\mathcal{ O}(10^{-39})$ cm$^2$~\cite{XENON:2019gfn}, which translates into a very weak bound on $\Lambda$ and is not relevant for our study. Even the future ambitious proposals like DARKSPHERE can only reach up to ${\cal O}(10^{-42})\:{\rm cm}^2$~\cite{Hamaide:2021hlp}, still 5 orders of magnitude weaker than that needed to probe a TeV-scale $\Lambda$ value [cf.~Eq.~\eqref{eq:direct-e}]. 

\begin{figure*}[htb]
\centering
 \includegraphics[width=0.9\linewidth]{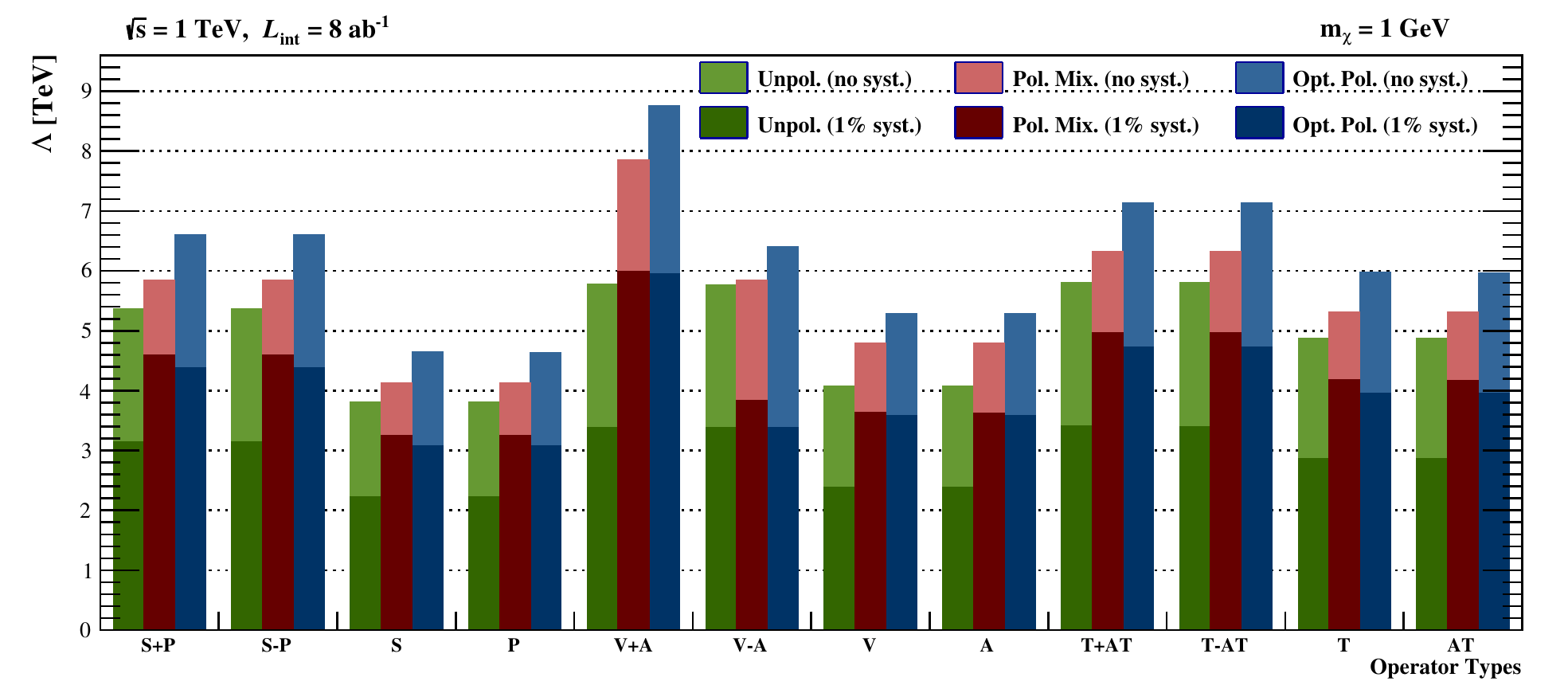}
 \caption{$3\sigma$ sensitivity reach of different operators in the monophoton channel with DM mass of $1$ GeV at $\sqrt{s}=1$ TeV and $\mathcal{L}_{\rm int}=8~{\rm ab}^{-1}$. The green, red and blue bars show the corresponding sensitivities with unpolarized, mixed polarized and optimally polarized beams respectively, and the lighter (darker) shade corresponds to zero (1\%) background systematics.}
 \label{fig:barG}
\end{figure*}

However, more stringent limits can be derived from DM-nucleon scattering searches. Even for a leptophilic DM as in our case, DM-nucleon couplings are necessarily induced at loop level from photon exchange between virtual leptons and the quarks. In fact, as shown in Ref.~\cite{Kopp:2009et}, the loop-induced DM-nucleon scattering almost always dominates over the DM-electron scattering. The analytic expressions for the one- and two-loop DM-nucleon scattering cross sections can be found in Ref.~\cite{Kopp:2009et}. The typical size of the one-loop cross section is~\cite{Barman:2021hhg} 
\begin{align}
   \sigma^1_{\chi N} &\simeq  \frac{\mu_N^2 c_e^2}{9\pi A^2}\left(\frac{\alpha_{\rm em}Z}{\pi \Lambda^2}\right)^2\left[\log\left(\frac{\Lambda^2}{\Lambda_{\rm NR}^2}\right)\right]^2 \nonumber\\
   &\approx \, 1.5\times 10^{-45}{\rm cm}^2\left(\frac{\mu_N}{10~{\rm GeV}}\right)^2\left(\frac{1~{\rm TeV}}{\Lambda}\right)^4
   \label{eq:direct}
\end{align}
where $\mu_N=m_Nm_\chi/(m_N+m_\chi)$ is the reduced mass of the DM-nucleon system, $Z$ and $A$ are, respectively, the atomic number and atomic mass of the given nucleus, $\alpha_{\rm em}$ is the fine-structure constant, $\Lambda_{\rm NR}$ is the direct detection scale, and $c_e$ is the operator coefficient value at the scale $\Lambda_{\rm NR}$. In the last step of Eq.~\eqref{eq:direct}, we have used $\Lambda_{\rm NR}=2$ GeV and $c_e=0.965$ following Ref.~\cite{Barman:2021hhg}. Since the most stringent DM-nucleon spin-independent cross section limits come from XENON1T~\cite{XENON:2018voc} and PANDAX-4T~\cite{PandaX-4T:2021bab} experiments, we have also set $Z=54$ and $A=131$ for the xenon nucleus and have translated the experimental upper limits onto the $(m_\chi,\Lambda)$ plane, as shown by the yellow- and grey-shaded regions, respectively, in \autoref{figure:ContourG}. Note that these limits are only applicable for the vector and tensor lepton currents, i.e. $\Gamma_\ell=\gamma_\mu,\ \sigma_{\mu\nu}$ in Eq.~\eqref{eq:EFT}. For the scalar lepton current, $\Gamma_\ell=1$, the one-loop DM-nucleon coupling vanishes, and one has to go to two loops, which is suppressed by $\alpha_{\rm em}^2$ for the scalar-scalar-type coupling and $\alpha_{\rm em}^2v^2$ (where $v\sim 10^{-3}$ is the DM velocity) for the pseudoscalar-scalar-type coupling. In contrast, for pseudoscalar and axial-vector lepton currents, i.e., $\Gamma_\ell=\gamma_5, \ \gamma_\mu \gamma_5$, the DM-nucleon coupling vanishes to all orders. Therefore, we have not shown the XENON1T and PANDAX-4T limits for the SP-type operator on the top left panel of \autoref{figure:ContourG}.          
The same effective operator given in Eq.~\eqref{eq:EFT} also enables DM annihilation into electrons $\chi \overline{\chi} \to e^+ e^-$. The exact analytic expressions for these cross sections in our EFT framework can be found in Appendix C of Ref.~\cite{Guha:2018mli} for all the operator types. Using these, we calculate the thermal-averaged cross section times relative velocity $\langle \sigma v\rangle$, which goes as $m_\chi^2/\Lambda^{4}$, and compare it with the existing indirect detection upper limits on $\langle \sigma v\rangle$ in the $e^+e^-$ channel to put a lower bound on $\Lambda$ as a function of the DM mass. This is shown in \autoref{figure:ContourG} by the red- and brown-shaded regions, respectively, for the Fermi-LAT~\cite{Leane:2018kjk} and AMS-02~\cite{John:2021ugy} constraints on $\langle \sigma v\rangle$. Similar constraints on $\langle \sigma v\rangle$ can be derived using CMB anisotropies~\cite{Leane:2018kjk}, which is shown by the cyan-shaded region in \autoref{figure:ContourG}, assuming an $s$-wave annihilation (for $p$-wave annihilation, the CMB bound will be much weaker).  

Along the dot-dashed line in \autoref{figure:ContourG}, the observed relic density can be reproduced for a WIMP DM. In principle, the region to the left and above of this line is disfavored for a thermal WIMP, because in this region $\langle \sigma v\rangle$ is smaller than the observed value of $\sim (2-5)\times 10^{-26}$ cm$^3$ sec$^{-1}$ (depending on the DM mass~\cite{Steigman:2012nb}), which leads to an overabundance of DM, since $\Omega_\chi h^2 \propto 1/\langle \sigma v\rangle$. However, this problem can be circumvented by either opening up additional leptonic annihilation channels (like $\mu^+\mu^-$,  $\tau^+\tau^-$, and $\nu\bar{\nu}$) or even going beyond the WIMP paradigm and invoking, e.g., the freeze-in mechanism~\cite{Hall:2009bx}. This will not affect the main results of our paper, since the collider phenomenology discussed here only depends on the DM coupling to electrons. 
Also shown in \autoref{figure:ContourG} is the supernova constraint, which excludes the magenta-shaded region from consideration of energy loss and optical depth criteria from the observation of SN1987A~\cite{Guha:2018mli}. Here we have used an average supernova core temperature of 30 MeV. Note that the supernova bound is only applicable for DM mass below $\sim$ 200 MeV or so, and for a certain range of $\Lambda$ values, above which the DM particles cannot be efficiently produced in the supernova core, and below which they will no longer free stream.  

From \autoref{figure:ContourG}, we find that, in spite of a large irreducible background, the accessible range of the cutoff scale $\Lambda$ at $\sqrt s=1$ TeV ILC looks quite promising in the monophoton channel, especially for low-mass DM, where the collider sensitivity is almost flat, whereas the existing direct and indirect detection constraints are much weaker. This complementarity makes the collider searches for DM very promising. With unpolarized beams, the 3$\sigma$ reach for the SP-type operator can be up to 5.4 TeV, while for the VA- and TAT-type operators, it can be up to 5.8 TeV. With optimally polarized beams, i.e., with $(+80\%,+30\%)$ for the SP- and TAT-types and $(+80\%,-30\%)$ for the VA type, the sensitivity reach can be extended to 6.6 (SP), 8.8 (VA), and 7.1 TeV (TAT), as shown in \autoref{figure:ContourG}. This increase in sensitivity results mainly from the background suppression due to the beam polarization. 
But the systematic uncertainties in background estimation play a crucial role here. It is clearly visible from \autoref{figure:ContourG} that the unpolarized case suffers the most due to the systematics, whereas the mixed polarization case is affected the least. As a result, we see that although the optimal polarization choice gives the best sensitivities without systematics, with systematic uncertainties included the mixed polarization case of the H20 scenario performs better. The latter gives $3\sigma$ sensitivities up to 4.6, 6.0, and 5.0 TeV for SP-, VA-, TAT-type operator, respectively, as compared to 3.4, 6.0, and 4.7 TeV in the optimal polarization case.\footnote{In Refs.~\cite{Habermehl:2018yul,Habermehl:2020njb}, it was shown that with a better treatment of the systematic uncertainties, their effect on the final sensitivity is much less ($\sim10\%$) in the case of H20 mixed polarization  scenario unlike in our case where it reduces the $\Lambda$-scale sensitivity by up to $\sim40\%\text{-}50\%$. This is mainly due to our simplified approach [cf.~Eq.~\eqref{eq:simp} and subsequent discussion] which results in a very conservative estimate. Even then, we find the collider sensitivities to be comparable to or better than the existing limits from direct and indirect detection searches, which makes ILC a promising tool for leptophilic DM.}

Now that we have the results for the SP-, VA-, and TAT-type assuming all the relevant coefficients in Eq.~\eqref{eq:operator} to be unity, it is instructive to see how the sensitivity reach differs for other choices of the coefficients.  This is shown in \autoref{fig:barG} for a fixed DM mass of 1 GeV. The green, blue and red bars show the $3\sigma$ ILC sensitivity with unpolarized, optimally polarized beams and mixed polarization case, respectively, whereas the lighter (darker) shade corresponds to \cmmnt{zero (1\%)} without (with) background systematics. The optimal choice of polarization is in general the $(+80\%,+30\%)$ with exceptions of V+A-, V-, A-type operators with $(+80\%,-30\%)$ and V$-$A with $(-80\%,+30\%)$. Here S+P, V+A, and T+AT refer to our default choice with all relevant coefficients equal to 1, whereas  S$-$P refers to the case with $c_S^\chi=c_S^e=1$ and $c_P^\chi=c_P^e=-1$ (similarly for V$-$A and T$-$AT), whereas pure S type (P type) means the pseudoscalar (scalar) coefficients are zero (and similarly for pure vector, axial-vector, tensor, and axial-tensor cases). We see that with the unpolarized beams, the best sensitivity is obtained for the T$\pm$AT operators, while with the optimally polarized beams as well as mixed polarization case, the V+A operator gives the best sensitivity.

In any case, all the operator types can be probed up to multi-TeV cutoff scales at $\sqrt s=1$ TeV with ${\cal L}_{\rm int}=8$ ab$^{-1}$ integrated luminosity.       
%
%
\section{Mono-$Z$ channel} \label{sec:4}
\begin{figure}[t!]
\centering
\includegraphics[width=0.36\textwidth]{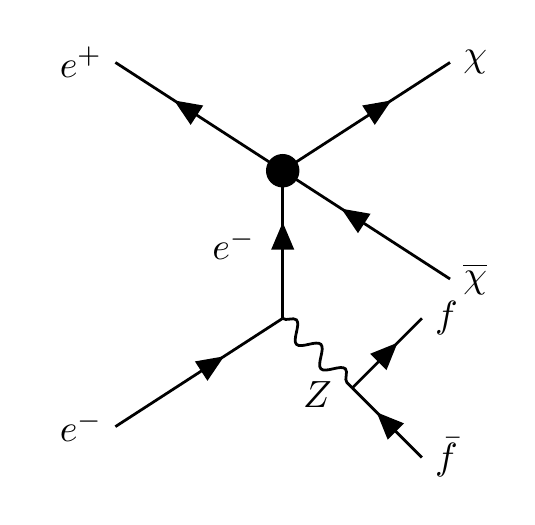}
\caption{\label{fig:mono-Z_Feyn} Feynman diagram for the mono-$Z$ channel. The $Z$ boson can also be emitted from the positron leg. Depending on the decay mode the particle $f$ can be leptons ($\ell^\pm\equiv e^\pm , \mu^\pm$) as well as jets ($j\equiv u,c,d,s,b$). }

\end{figure}

In addition to the monophoton channel discussed in the previous section, another useful channel for leptophilic DM search at lepton colliders is the mono-$Z$ channel, where the $Z$ boson is emitted from one of the initial states as shown in the Feynman diagram in \autoref{fig:mono-Z_Feyn}. 

\begin{figure*}[ht!]
\includegraphics[width=0.48\linewidth]{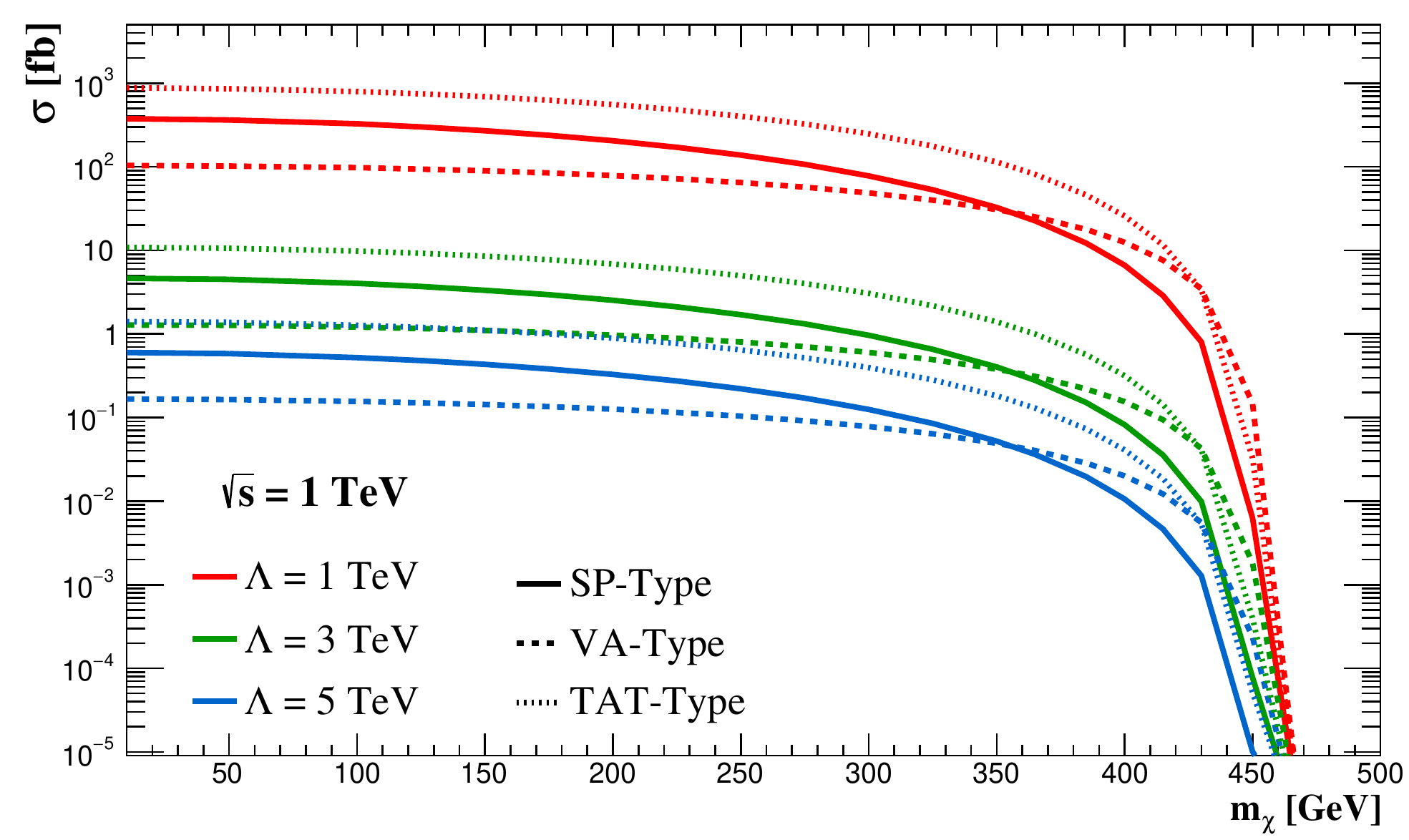}
\includegraphics[width=0.48\linewidth]{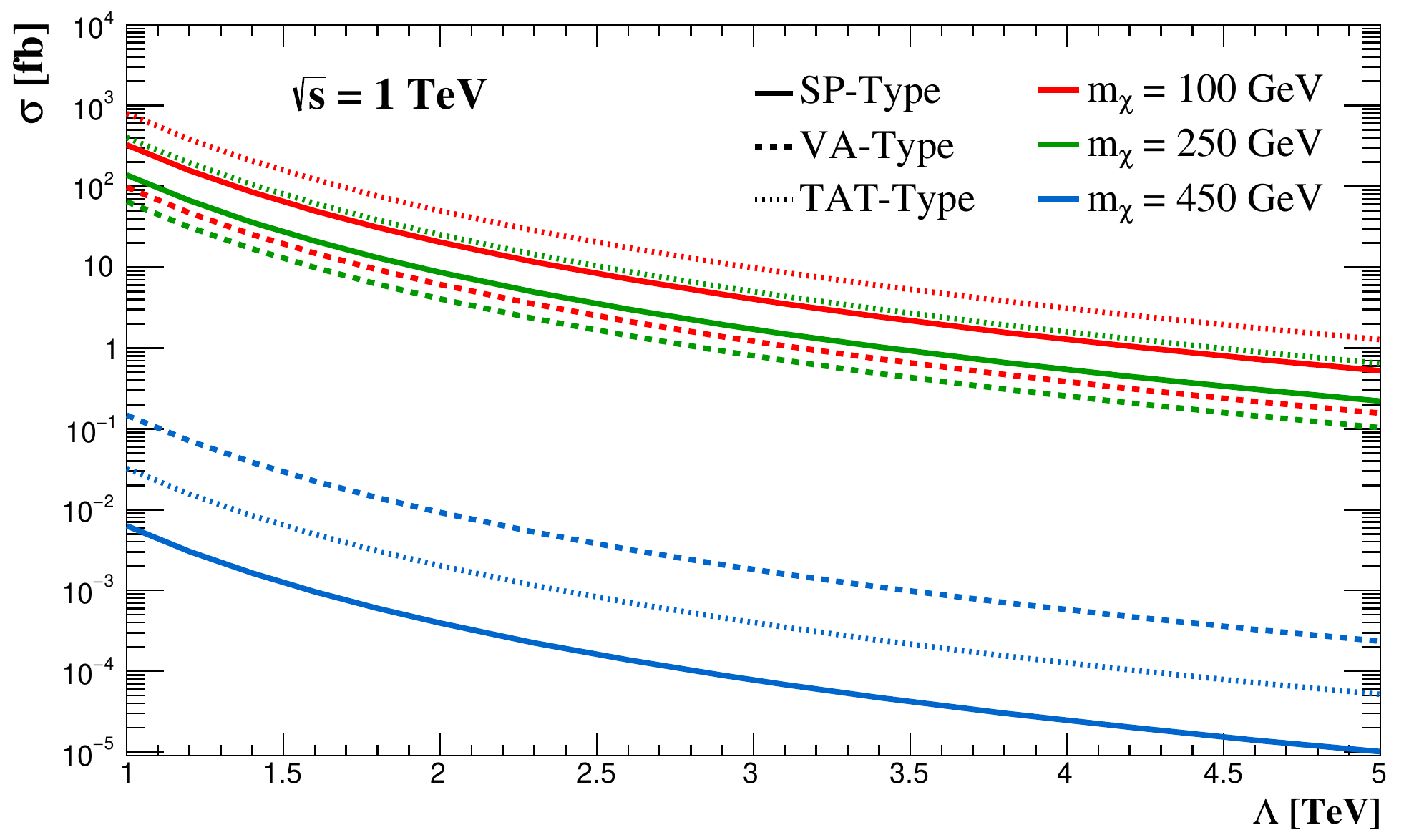}
\caption{\label{fig:zCS} Variation of the mono-$Z$ production cross section with the DM mass (left) and cutoff scale (right) at $\sqrt s=1$ TeV ILC. The solid, dashed and dotted lines are for the SP-, VA- and TAT-type operators, respectively. In the left panel, the red, green and blue curves correspond to different values of the cutoff scale $\Lambda=1$, 3, and 5 TeV respectively, while in the right panel, they correspond to different values of the DM mass $m_\chi=$100, 250, and 450 GeV, respectively.}
\end{figure*}

In \autoref{fig:zCS}, we show the variations of the mono-$Z$ production cross section for the signal as a function of the DM mass and the cutoff scale in the left and right panels, respectively, for all three operator types, namely, SP (solid), VA (dashed) and TAT (dotted) types. We find that the unpolarized cross section is the smallest (largest) for the VA (TAT)-type operator at any given DM mass. In the left panel, the cross section drops rapidly as $m_\chi$ approaches $\sqrt s/2-m_Z$ due to phase-space suppression. Otherwise, for smaller DM masses, the cross section for a given operator type and a given cutoff scale is almost independent of the DM mass, like in the monophoton case (cf.~\autoref{fig:gCS}). In the right panel, we see that for a given DM mass the cross section drops as $\Lambda^{-4}$, as expected. For subsequent decay of the $Z$ boson the nature of the graphs remains the same, and only the numbers are shifted as per the branching fractions to different channels. In the following, we discuss and compare how the leptonic and hadronic decay channels perform as a probe of the fermionic DM scenario, in comparison to the monophoton case discussed above.

%

\subsection{Leptonic mode} \label{sec:4.1}
For the leptonic decay of the $Z$ boson, we examine the process $e^+e^-\to\chi\overline{\chi}Z(\to \ell^-\ell^+)$, as shown in  \autoref{fig:mono-Z_Feyn}. We will only consider $\ell=e,\mu$ for simplicity and use the lepton pair as the visible particles for tagging. The main SM background for this channel is  $e^+e^-\to\nu\overline{\nu}\;\ell^+\ell^-$, and it is polarization dependent. 
\subsubsection{Unpolarized and polarized cross sections}
For the signal and background simulation, we generated the UFO library for our EFT framework using \texttt{FeynRules}~\cite{Alloul:2013bka} and then generated events for both signal and background using \texttt{MadGraph5}~\cite{Alwall:2014hca}, with the following basic baseline cuts to the parameter space:

\begin{equation}
p_T(\ell) > 2\text{ GeV},\;\;\;\; |\eta_\ell| \le 3.5,\;\;\;\; \Delta R_{\ell\ell} \ge 0.2  \, .
\end{equation}

For the signal, the $Z$ bosons are decayed into the charged lepton pairs via the \texttt{MadSpin}~\cite{Frixione:2007zp, Artoisenet:2012st} package, which is implemented in \texttt{MadGraph5}, to take care of the spin-correlation effects of the lepton pairs. A fast detector simulation to these events is done using \texttt{Delphes3}~\cite{deFavereau:2013fsa} with the same configuration card (\texttt{ILCgen}) as discussed in \autoref{sec:3.1}.

Similar to the monophoton case, we also examine the effect of polarization on the signal and background cross sections, as shown in \autoref{table:PolZ}. The neutrino background can be reduced to 28\% of its original value by making the electron beam $+80\%$ polarized and further reduced to 21\% of its original value by additionally making the positron beam $-30\%$ polarized. The $(+80\%, -30\%)$ polarization configuration also enhances the VA-type signal by a factor of 2.4. However, the $(+80\%, +30\%)$ configuration is better for the SP- and TAT-type signals. For ease of comparison between different operator types, we choose to work with the $(+80\%, +30\%)$ configuration democratically for all operator types, as well as for the background, unless otherwise specified.
%
\begin{table*}[t!]
\caption{\label{table:PolZ}%
Comparison of the leptonic mono-$Z$  background and signal cross sections for different choices of beam polarization for $m_{\chi}=100\text{ GeV}$ and $\Lambda=3\text{ TeV}$ at $\sqrt{s}=1\text{ TeV}$ ILC. The numbers in bold highlight the optimal polarization choice for a given operator type.}
\begin{ruledtabular}
\begin{tabular}{lcc cccc}
   &  & & \multicolumn{4}{c}{Polarized cross section (fb)} \\ \cline{4-7} 
 Process type & Unpolarized cross section (fb) & Polarization $P(e^-,e^+)$ & $(+,+)$ & $(+,-)$ & $(-,+)$ & $(-,-)$ \\
  \colrule
  \multirow{3}{3.2em}{ $\nu\overline{\nu}\ell^-\ell^+$} & & $(80, 0)$ & $2.60\times10^2$\phantom{1} & \phantom{1}$2.60\times10^2$\phantom{1} & \phantom{1}$1.26\times10^3$\phantom{1} & \phantom{1}$1.26\times10^3$ \\
   &$7.66\times10^2$ & $(80,20)$ & $3.01\times10^2$\phantom{1} & \phantom{1}$2.21\times10^2$\phantom{1} & \phantom{1}$1.48\times10^3$\phantom{1} & \phantom{1}$1.06\times10^3$  \\
   & & $(80,30)$ & $3.22\times10^2$\phantom{1} & \phantom{1}$2.01\times10^2$\phantom{1} & \phantom{1}$1.59\times10^3$\phantom{1} & \phantom{1}$9.53\times10^2$ \\
  \colrule
  \multirow{3}{5.2em}{SP} &  & (80, 0) & $0.28$ & $0.28$ & $0.28$ & $0.28$  \\
   &$0.28$ & (80,20) & $0.32$ & $0.23$ & $0.23$ & $0.32$ \\
   &  & $(80,30)$ & \boldmath{$0.35$} & $0.21$ & $0.21$ & $0.34$  \\
  \hline
  \multirow{3}{5.2em}{VA} &  & (80, 0) & $0.15$ & $0.15$ & $0.02$ & $0.02$  \\
   &$0.08$ & (80,20) & $0.12$ & $0.18$ & $0.01$ & $0.02$  \\
   &  & $(80,30)$ & $0.10$ & \boldmath{$0.19$} & $0.01$ & $0.02$  \\
  \colrule 
  \multirow{3}{5.2em}{TAT} &  & (80, 0) & $0.67$ & $0.67$ & $0.68$ & $0.68$ \\
   &$0.68$ & (80,20) & $0.79$ & $0.57$ & $0.57$ & $0.79$  \\
   &  & $(80,30)$ & \boldmath{$0.84$} & $0.51$ & $0.51$ & $0.84$  \\
\end{tabular}
\end{ruledtabular}
\end{table*}
%
\subsubsection{Cut-based analysis}
Now we proceed with our cut-based analysis to enhance the signal-to-background ratio.  
\paragraph*{Baseline selection cuts:} 
We define our signals by those events that pass through the baseline selection criteria as defined below where the $Z$ boson is reconstructed by the condition that all final state lepton-pairs are oppositely charged and of same flavor (OSSF).
\begin{equation}
 p_{T, \ell} > 5\text{ GeV, }\;\;\; |\eta_\ell| < 2.8 \, 
 \label{eq:baseZ}
\end{equation}
Other selection criteria are dynamic with respect to different BPs, as defined in \autoref{table:BPs&Cuts}. We have taken the same three BPs as in the monophoton case to probe  different regions of the parameter space, namely, BP1 essentially represents all light DM region, BP3 represents the region close to the kinematic limit of $\sqrt s/2-m_Z$, whereas BP2 captures the intermediate DM mass region. 

After implementing the baseline selection criteria given in Eq.~\eqref{eq:baseZ}, the signal and background events are given in the first row of \autoref{table:CutBGZ}. We find that the background is reduced to about $31\%$ of its original value in \autoref{table:PolZ} for the unpolarized case, whereas the signals are reduced to about 60\%-80\% of their original values across VA-type to TAT-type operator.

\begin{figure*}[htbp]
\centering 
\includegraphics[width=0.32\linewidth]{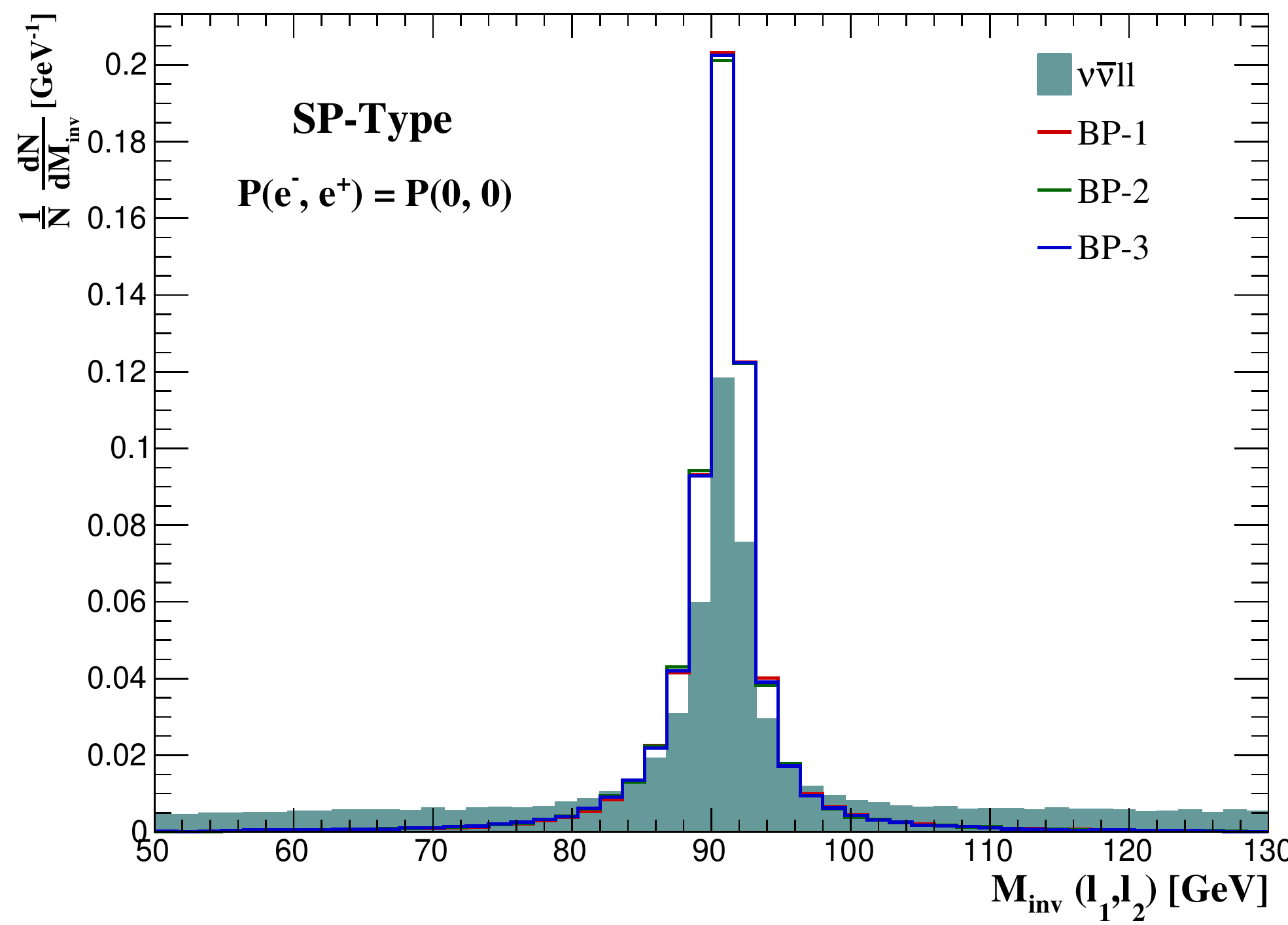}
\includegraphics[width=0.32\linewidth]{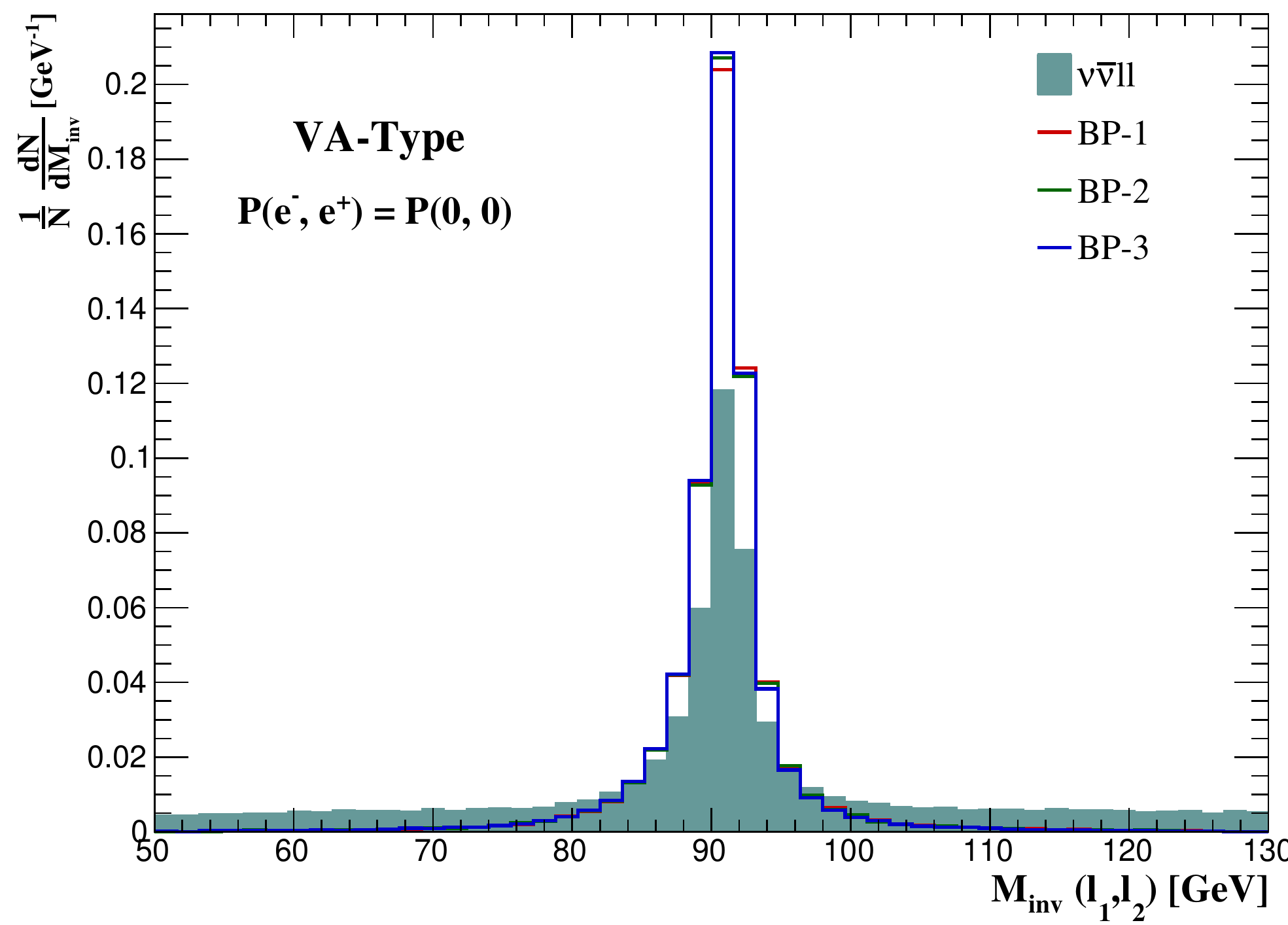}
\includegraphics[width=0.32\linewidth]{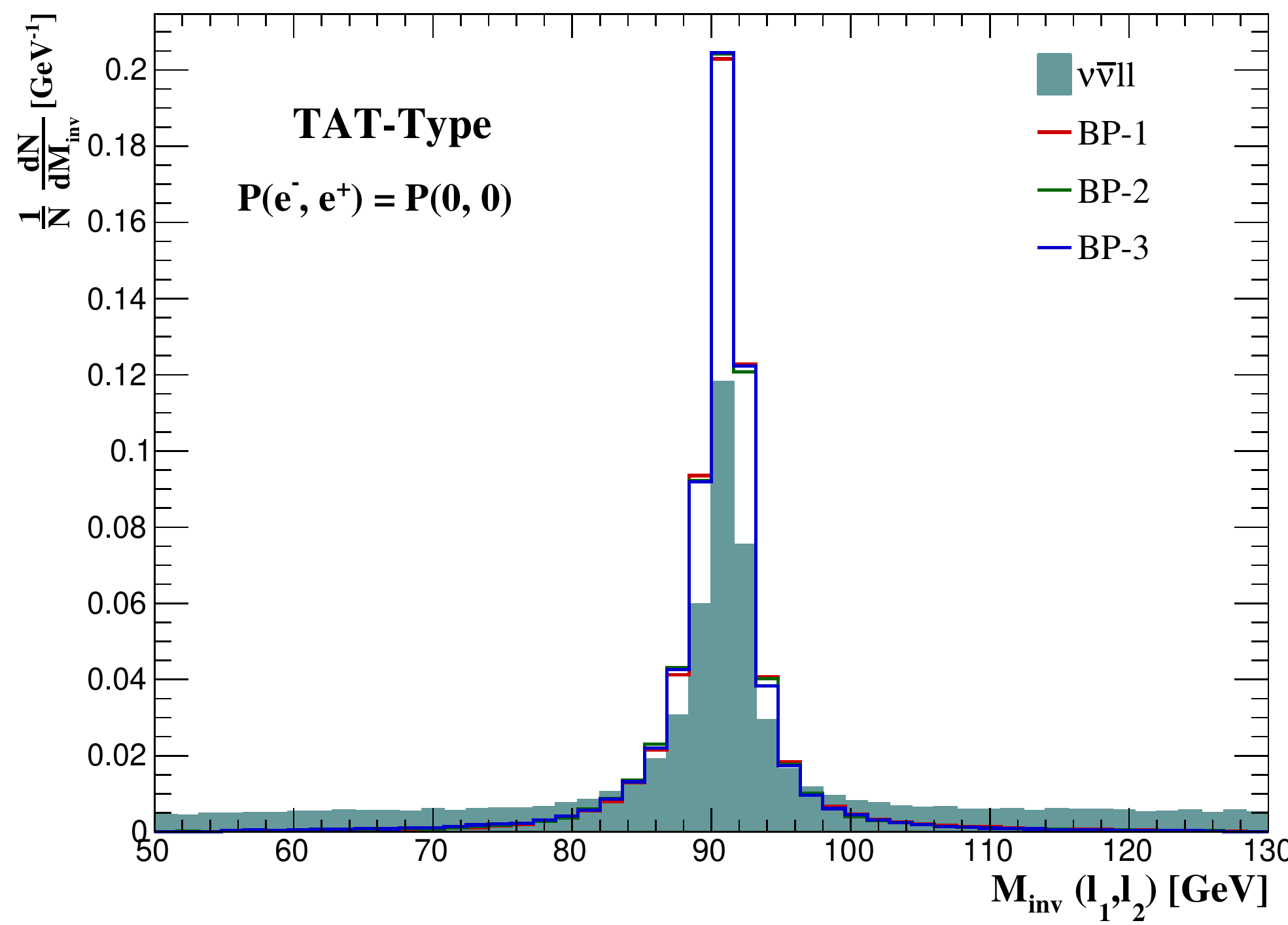}
\includegraphics[width=0.32\linewidth]{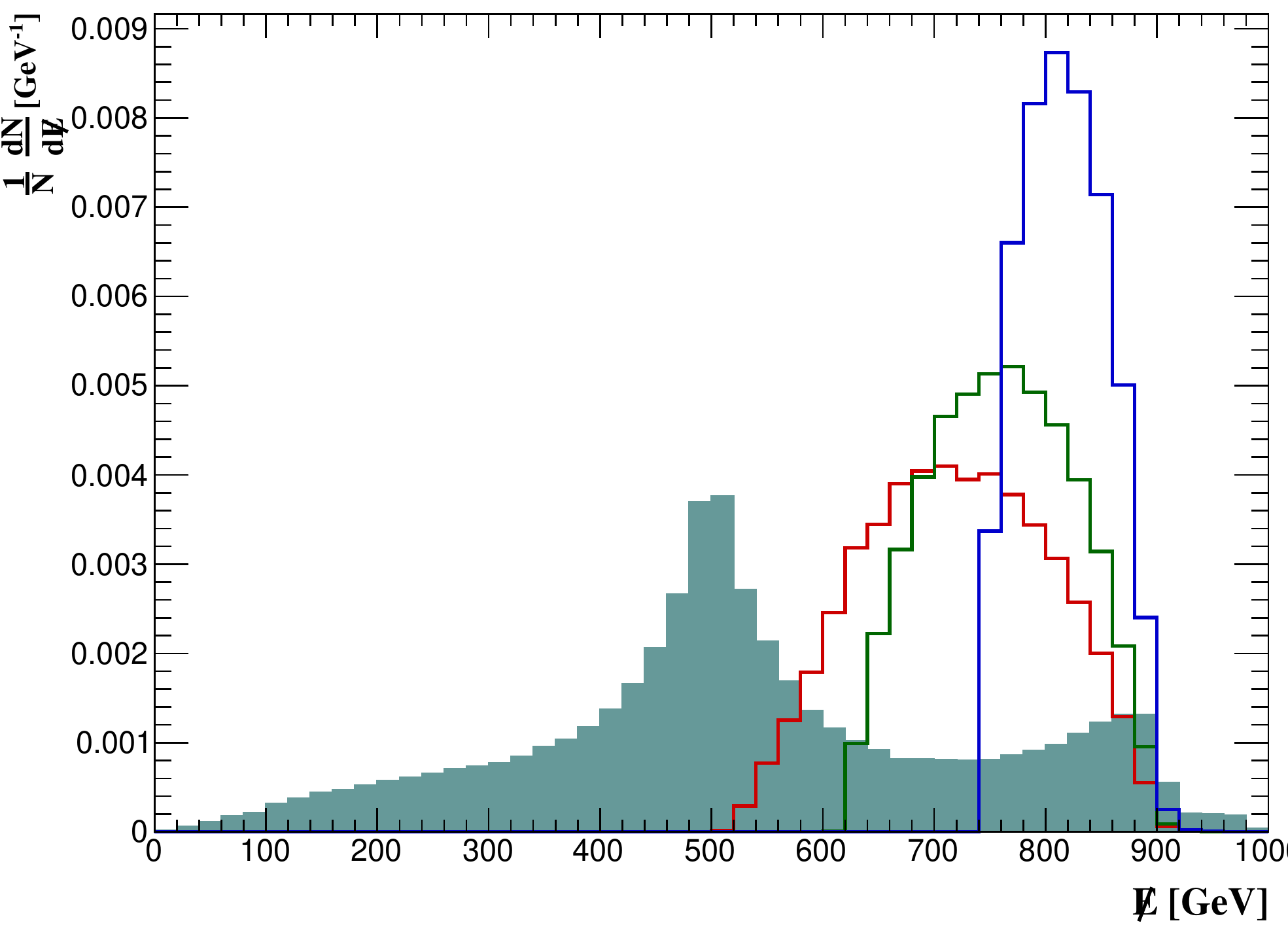}
\includegraphics[width=0.32\linewidth]{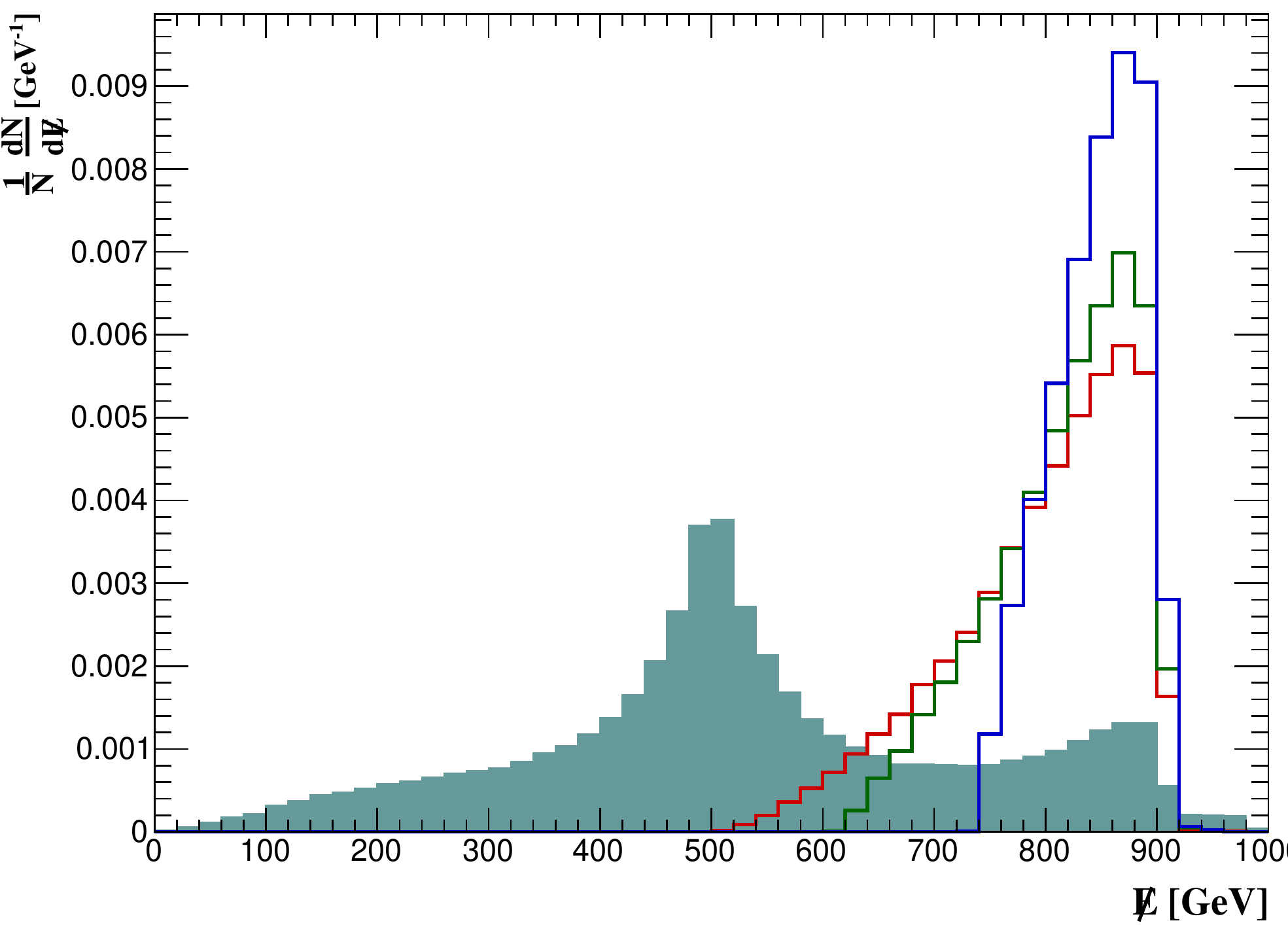}
\includegraphics[width=0.32\linewidth]{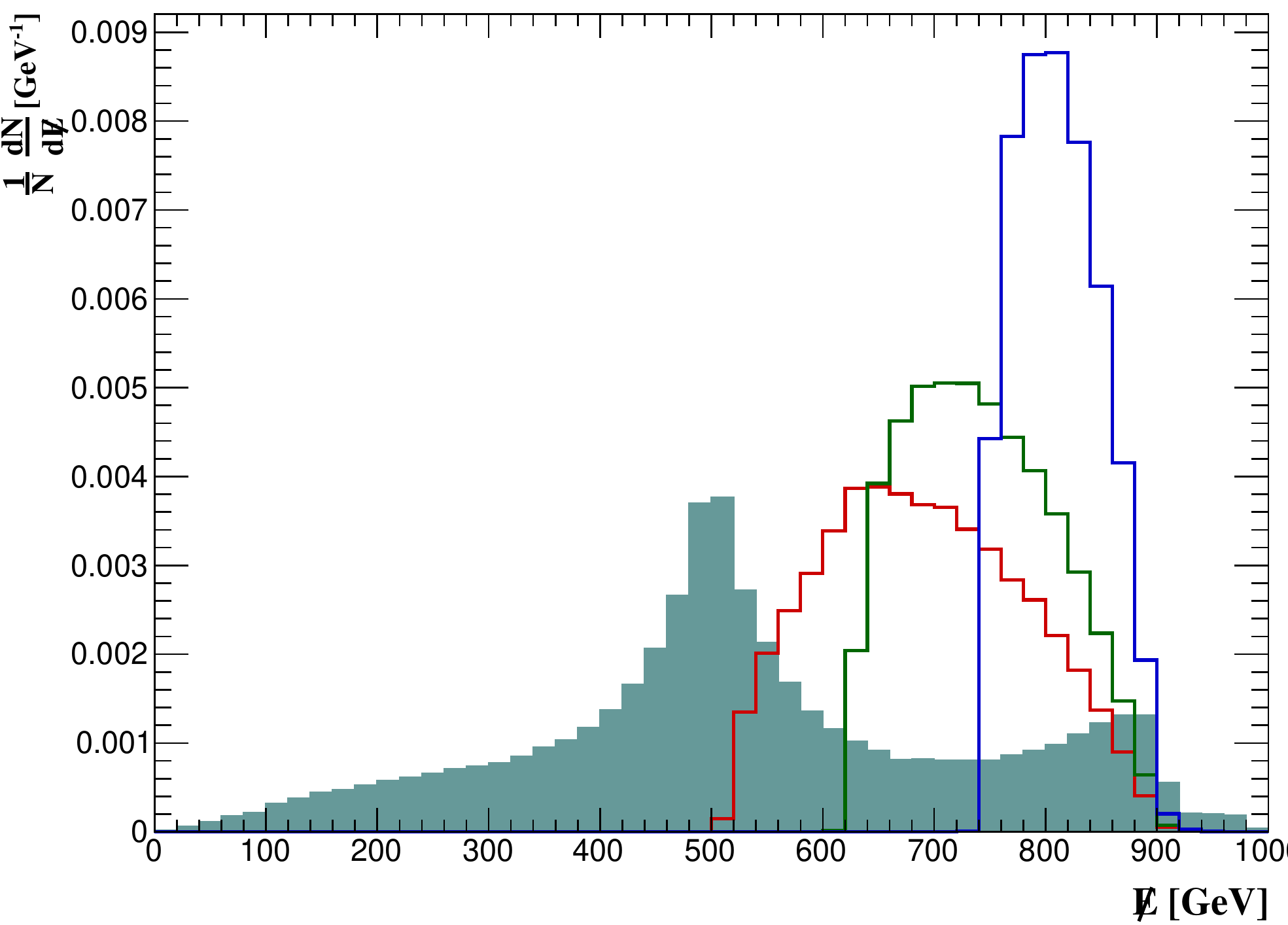}
\includegraphics[width=0.32\linewidth]{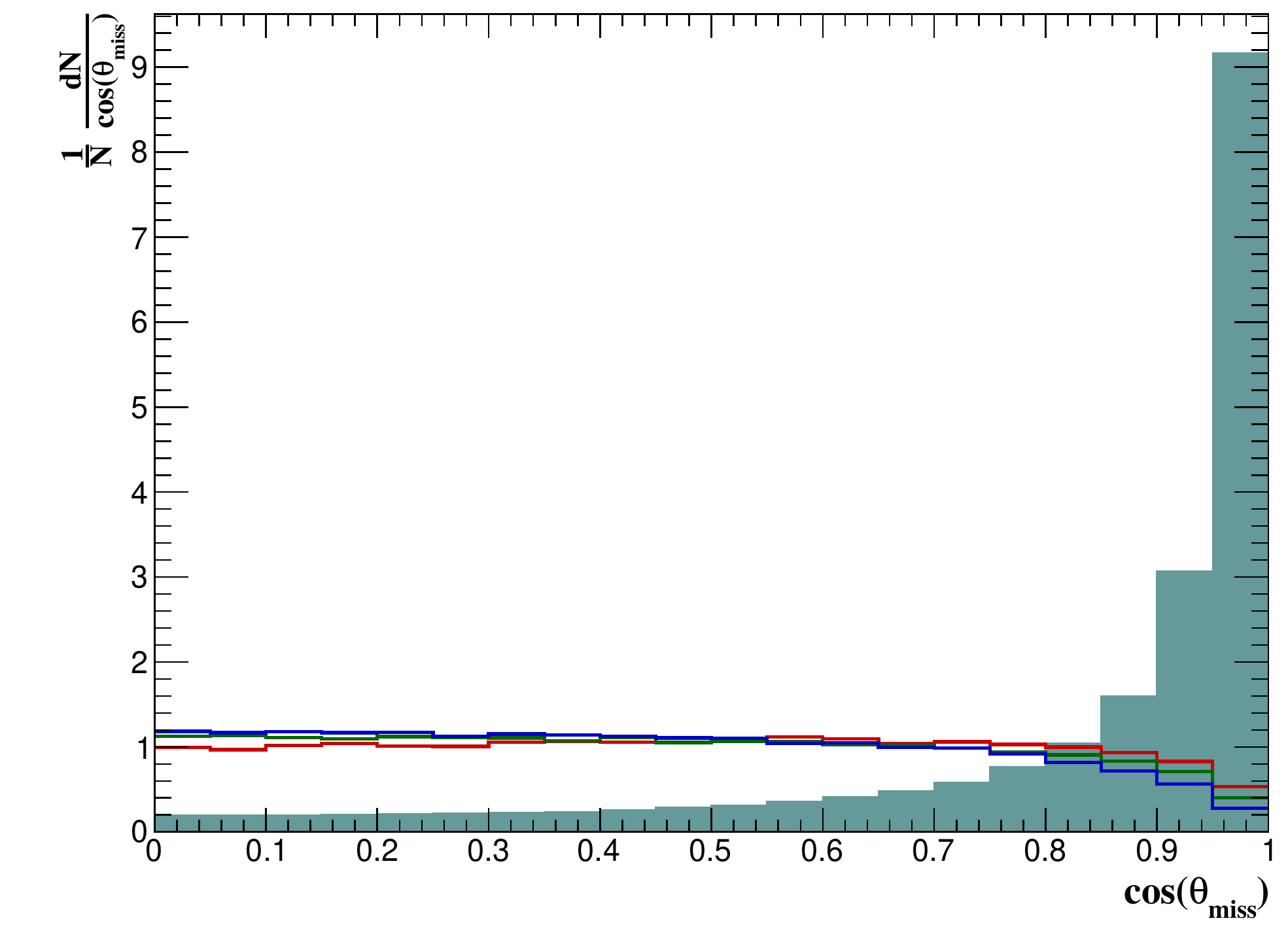}
\includegraphics[width=0.32\linewidth]{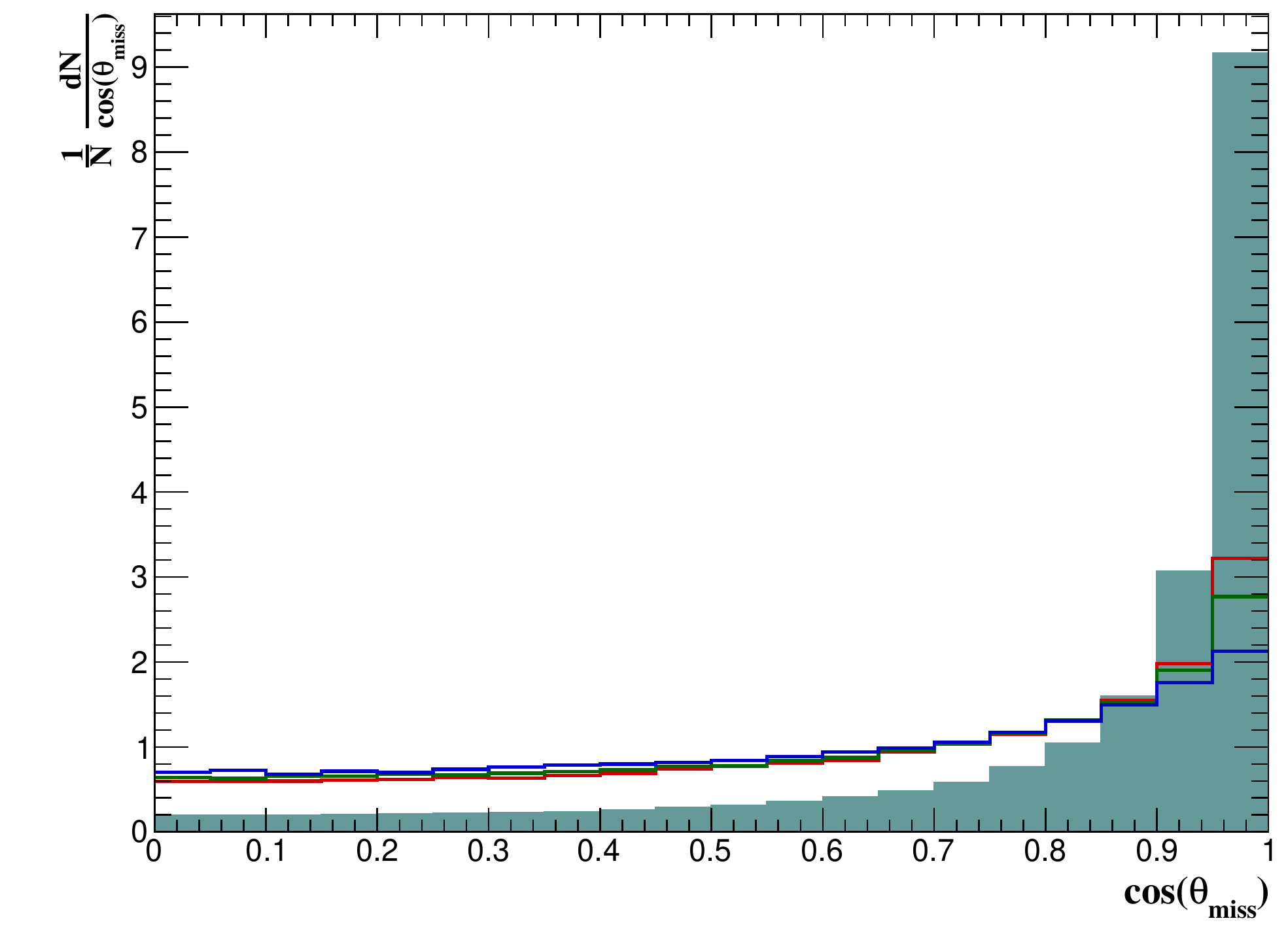}
\includegraphics[width=0.32\linewidth]{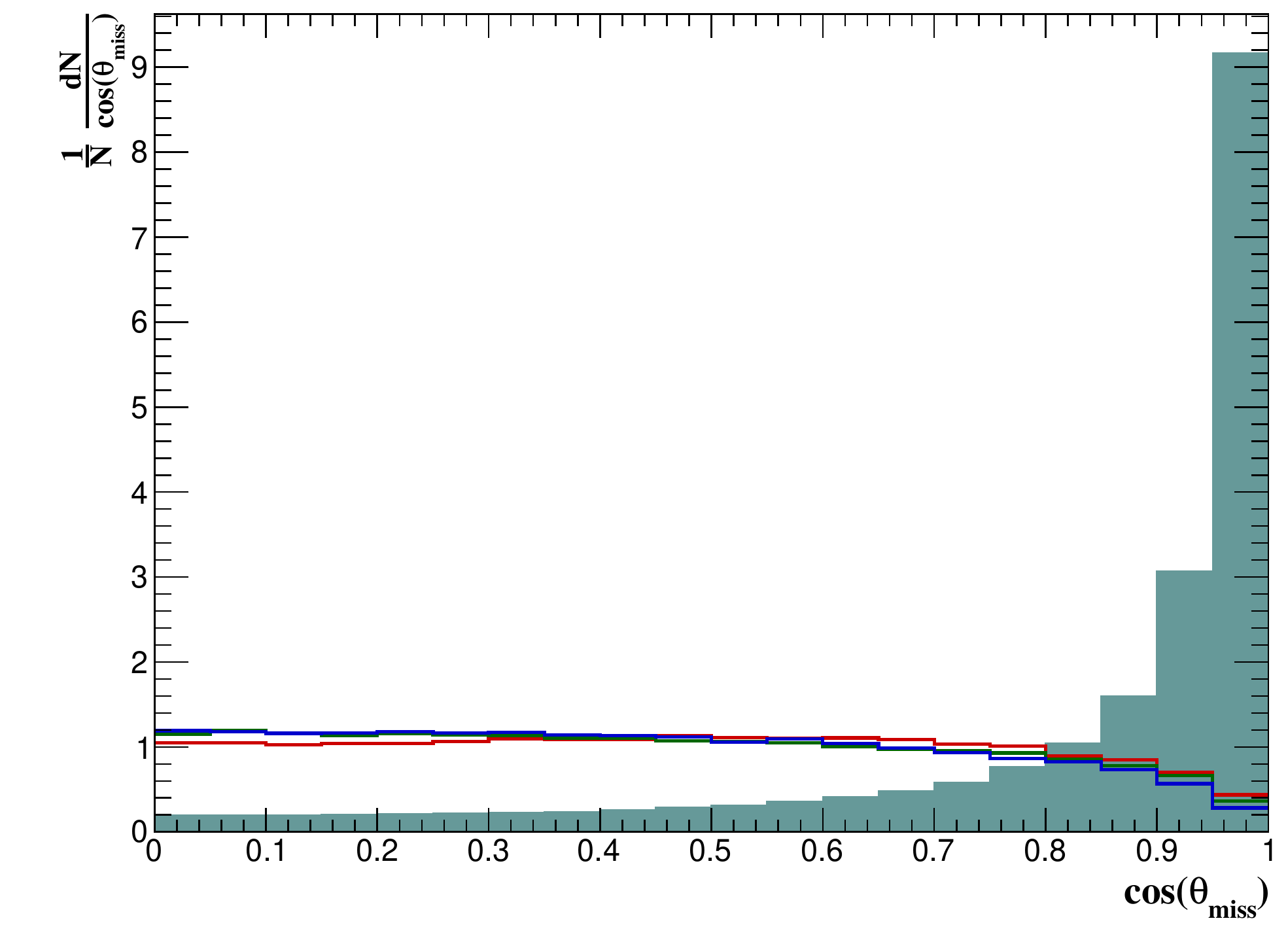}
\includegraphics[width=0.32\linewidth]{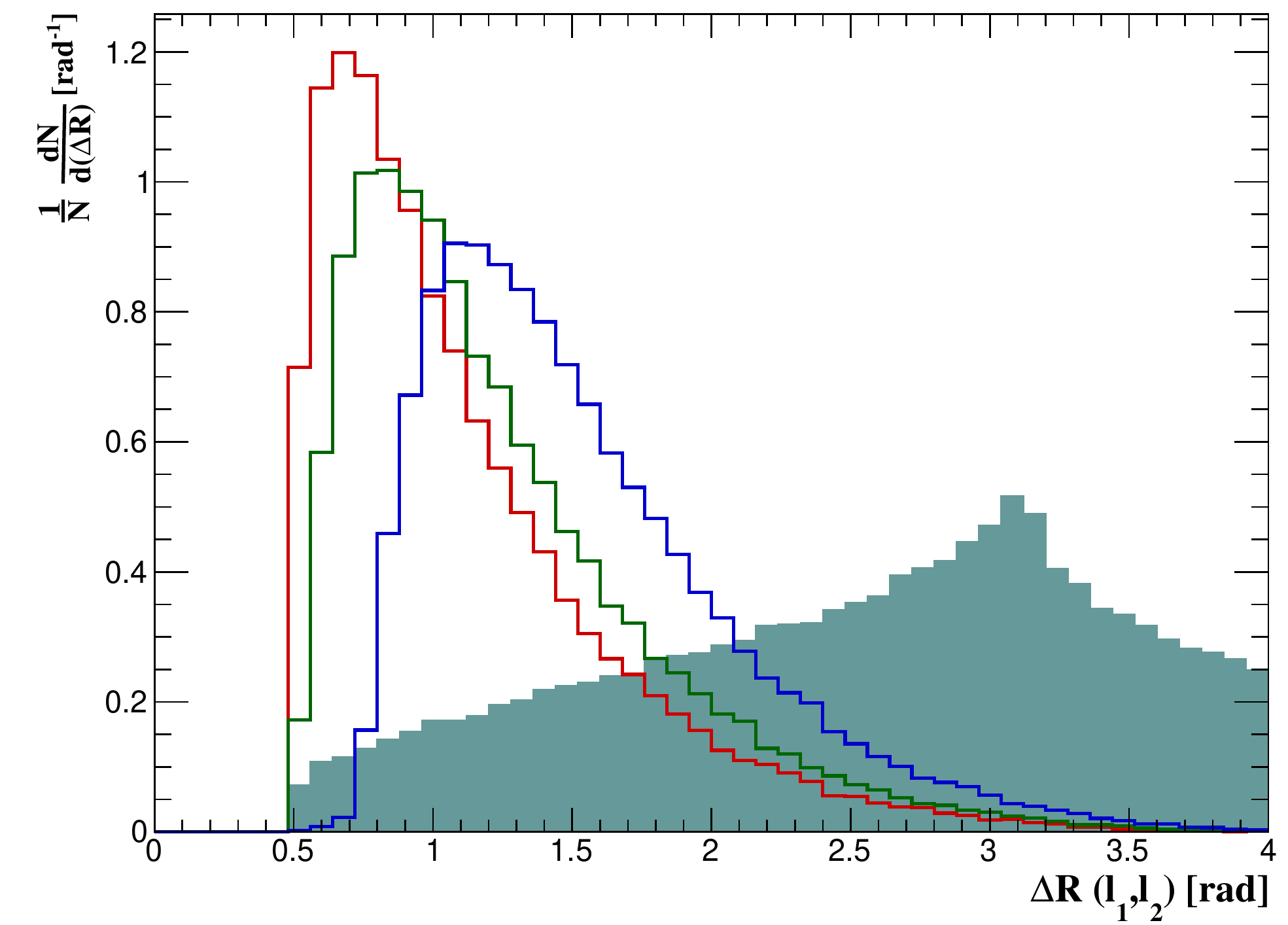}
\includegraphics[width=0.32\linewidth]{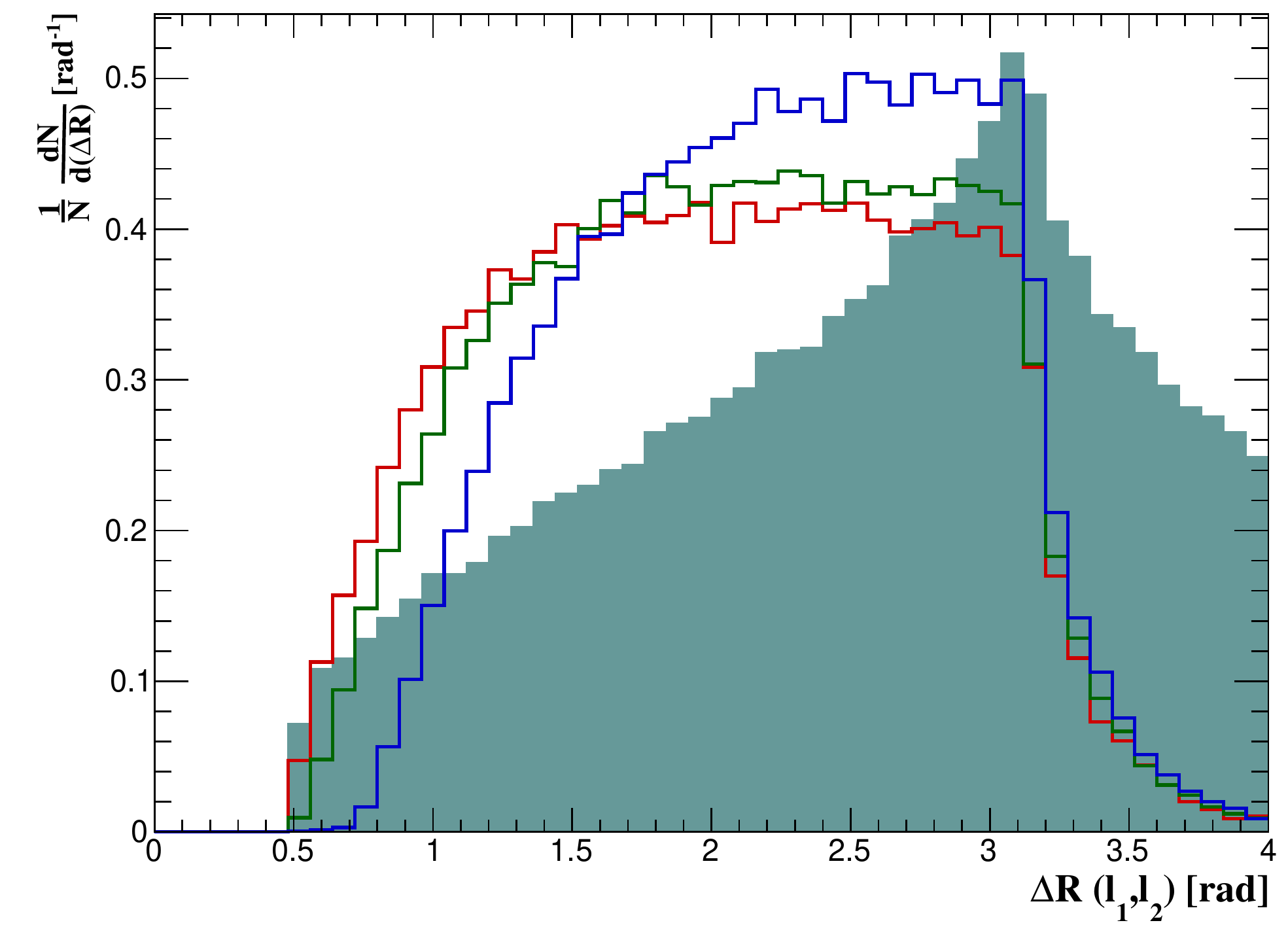}
\includegraphics[width=0.32\linewidth]{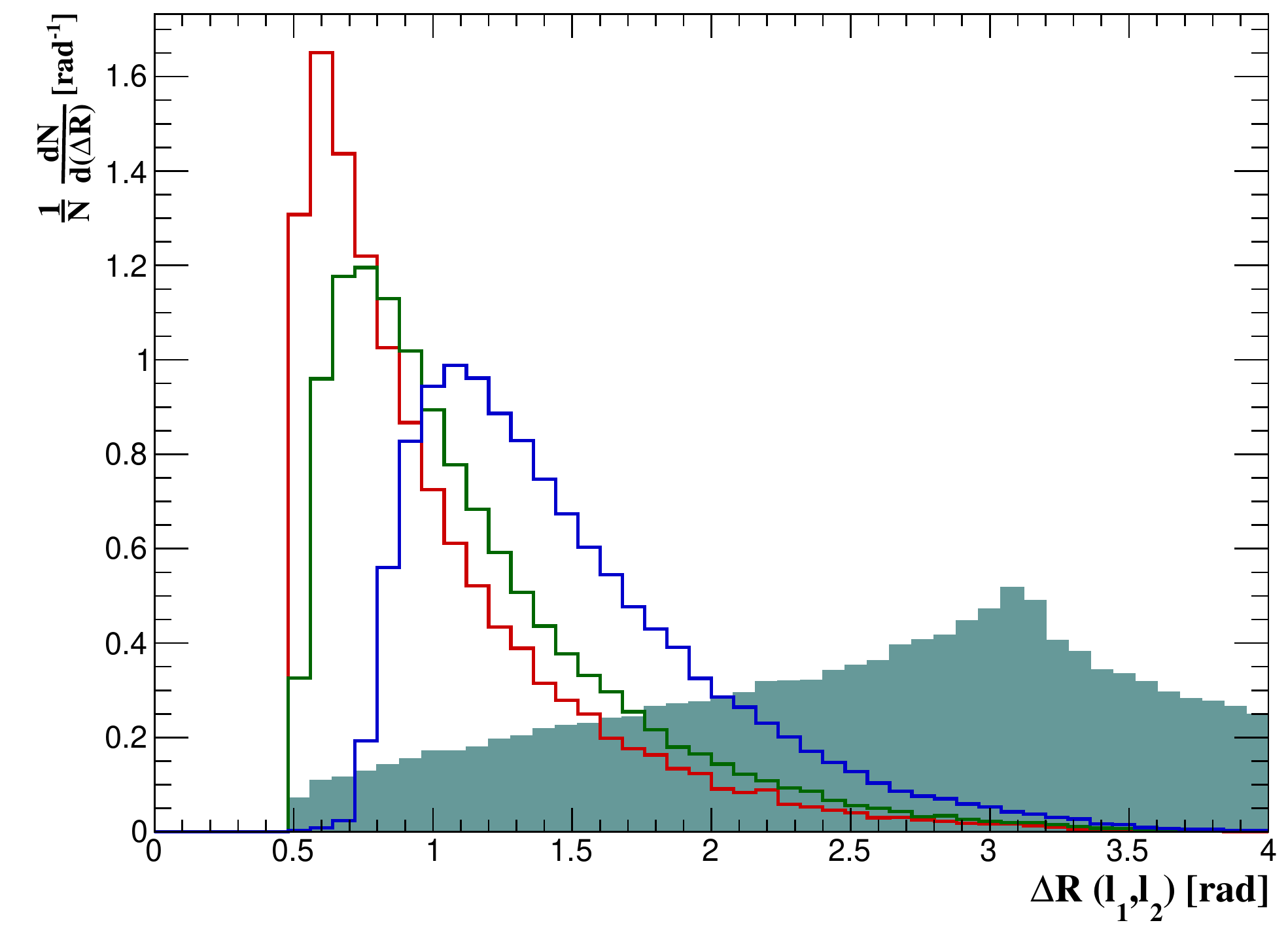}
\caption{Normalized differential distributions for leptonic mono-$Z$ background (shaded) and signals after baseline selection cuts for the kinematic variables shown in \autoref{table:BPs&Cuts} with unpolarized beams. The corresponding distributions with polarized beams are given in \autoref{figure:HistosZpol} in the Appendix.}
\label{figure:HistosZ}
\end{figure*}

We now consider various kinematic distributions for the signal and background, as shown in \autoref{figure:HistosZ} for the unpolarized case. 
Based on these histograms, some specialized selection cuts are chosen as follows (also summarized in \autoref{table:BPs&Cuts}) which enhance the signal significance: 
%
\begin{table*}[t]
\caption{\label{table:BPs&Cuts}%
Leptonic mono-$Z$ selection cuts for different BPs across operator types.}
\begin{ruledtabular}
 \begin{tabular}{l ccc} 
  & BP1 & BP2 & BP3 \\
  \cline{2-4}
  Definition & $m_{\chi} = 100\text{ GeV, }$ $\Lambda = 3\text{ TeV}$ & $m_{\chi} = 250\text{ GeV, }$ $\Lambda = 3\text{ TeV}$ & $m_{\chi} = 350\text{ GeV, }$ $\Lambda = 3\text{ TeV}$ \\
  \colrule
  Baseline selection & \multicolumn{3}{c}{OSSF lepton-pairs with $\;\; p_{T, \ell} > 5\text{ GeV, }\;\;\; |\eta_\ell| < 2.8$} \\
  \hline
  \multicolumn{4}{l}{SP type} \\
  \hline
  Cut 1 & \multicolumn{3}{c}{$80\text{ GeV}\le M_{\rm inv}(\ell^-\ell^+) \le 100\text{ GeV}$} \\
  Cut 2 & $ 540\text{ GeV} < \slashed{E} < 780\text{ GeV} $\phantom{00} & \phantom{0}$ 620\text{ GeV} < \slashed{E} < 850\text{ GeV} $\phantom{00} & \phantom{0}$ 740\text{ GeV} < \slashed{E} < 900\text{ GeV} $ \\
  Cut 3 &  \multicolumn{3}{c}{$\cos\theta_{\rm miss} < 0.79$} \\
  Cut 4 & $\Delta{R}_{\ell\ell} < 1.1\text{ rad} $ & $\Delta{R}_{\ell\ell} < 1.3\text{ rad} $ & $\Delta{R}_{\ell\ell} < 1.6\text{ rad} $\\
  \colrule
\multicolumn{4}{l}{VA type} \\
  \colrule
  Cut 1  & \multicolumn{3}{c}{$80\text{ GeV}\le M_{\rm inv}(\ell^-\ell^+) \le 100\text{ GeV}$} \\
  Cut 2 & $ 610\text{ GeV} < \slashed{E} $ & $ 650\text{ GeV} < \slashed{E} $ & $ 750\text{ GeV} < \slashed{E} $ \\
  Cut 3 &  \multicolumn{3}{c}{$\cos{\theta_{\rm miss}} < 0.95$} \\
  \colrule

  \multicolumn{4}{l}{TAT type} \\
  \colrule
  Cut 1 & \multicolumn{3}{c}{$80\text{ GeV}\le M_{\rm inv}(\ell^-\ell^+) \le 100\text{ GeV}$} \\
  Cut 2) & $ 520\text{ GeV} < \slashed{E} < 800\text{ GeV} $ & $ 620\text{ GeV} < \slashed{E} < 780\text{ GeV} $ & $ 740\text{ GeV} < \slashed{E} < 930\text{ GeV} $ \\
  Cut 3 &  \multicolumn{3}{c}{$\cos\theta_{\rm miss} < 0.8$} \\
  Cut 4  & $\Delta{R}_{\ell\ell} < 1.3\text{ rad} $ & $\Delta{R}_{\ell\ell} < 1.3\text{ rad} $ & $\Delta{R}_{\ell\ell} < 1.5\text{ rad} $\\
 
 \end{tabular}%
\end{ruledtabular}
\end{table*}

\begin{table*}[t]
\caption{\label{table:CutBGZ}%
Cut-flow table for different operators and BPs for the leptonic mono-$Z$ signal and background. The cut efficiencies are calculated with respect to the baseline selection cuts given in Eq.~\eqref{eq:baseZ}, with the corresponding event numbers shown in the first row. The background event yields vary across operator types because of the dynamic nature of the cuts chosen (cf.~\autoref{table:BPs&Cuts}). }
\begin{ruledtabular}
  \begin{tabular}{l ccc ccc} 
   & \multicolumn{6}{c}{Event Numbers (Cut efficiencies)}\\
  \cline{2-7}
  & \multicolumn{3}{c}{$ \nu\bar{\nu}\ell^{+}\ell^{-} $} & \multicolumn{3}{c}{Signal}\\
  \cline{2-7}
  Selection cuts & BP1 & BP2 & BP3 & BP1 & BP2 & BP3  \\
  \colrule
  \multicolumn{7}{l}{\bf{SP-type:}}\\ \hline 
  \multirow{2}{3.em}{Baseline selection} & \multicolumn{3}{c}{$1.92\times10^6$} & $1.60\times10^3$ & $7.14\times10^2$ & $1.71\times10^2$\\
  & \multicolumn{3}{c}{$(100\%)$} & $(100\%)$ & $(100\%)$ & $(100\%)$ \\
  \multirow{2}{4.0em}{After Final cut} 
  & $2.72\times10^4$ & $4.37\times10^4$ & $4.77\times10^4$ & $7.79\times10^2$ & $4.05\times10^2$ & $9.7\times10^1$  \\
  & $(1.42\%)$ & $(2.28\%)$ & $(2.49\%)$ & $(48.72\%)$ & $(56.72\%)$ & $(56.73\%)$  \\
  \colrule
  \multicolumn{7}{l}{\bf{VA-type:}}\\ \hline 
  \multirow{2}{3.em}{Baseline selection} & \multicolumn{3}{c}{$1.92\times10^6$} & $5.14\times10^2$ & $3.43\times10^2$ & $1.64\times10^2$\\
  & \multicolumn{3}{c}{$(100\%)$} & $(100\%)$ & $(100\%)$ & $(100\%)$ \\
  \multirow{2}{4.0em}{After Final cut} 
  & $2.75\times10^5$ & $2.64\times10^5$ & $2.15\times10^5$ & $3.98\times10^2$ & $2.76\times10^2$ & $1.38\times10^2$  \\
  & $(14.36\%)$ & $(13.79\%)$ & $(11.23\%)$ & $(77.43\%)$ & $(80.47\%)$ & $(84.15\%)$  \\
  \colrule
  \multicolumn{7}{l}{\bf{TAT-type:}}\\ \hline 
  \multirow{2}{3.em}{Baseline selection} & \multicolumn{3}{c}{$1.92\times10^6$} & $3.54\times10^3$ & $2.07\times10^3$ & $5.99\times10^2$\\
  & \multicolumn{3}{c}{$(100\%)$} & $(100\%)$ & $(100\%)$ & $(100\%)$ \\
  \multirow{2}{4.0em}{After Final cut} 
  & $4.10\times10^4$ & $3.00\times10^4$ & $4.06\times10^4$ & $2.30\times10^3$ & $1.10\times10^3$ & $3.29\times10^2$  \\
  & $(2.14\%)$ & $(1.56\%)$ & $(2.12\%)$ & $(65.03\%)$ & $(53.11\%)$ & $(54.92\%)$  \\

\end{tabular}
\end{ruledtabular} 
\end{table*}

Cut 1: 
For the $\nu\overline{\nu}\ell^-\ell^+$ background, the lepton pairs could come from different channels (such as from the $t$-channel $W$ and $Z$ boson exchange, in addition to the on/off shell $WW$ and $ZZ$ decays), the invariant mass of the lepton pairs are distributed over a wide range. But for the signals, the lepton pairs come as decay products of the on shell $Z$ boson giving a sharp peak at the $Z$ boson mass, as is evident in \autoref{figure:HistosZ} (top row). Thus, selecting events that fall within a narrow window of dilepton invariant mass around the $Z$ mass, i.e. $M_{\rm inv}(\ell^+\ell^-)\in [80,\,100]\text{ GeV}$, suppresses the background significantly (to 22\% of the baseline value with the unpolarized beams) while keeping most of the signals ($\sim94\%$).

Cut 2: 
The next variable is the missing energy which is reconstructed using the initial and detected visible 4-momentum of the events. It is basically the energy of the 4-momentum, $\mathbf{p}_{\rm miss}=\mathbf{p}_{\rm initial}-\mathbf{p}_{\ell^+\ell^-}$, where $\mathbf{p}_{\rm initial}=(1000,0,0,0)$~GeV (neglecting the 14~mrad crossing angle since \texttt{Delphes} does not consider it) and $\mathbf{p}_{\ell^+\ell^-}$ is the momentum of the dielectron system. We see that as the DM mass increases the curve gets narrower with the peak shifting toward the total CM energy, while for the lighter DM it is more spread. So we use a dynamic cut across the benchmark points.

Cut 3: 
The third variable is the cosine of the missing polar angle, $\theta_{\rm miss}$. This variable appears to be very effective since the heavier candidates of the missing momentum tend to scatter at a larger angle than the SM ones. We see that this cut reduces the backgrounds to $\sim2.5\%$ for SP- and TAT-type operators to $\sim11\%\text{-}14\%$ for the VA-type operators while keeping most of the signal intact.

Cut 4: 
Our final selection cut is on the distance of the two leptons in the $(\eta,\phi)$, defined as $R=\sqrt{(\Delta\eta)^2+(\Delta\phi)^2}$, where $\Delta\eta\; (\Delta\phi)$ is the difference in pseudorapidity (azimuthal angle) of the two leptons. The lepton pairs from the $Z$ boson decay are more collinear than the ones in the SM case. So a strict cut on this variable enhances the signal significance. We do not use this cut for the VA-type operator since (as evident from the distribution) it is not sensitive for this particular case. The cut efficiencies and the selected events after this cut are reported in \autoref{table:CutBGZ}. The corresponding number for the case of optimally polarised beams are given in \autoref{table:CutBGZpol} in the Appendix. 
\par
In \autoref{table:CutBGZ}, we tabulate the number of events and efficiencies after baseline selection and final selection cut for both background as well as signal for different operators. We find that, after applying cuts 1--4, we can still retain about $48\%-65\%$ of the signal (77\%-84\% for VA-type), whereas the background is reduced to below percent level (3.5\% in the case of VA-type operator) of the original values given in \autoref{table:PolZ}.

\begin{table*}[!t]
\caption{\label{table:SigZ}%
Signal significance in the mono-$Z$ leptonic channel at $\sqrt{s}=1$ TeV and $\mathcal{L}_{\rm int} = 8\, {\rm ab}^{-1}$ after implementing all the cuts mentioned in the text. The values in the parentheses denote the significances with a 1\% background systematic  uncertainty. }
\begin{ruledtabular}
\begin{tabular}{l ccc ccc ccc} 
  & \multicolumn{9}{c}{Signal significance for ${\cal L}_{\rm int}=8\,{\rm ab}^{-1}$}\\
   \cline{2-10}
   & \multicolumn{3}{c}{Unpolarized beams} & \multicolumn{3}{c}{H20 scenario} & \multicolumn{3}{c}{Optimally polarized beams}\\
   \cline{2-10}
   Operator Type & BP1 & BP2 & BP3 & BP1 & BP2 & BP3 & BP1 & BP2 & BP3 \\
  \colrule
  SP & $4.7\;(2.4)$ & $1.9\;(0.8)$ & $0.4\;(0.2)$ & 
  $6.7\;(6.2)$ & $2.9\;(2.6)$ & $0.7\;(0.6)$ 
  & $10.6\;(8.2)$ & $4.5\;(3.1)$ & $1.1\;(0.7)$ \\
  VA & $0.8\;(0.14)$ & $0.5\;(0.10)$ & $0.3\;(0.06)$ & 
  $3.0\;(1.9)$ & $2.1\;(1.4)$ & $1.9\;(0.8)$
  & $4.6\;(2.1)$ & $3.3\;(1.5)$ & $1.8\;(0.9)$ \\
  TAT & $11.1\;(5.0)$ & $6.2\;(3.2)$ & $1.6\;(0.7)$ & 
  $15.3\;(13.8)$ & $8.9\;(8.2)$ & $2.4\;(2.2)$ 
  & $24.1\;(17.6)$ & $14.0\;(10.7)$ & $3.9\;(2.7)$ \\

\end{tabular}
\end{ruledtabular}
\end{table*}
%
%
\begin{figure*}[t!]
\centering 
\includegraphics[width=0.51\linewidth]{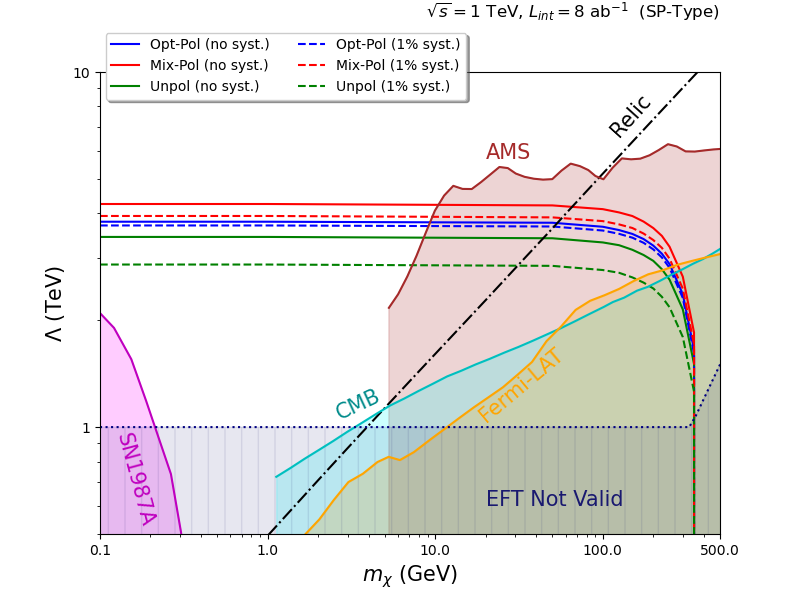}
\hspace{-0.7cm}
\includegraphics[width=0.51\linewidth]{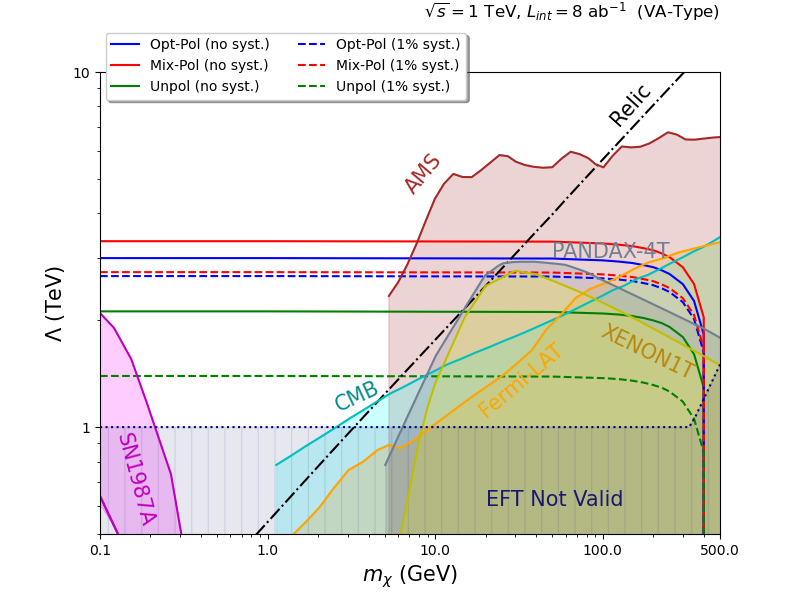}
\includegraphics[width=0.51\linewidth]{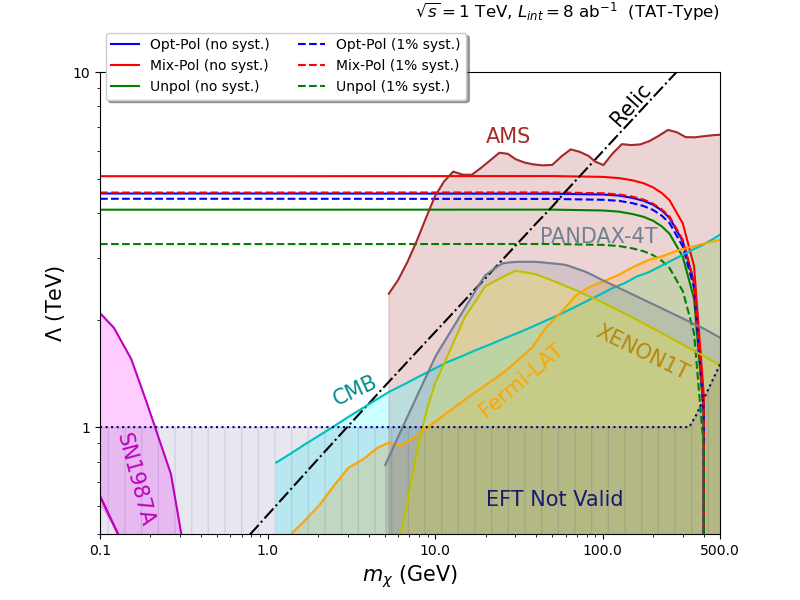}
\caption{$3\sigma$ sensitivity contours in the  mono-$Z$ leptonic channel for the SP-type (top left panel), VA-type (top right panel) and TAT-type (bottom panel) operators with unpolarized (green lines), mixed polarized (red lines) and optimally polarized (blue lines) $e^+e^-$ beams at $\sqrt s=1$ TeV center-of-mass energy and with  ${\cal L}_{\rm int}=8$ ab$^{-1}$ integrated luminosity. The solid (dashed) contours are assuming zero (1\%) background systematics. The various shaded regions are excluded by direct detection (XENON1T, PANDAX-4T), indirect detection (Fermi-LAT, AMS), astrophysics (SN1987A) and cosmology (CMB) constraints. In the shaded region below $\Lambda={\rm max}\{\sqrt s, 3m_\chi\}$, our EFT framework is not valid. Along the dot-dashed line, the observed DM relic density is reproduced for a thermal WIMP assuming only DM-electron effective coupling. }
\label{figure:ContourZ}
\end{figure*}
\subsubsection{Results}
After implementing all these cuts, we calculate the final signal significance for the three BPs using Eq.~\eqref{eq:significance}. Our results are given in \autoref{table:SigZ} for an integrated luminosity of ${\cal L}_{\rm int}=8$ ab$^{-1}$. We see that as we go higher up in the DM mass the signal significance drops. We also find that the best-performing operator type is the TAT-type, for which more than $97\%$ of the background events are removed after all the selection cuts. For the signal we retain $54\%$--$65\%$ of the events. 
We also analyzed the effect of beam polarization, in particular, the mixed polarization as per the H20 operating scenario. We note, unlike the monophoton case, the optimal polarization case gives the best significance both with and without systematic uncertainty. The systematic uncertainty affects more if the background is large. Compared to monophoton case, the background here is not large enough and hence its effect on the signal-to-background ratio is relatively small, resulting in less suppression of the significance for the optimally polarized case. Nevertheless, we observe positive effect of beam polarization with excess of 2 times increase in significance across the two cases discussed.

\begin{figure*}[!ht]
 \includegraphics[width=0.9\linewidth]{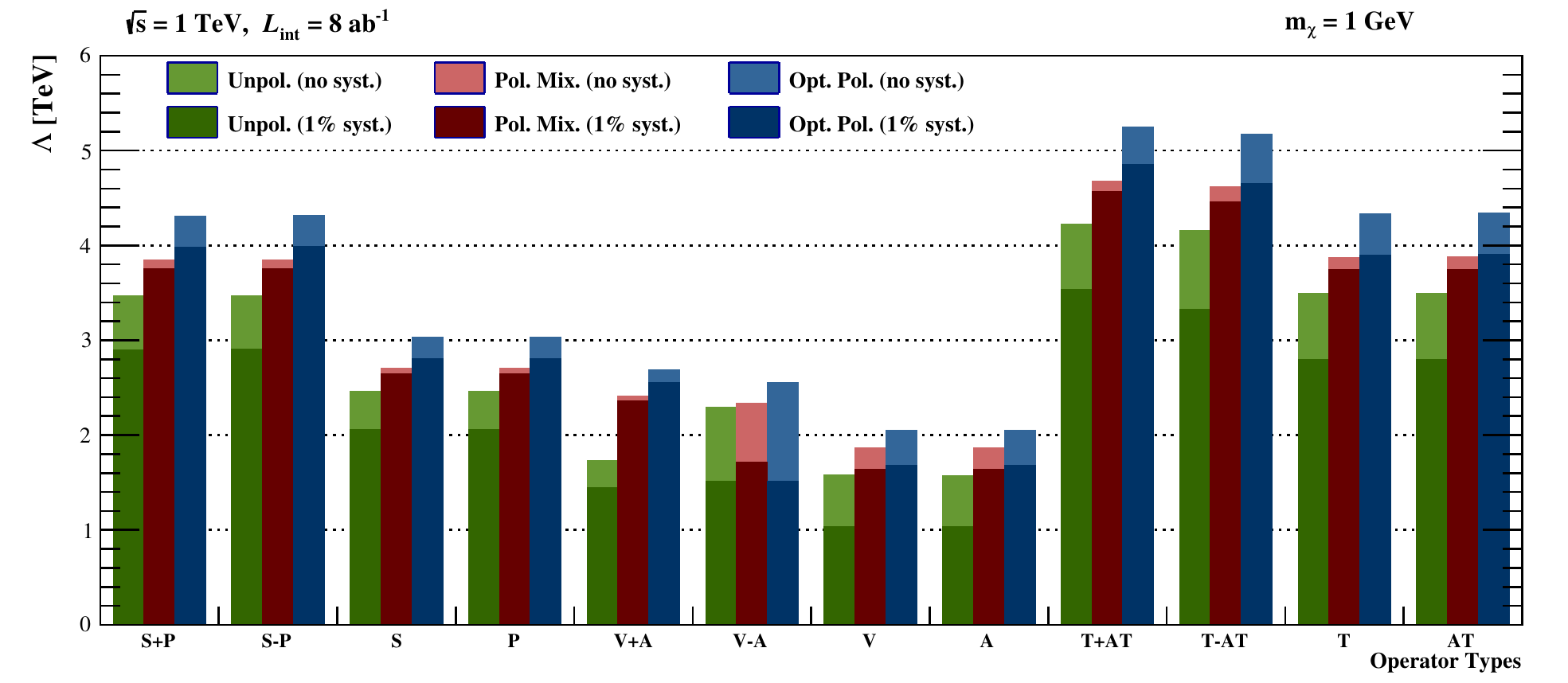}
 \caption{$3\sigma$ sensitivity reach of different operators in the leptonic mono-$Z$ channel with DM mass of $1$ GeV at $\sqrt{s}=1$ TeV and $\mathcal{L}_{\rm int}=8\, {\rm ab}^{-1}$. The green, red, and blue bars show the sensitivities with unpolarized, mixed and optimally polarized beams respectively and the lighter (darker) shade corresponds to zero (1\%) background systematics. 
}
 \label{fig:barZll}
\end{figure*}

Going beyond the three BPs, we now vary the DM mass and present the $3\sigma$ sensitivity reach for this channel in \autoref{figure:ContourZ} for all the operators. The labels and shaded regions are the same as in the monophoton case (cf.~\autoref{figure:ContourG}). 
We see that the accessible range of the cutoff scale $\Lambda$ for the unpolarized beams can reach up to 4.2~TeV for the TAT-type operator, whereas for the SP and VA-type, it can reach up to 3.5 and 1.7~TeV, respectively. But with the application of optimally polarized beams as discussed earlier, we see an increase by about $20\%\text{--}25\%\;(55\%)$ of the $3\sigma$ reach on the $\Lambda$ scale of SP and TAT-type (VA-type), up to 4.3, 2.7, and 5.2~TeV for the for SP, VA, and TAT-type operators, respectively. With the mixed polarization case we obtain 3.8, 2.4, and 4.7~TeV for the three operators, respectively. 

We also compare the leptonic mono-$Z$  sensitivities for different operator types in \autoref{fig:barZll} for a fixed DM mass of 1 GeV. As in \autoref{fig:barG} for the monophoton case, the green, blue and red bars show the $3\sigma$ sensitivity with unpolarized, optimally polarized and mixed polarization cases and the lighter (darker) shade corresponds to zero ($1\%$) background systematics. We again find the best sensitivity for T$\pm$ AT operators, and the S$\pm$ P operators also perform well, while, unlike the monophoton case, the vector- and axial-vector-type operators have the weakest sensitivity in this channel mainly because of its similarity to the background in the event distributions. Also, for the reasons discussed above the optimal polarization case does best even with background systematics.
\subsection{Hadronic mode} \label{sec:4.2}
Now we look at the process $e^+e^-\to\chi\overline{\chi}Z(\to jj)$, where $j\equiv u,d,c,s,b$ quarks (as shown in \autoref{fig:mono-Z_Feyn}). The relevant SM background processes for this channel are $e^+e^-\to\nu\overline{\nu}jj$ and $e^+e^-\to jj\ell \nu$ (with one charged lepton escaping the detector) where the jets and leptons in the final state can come from any possible source (not necessarily from an on shell $Z$). %

\subsubsection{Unpolarized and polarized cross sections}
We use the same UFO library as before which is implemented using \texttt{FeynRules}~\cite{Alloul:2013bka} and simulate the events for the signal and backgrounds via \texttt{MadGraph5}~\cite{Alwall:2014hca} with the following basic cuts to the parameter space:

\begin{align}
&p_T(j) > 5\text{ GeV}, \;\;\;\;p_T(\ell) > 2\text{ GeV}, \nonumber\\
&|\eta_j| \le 3.2, \;\;\;|\eta_\ell| \le 3.5, \;\;\;\Delta R_{jj,\ell j} \ge 0.2 \, .
\end{align}
%
For the signals, as in the leptonic case, the on shell $Z$ bosons are decayed into the pairs of jets using the \texttt{MadSpin} package~\cite{Frixione:2007zp, Artoisenet:2012st}, implemented in \texttt{MadGraph5}. Both the signal and background samples are hadronized using \texttt{Pythia8.3}~\cite{Sjostrand:2014zea} and then the fast detector simulation is done using \texttt{Delphes3.5}~\cite{deFavereau:2013fsa} with the same configuration card as discussed in \autoref{sec:3.1}. The particle flow objects built with the reconstructed tracks and calorimeter towers and rejected to be isolated leptons and photons are clustered with the Valencia algorithm for linear colliders (usually referred to as the \texttt{VLC} algorithm) with $R=2,\; \beta=1,\; \gamma=0$ which mimics the \texttt{Durham} (\texttt{ee\_kt\_algorithm}) algorithm. It is important to mention here that a beam-induced background from $\gamma\gamma\to {\rm hadrons}$~\cite{CLIC_cdr}, where the photons are radiated from incoming beams (bremsstrahlung and beamstrahlung), are relevant in linear colliders and have been discussed in the full simulation studies. To consider that, one has to externally generate the events and overlay with detector simulation (as possibly pileup events)~\cite{Boronat:2016tgd}. At the ILC, with larger bunch spacing ($\sim500~{\rm ns}$), individual bunch crossings are better distinguishable, thus minimizing this background \cite{Boronat:2016tgd}; hence  it will not change our results much. \par
We then examine different choices of beam polarization on both the event samples for this channel, as shown in \autoref{table:PolZjj}. We find that both the backgrounds are polarization dependent and fall off significantly for a right-handed electron beam and with increasing degree of polarization for a left-handed positron beam. We observe as earlier that the optimal polarization choice is $P(e^-,e^+)=(+80\%,+30\%)$ for the SP- and TAT-type operators and $(+80\%,-30\%)$ for the VA-type operators.%

\subsubsection{Cut-based analysis}
After obtaining the signal and background cross sections as reported in \autoref{table:PolZjj}, we proceed with our cut-based analysis to optimize the signal significance, as follows.   
%
\begin{table*}[ht]
\caption{\label{table:PolZjj}%
Comparison of the hadronic mono-$Z$ signal and background cross sections for different choices of beam polarization with $m_\chi = 100~\rm{GeV}$ and $\Lambda = 3~\rm{TeV}$ at $\sqrt{s} = 1~\rm{TeV}$ ILC. The numbers in bold highlight the optimal polarization choice for a given operator type.}
\begin{ruledtabular}
 \begin{tabular}{l c c cccc} 
  &  &  &  \multicolumn{4}{c}{Polarized cross section (fb)}  \\
   \cline{4-7}
 Process type & Unpolarized cross section (fb) & Polarized $P(e^{-},e^{+})$ & $(+,+)$ & $(+,-)$ & $(-,+)$ & $(-,-)$ \\
\colrule
           &  & $(80,0)$ & $1.73\times10^2$\phantom{1} & \phantom{1}$1.73\times10^2$\phantom{1} & \phantom{1}$1.41\times10^3$\phantom{1} & \phantom{1}$1.41\times10^3$ \\
$\nu\overline{\nu} j j$ & $8.39\times10^2$ & $(80,20)$ & $1.99\times10^2$\phantom{1} & \phantom{1}$1.47\times10^2$\phantom{1} & \phantom{1}$1.68\times10^3$\phantom{1} & \phantom{1}$1.08\times10^3$ \\
           &  & $(80,30)$ & $2.31\times10^2$\phantom{1} & \phantom{1}$1.47\times10^2$\phantom{1} & \phantom{1}$1.93\times10^3$\phantom{1} & \phantom{1}$1.04\times10^3$ \\

\colrule
           &  & $(80,0)$ & $3.26\times10^2$\phantom{1} & \phantom{1}$3.26\times10^2$\phantom{1} & \phantom{1}$2.22\times10^3$\phantom{1} & \phantom{1}$2.22\times10^3$ \\           
 $\ell\nu j j$ & $2.14\times10^3$ & $(80,20)$ & $3.85\times10^2$\phantom{1} & \phantom{1}$2.66\times10^2$\phantom{1} & \phantom{1}$2.65\times10^3$\phantom{1} & \phantom{1}$1.83\times10^3$ \\
           &  & $(80,30)$ & $8.39\times10^2$\phantom{1} & \phantom{1}$4.83\times10^2$\phantom{1} & \phantom{1}$4.56\times10^3$\phantom{1} & \phantom{1}$2.69\times10^3$ \\

\colrule

           &  & $(80,0)$ & 2.77 & 2.77 & 2.78 &  2.78 \\
SP & $2.78$ & $(80,20)$ & 3.23 & 2.34 & 2.34 &  3.22 \\
           &  & $(80,30)$ & \bf{3.45} & 2.11 & 2.11 &  3.44 \\

\colrule

           &  & $(80,0)$ & 1.50 & 1.50 & 0.17 &  0.17 \\
VA & $0.83$ & $(80,20)$ & 1.20 & 1.79 & 0.13 &  0.20 \\
           &  & $(80,30)$ & 1.05 & \bf{1.95} & 0.12 &  0.22 \\

\colrule

           &  & $(80,0)$ & 6.74 & 6.74 & 6.78 & 6.21 \\
TAT & $6.77$ & $(80,20)$ & 7.86 & 5.70 & 5.70 & 7.88 \\
           &  & $(80,30)$ & \bf{8.40} & 5.14 & 5.14 & 8.38 \\

\end{tabular}
\end{ruledtabular}
\end{table*}  
%
Baseline selection cuts: 
We select the events that contain at least two jets with the following transverse momentum and pseudorapidity requirements: 
\begin{equation}
p_{T, j} > 10\text{ GeV, }\;\;\;\; |\eta_j| < 2.6 \, ,
\label{eq:BSjj}
\end{equation}
and do not have any isolated lepton and photon, i.e., electromagnetic (EM) particles, with $p_T>5$~GeV and $|\eta|<2.8$. The two highest $p_T$ jets are required to reconstruct the $Z$ boson. Further selection cuts are applied, some of which depend on the DM mass. So, as in the leptonic channel, we have taken the same three BPs with varying DM mass and imposed dynamic cuts. The benchmark parameters and the different selection cuts are defined in \autoref{table:BPs&Cutsjj}. The number of signal and background events after implementing the baseline selection criteria are given in the first row of \autoref{table:CutBZjj}.
%

\begin{table*}[!t]
\caption{\label{table:BPs&Cutsjj}%
Different BPs and selection cuts across operator types for mono-$Z$ hadronic channel.}
\begin{ruledtabular}
\begin{tabular}{lccc}
  & BP1 & BP2 & BP3 \\
\colrule
  Definition & $m_{\chi} = 100\text{ GeV, }$ $\Lambda = 3\text{ TeV}$ & $m_{\chi} = 250\text{ GeV, }$ $\Lambda = 3\text{ TeV}$ & $m_{\chi} = 350\text{ GeV, }$ $\Lambda = 3\text{ TeV}$ \\
\colrule
  Baseline selection & \multicolumn{3}{c}{ $\;\; p_{T, j} > 10\text{ GeV, }\;\;\; |\eta_j| < 2.6,\; \text{no EM particles}\;$ } \\
\colrule
  \multicolumn{4}{l}{SP type} \\
\colrule
  Cut 1 & \multicolumn{3}{c}{$85\text{ GeV}\le M_{\rm inv}(j j) \le 100\text{ GeV}$} \\
  Cut 2 & $ 510\text{ GeV} < \slashed{E} < 760\text{ GeV}\phantom{0} $ & $\phantom{0} 610\text{ GeV} < \slashed{E} < 840\text{ GeV} $\phantom{0} & \phantom{0}$ 710\text{ GeV} < \slashed{E} < 860\text{ GeV} $ \\
  Cut 3 & \multicolumn{3}{c}{$|\cos\theta_{\rm miss}| < 0.79 $} \\
   Cut 4 & $\Delta R_{jj} < 1.3\text{ rad} $ & $\Delta R_{jj} < 1.4\text{ rad} $ & $\Delta R_{jj} < 1.5\text{ rad} $ \\

\colrule

\multicolumn{4}{l}{VA type} \\
\colrule
  Cut 1 & \multicolumn{3}{c}{$85\text{ GeV}\le M_{\rm inv}(j j) \le 100\text{ GeV}$} \\
  Cut 2 & $ 590\text{ GeV} < \slashed{E} $ & $ 690\text{ GeV} < \slashed{E} $ & $ 750\text{ GeV} < \slashed{E} $ \\
  Cut 3 & \multicolumn{3}{c}{$|\cos\theta_{\rm miss}| < 0.9 $} \\

\colrule

  \multicolumn{4}{l}{TAT type} \\
\colrule
  Cut 1 & \multicolumn{3}{c}{$85\text{ GeV}\le M_{\rm inv}(j j) \le 100\text{ GeV}$} \\
  Cut 2 & $ 510\text{ GeV} < \slashed{E} < 770\text{ GeV} $\phantom{00} & \phantom{0}$ 610\text{ GeV} < \slashed{E} < 780\text{ GeV} $\phantom{00} & \phantom{0}$ 720\text{ GeV} < \slashed{E} < 850\text{ GeV} $ \\
  Cut 3 & \multicolumn{3}{c}{$|\cos\theta_{\rm miss}| < 0.79 $} \\
   Cut 4 & $\Delta R_{jj} < 1.3\text{ rad} $ & $\Delta R_{jj} < 1.4\text{ rad} $ & $\Delta R_{jj} < 1.6\text{ rad} $ \\

\end{tabular}%
\end{ruledtabular}
\end{table*}
\begin{figure*}[!t]
\centering 
\includegraphics[width=0.325\linewidth]{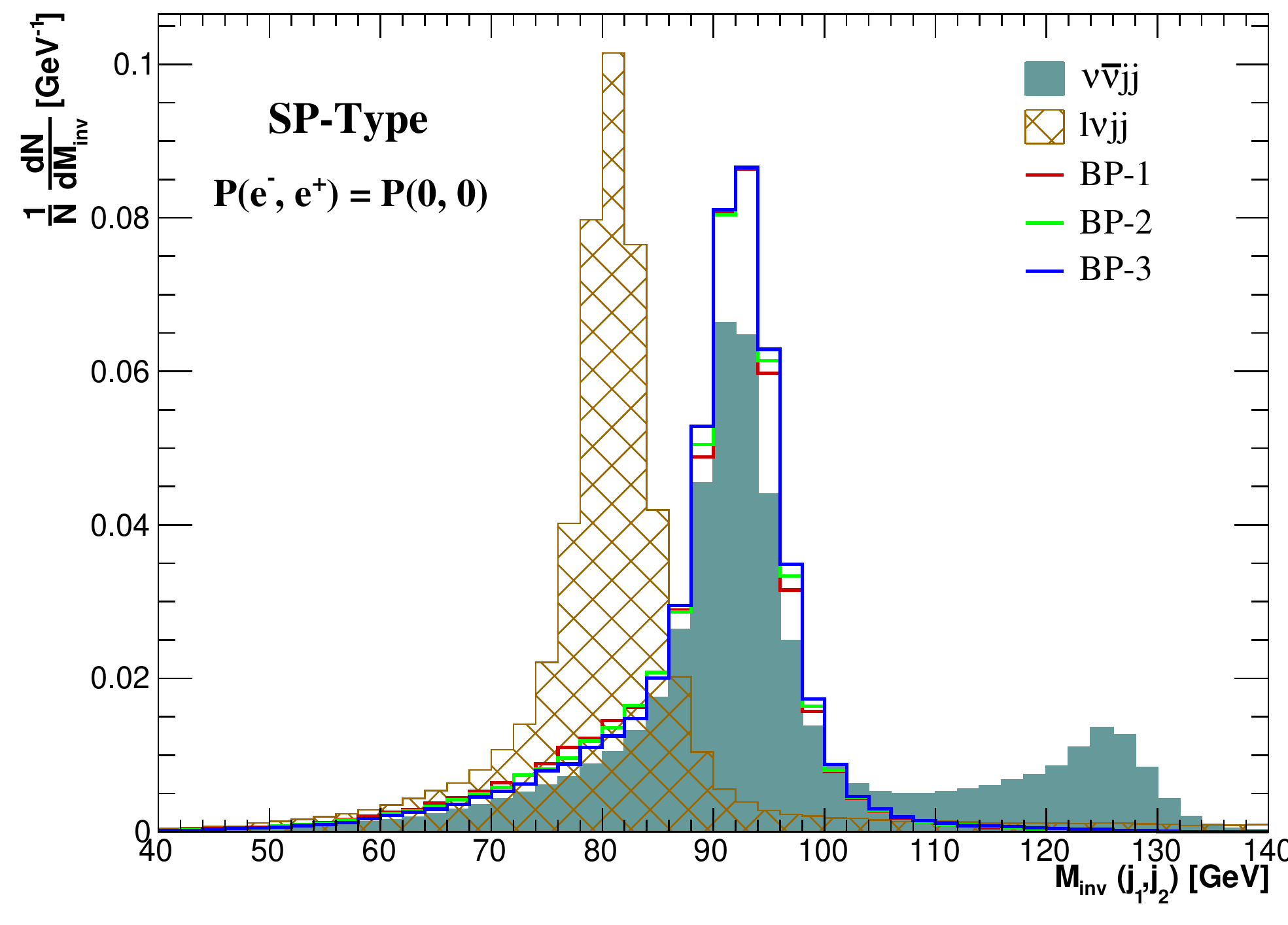}
\includegraphics[width=0.325\linewidth]{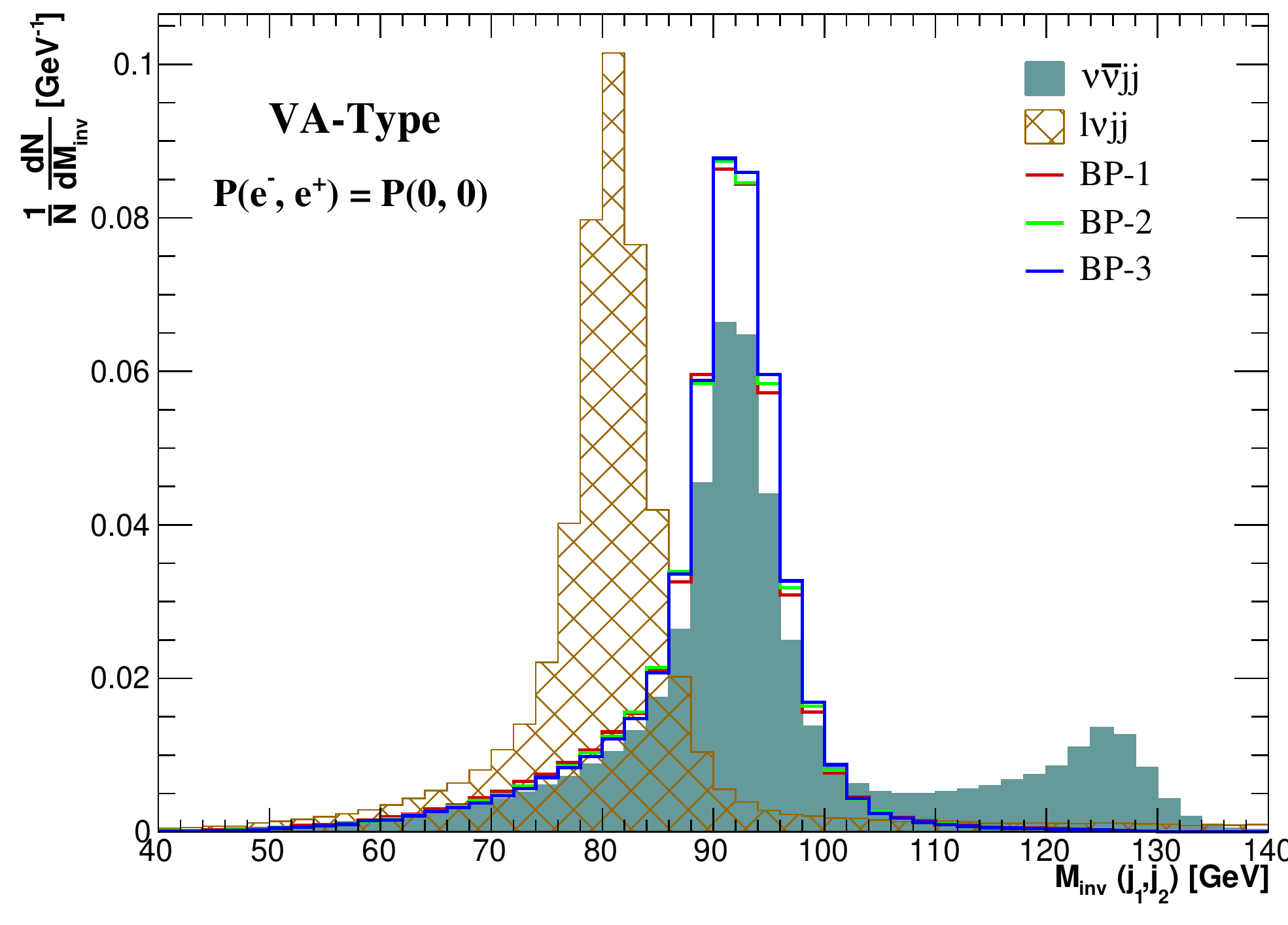}
\includegraphics[width=0.325\linewidth]{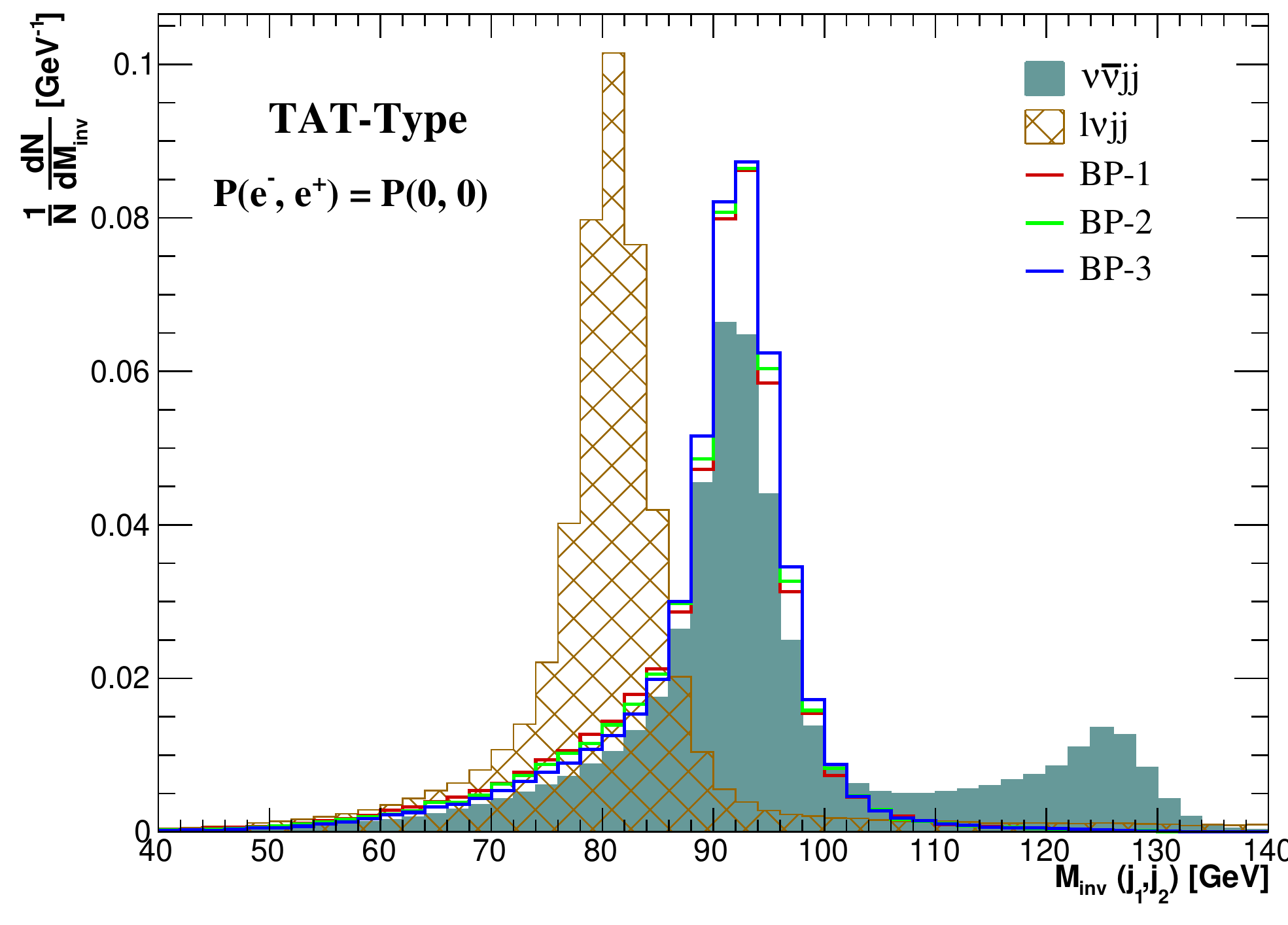}
\includegraphics[width=0.325\linewidth]{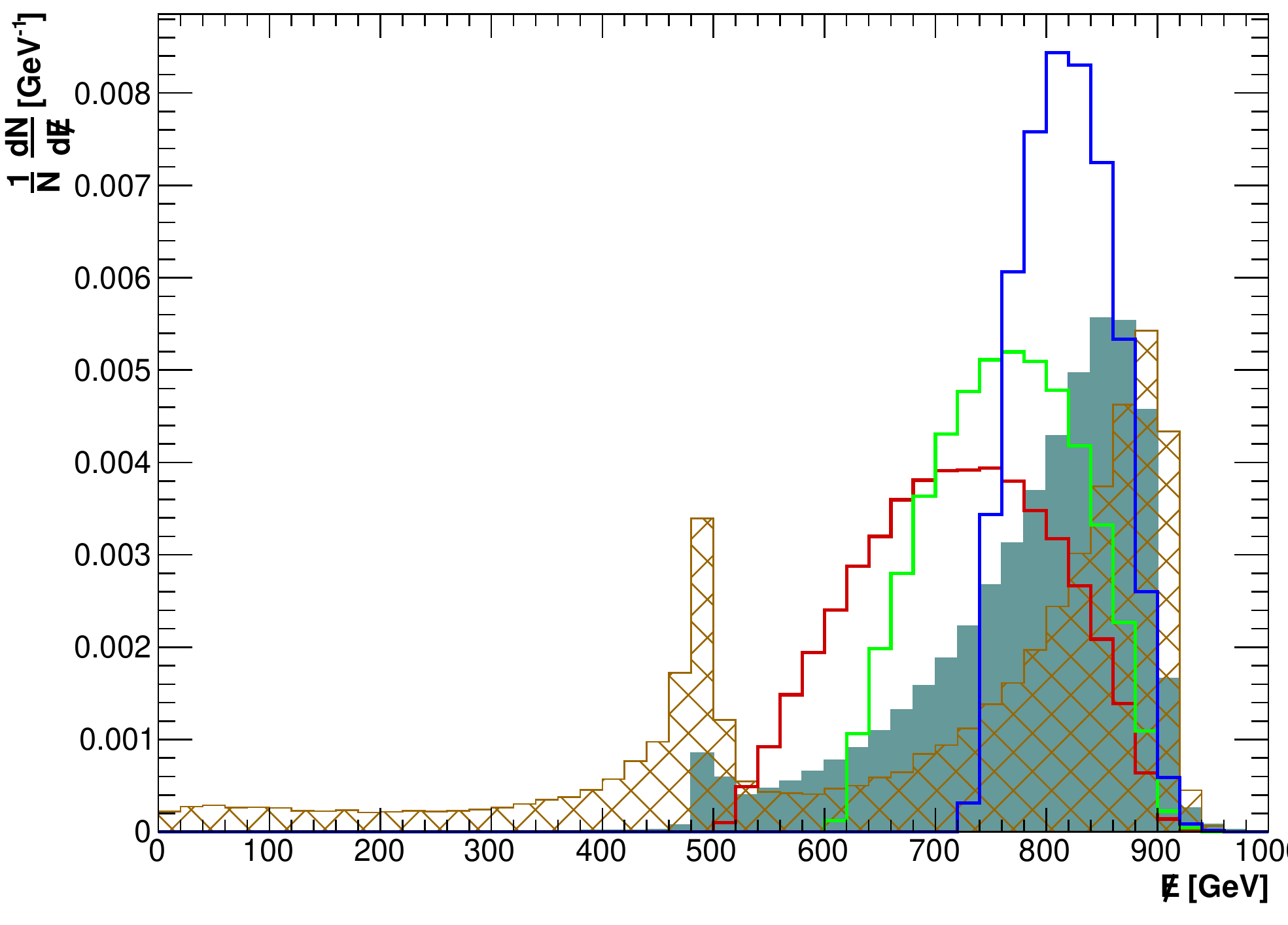}
\includegraphics[width=0.325\linewidth]{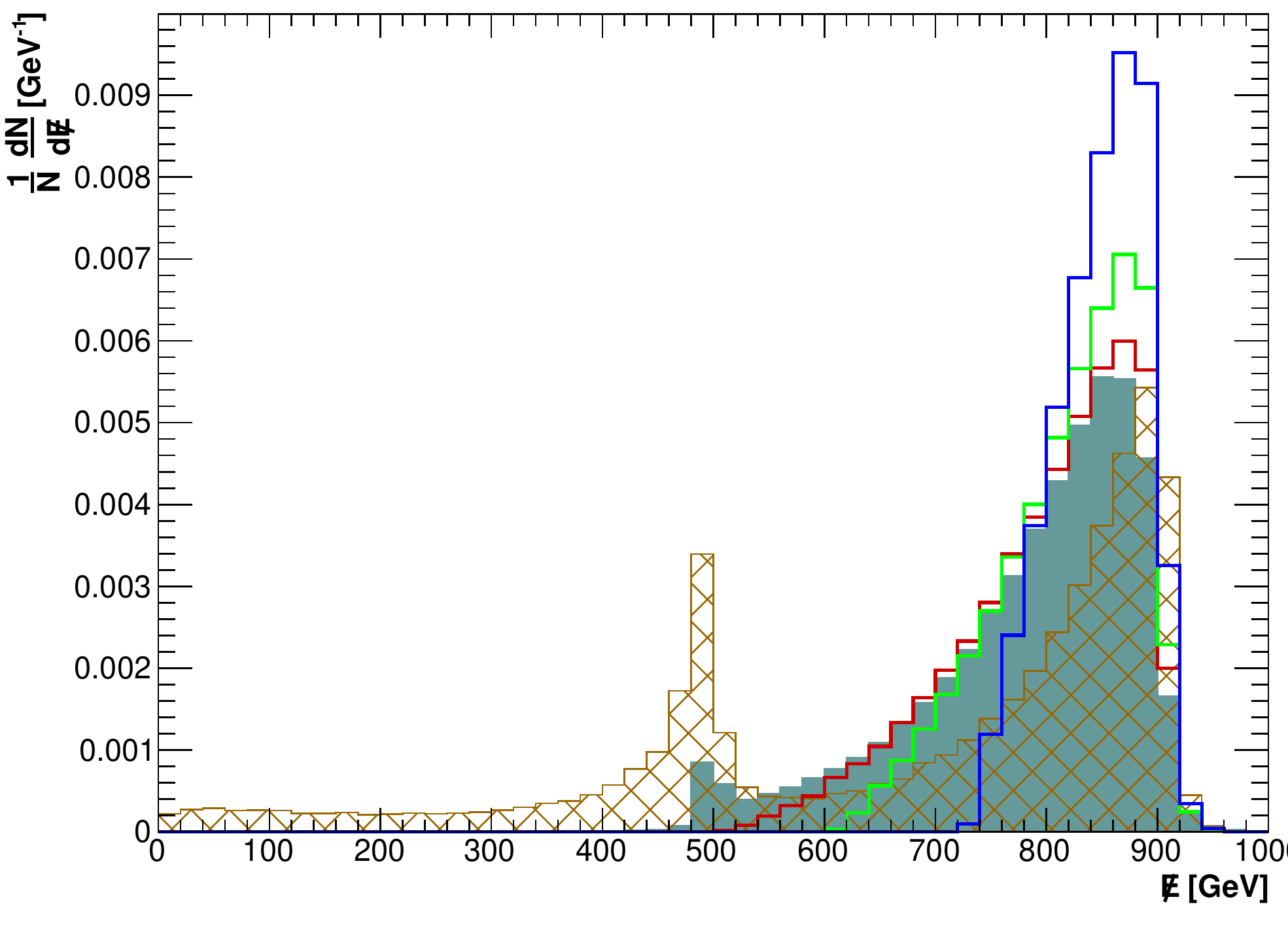}
\includegraphics[width=0.325\linewidth]{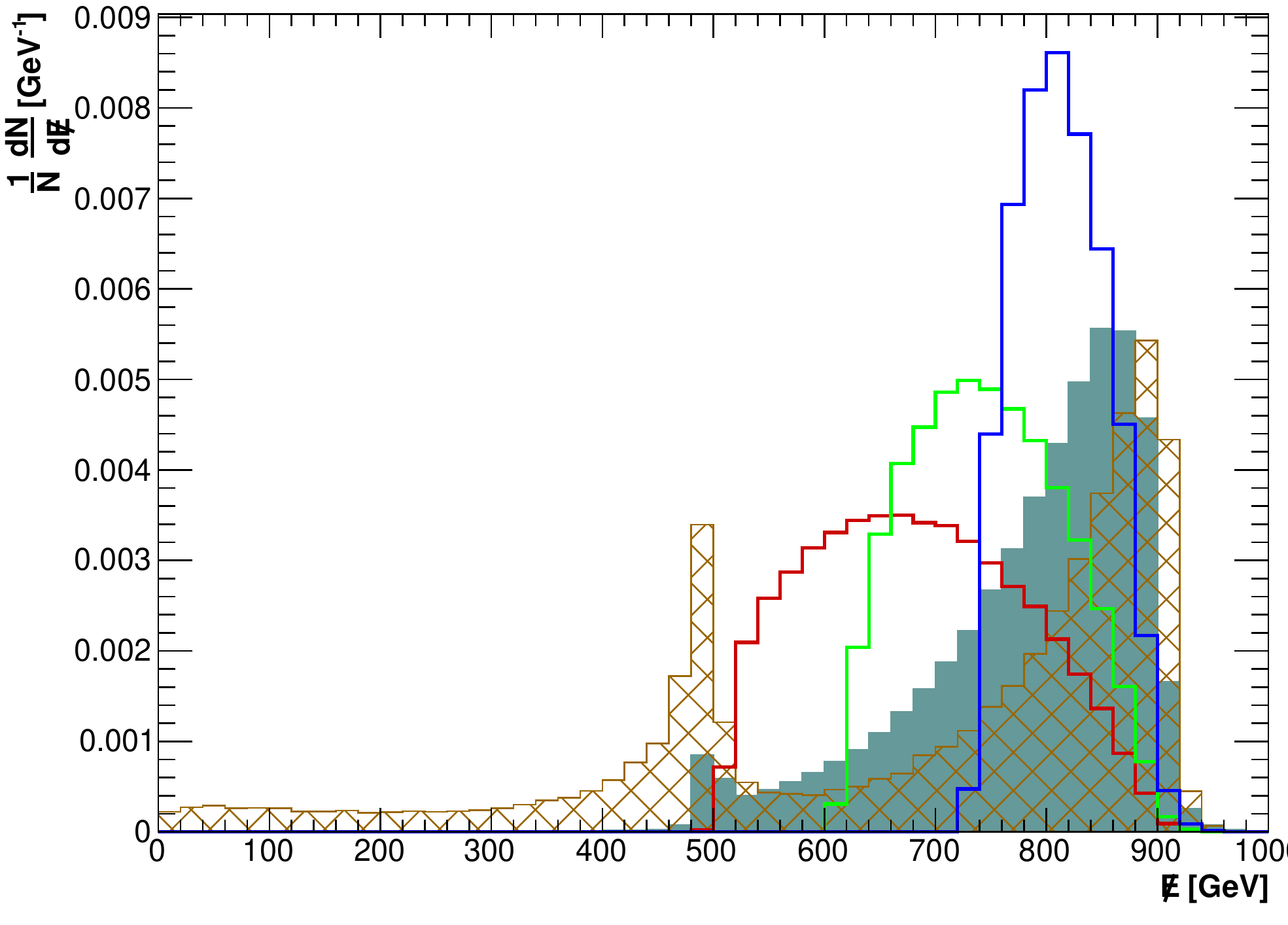}
\includegraphics[width=0.325\linewidth]{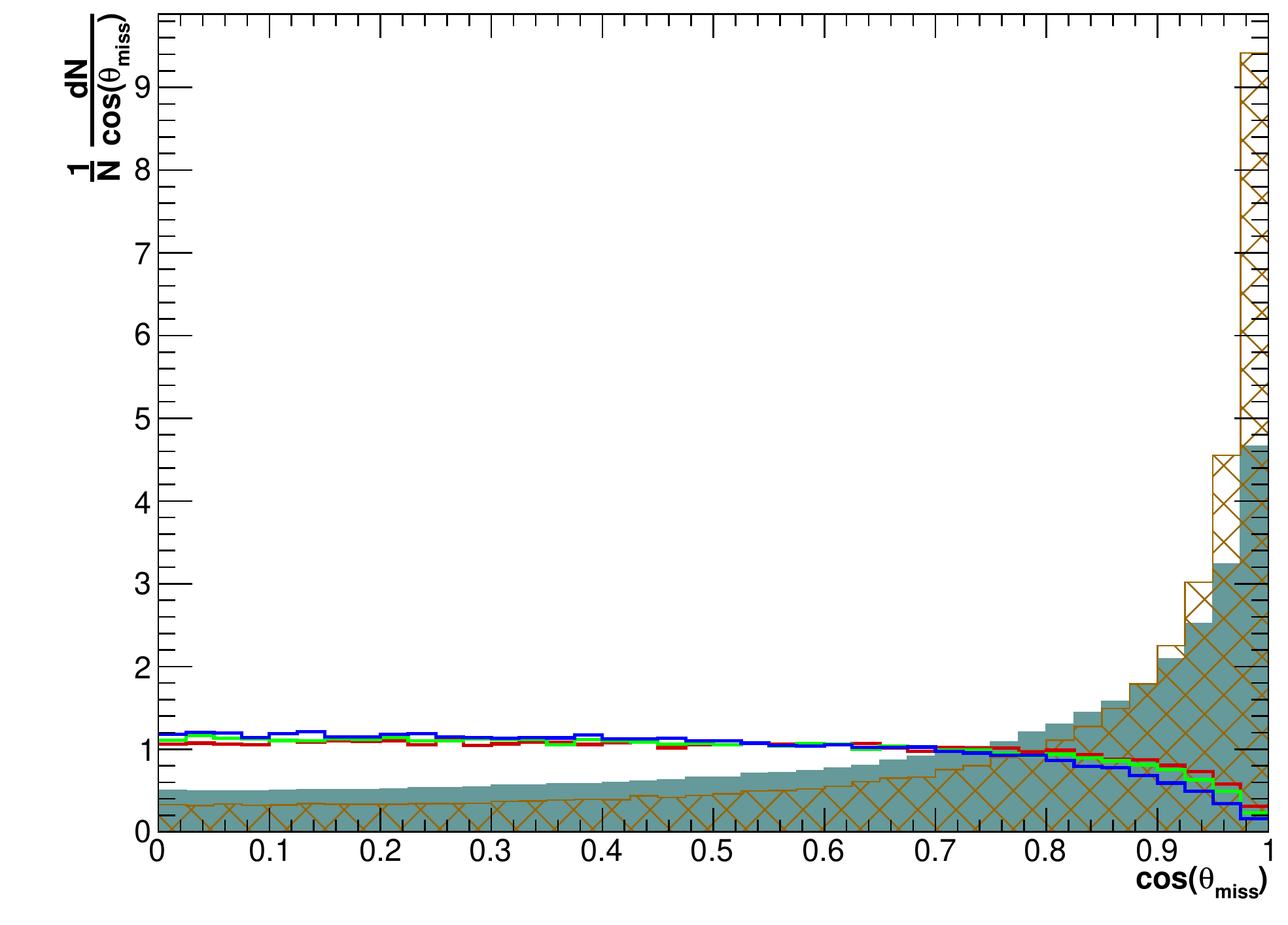}
\includegraphics[width=0.325\linewidth]{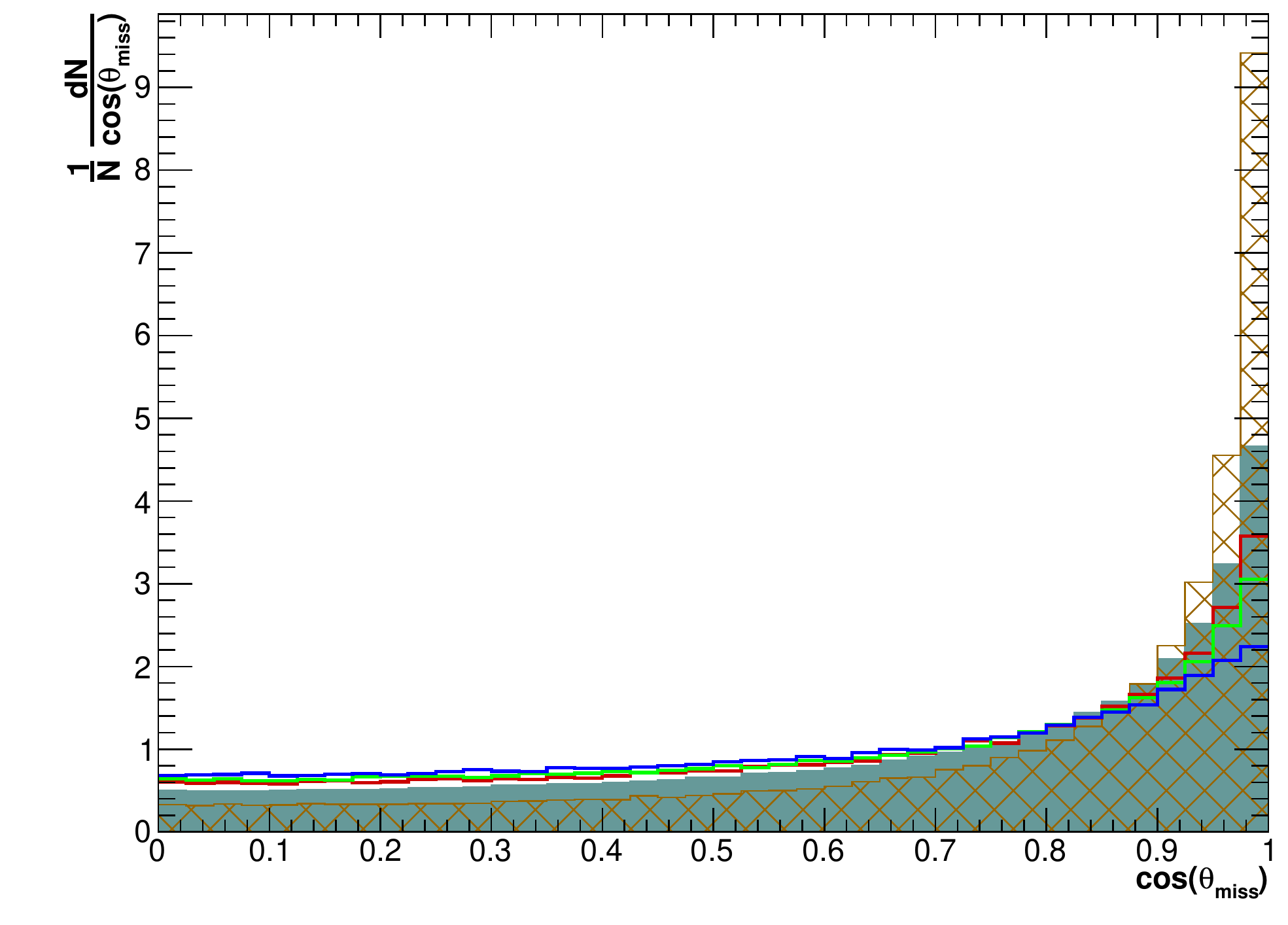}
\includegraphics[width=0.325\linewidth]{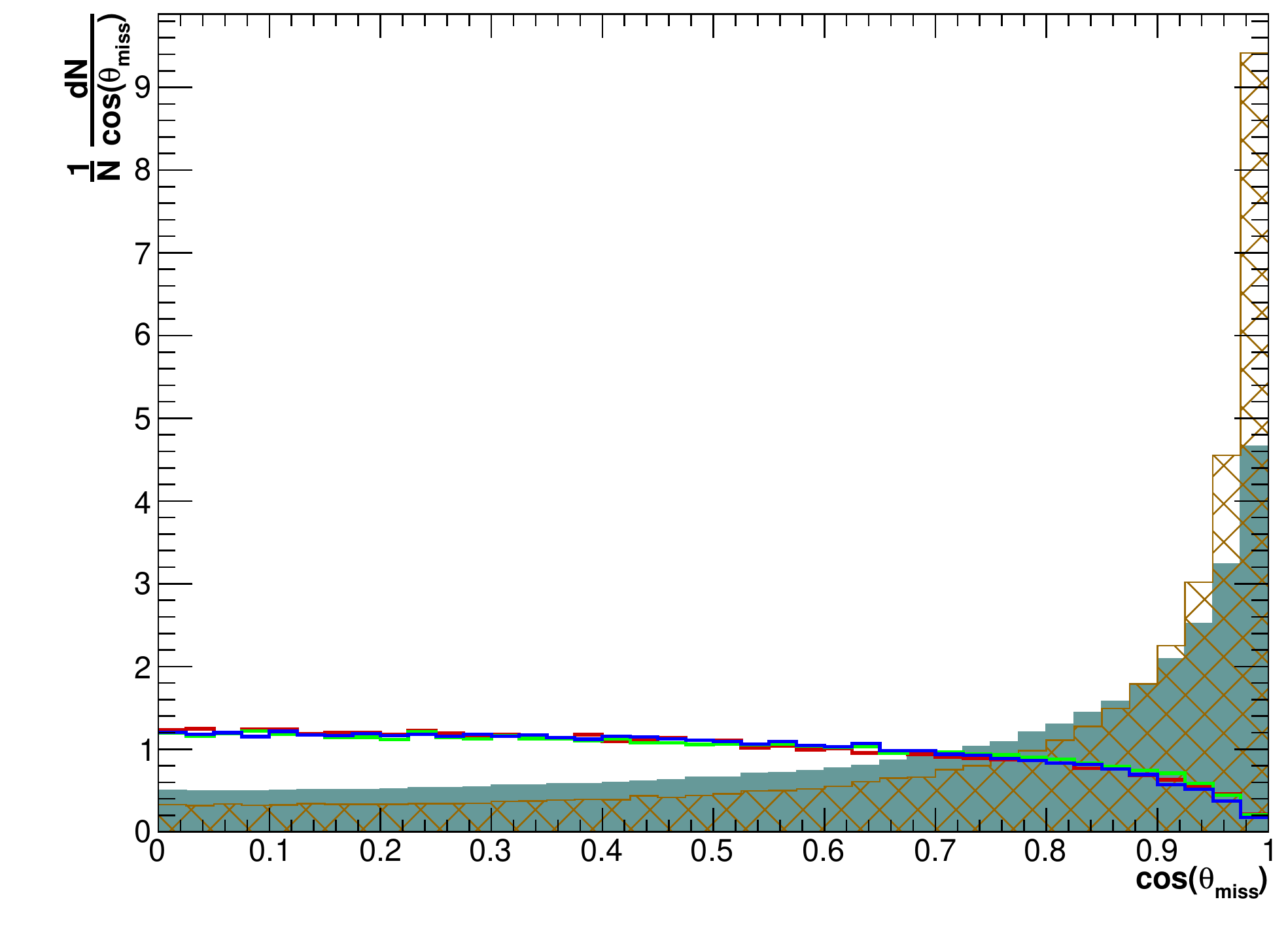}
\includegraphics[width=0.325\linewidth]{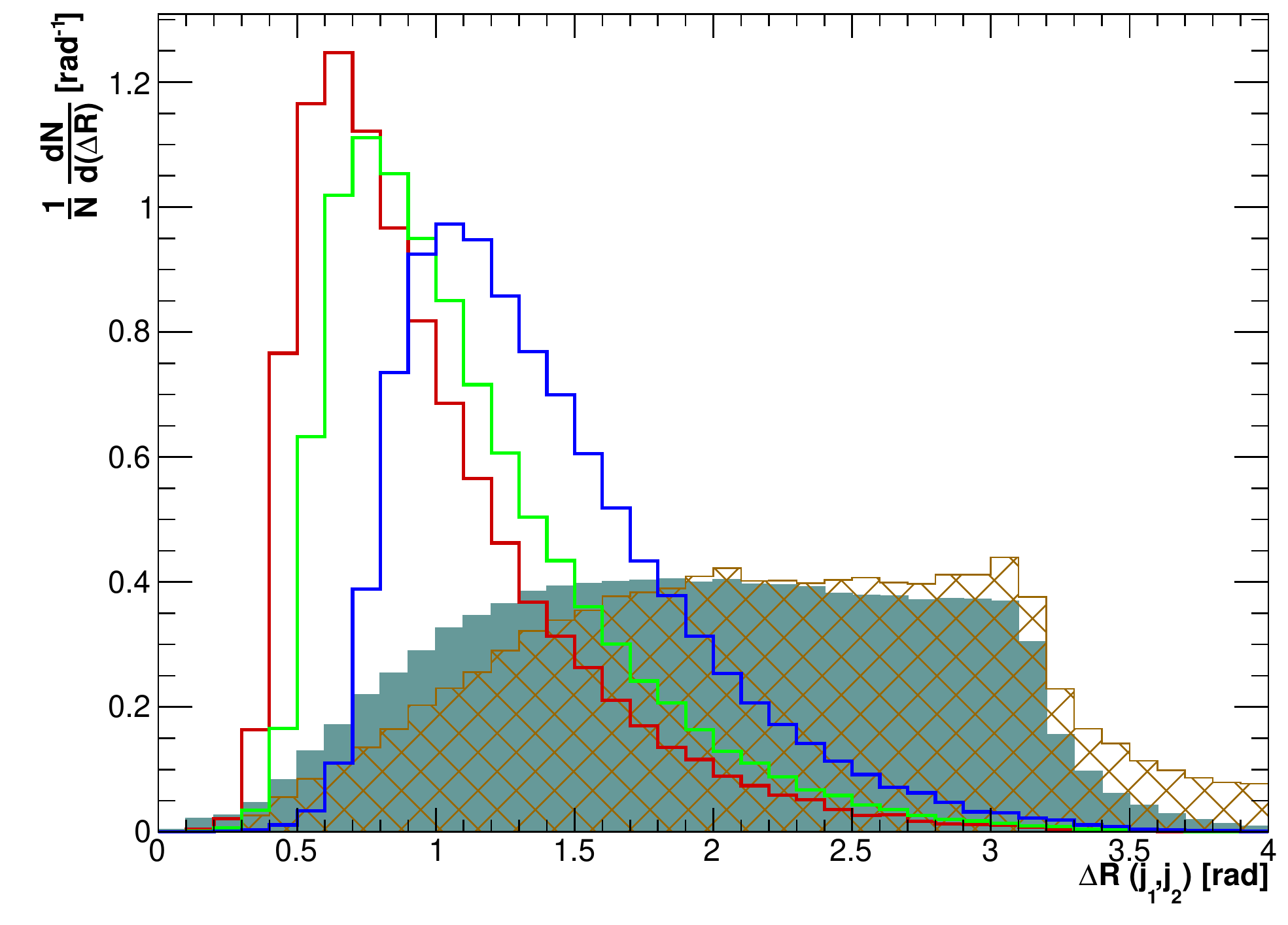}
\includegraphics[width=0.325\linewidth]{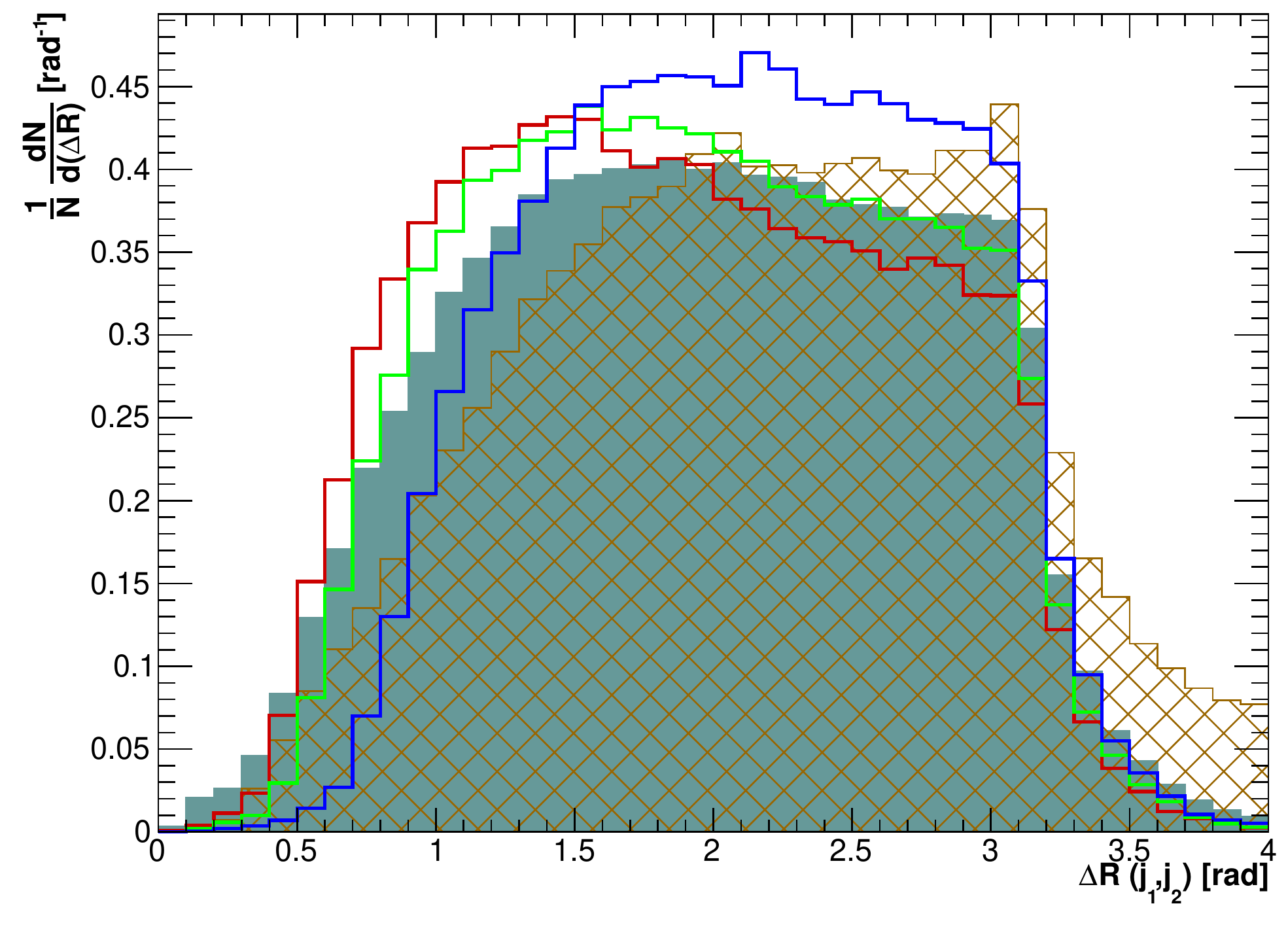}
\includegraphics[width=0.325\linewidth]{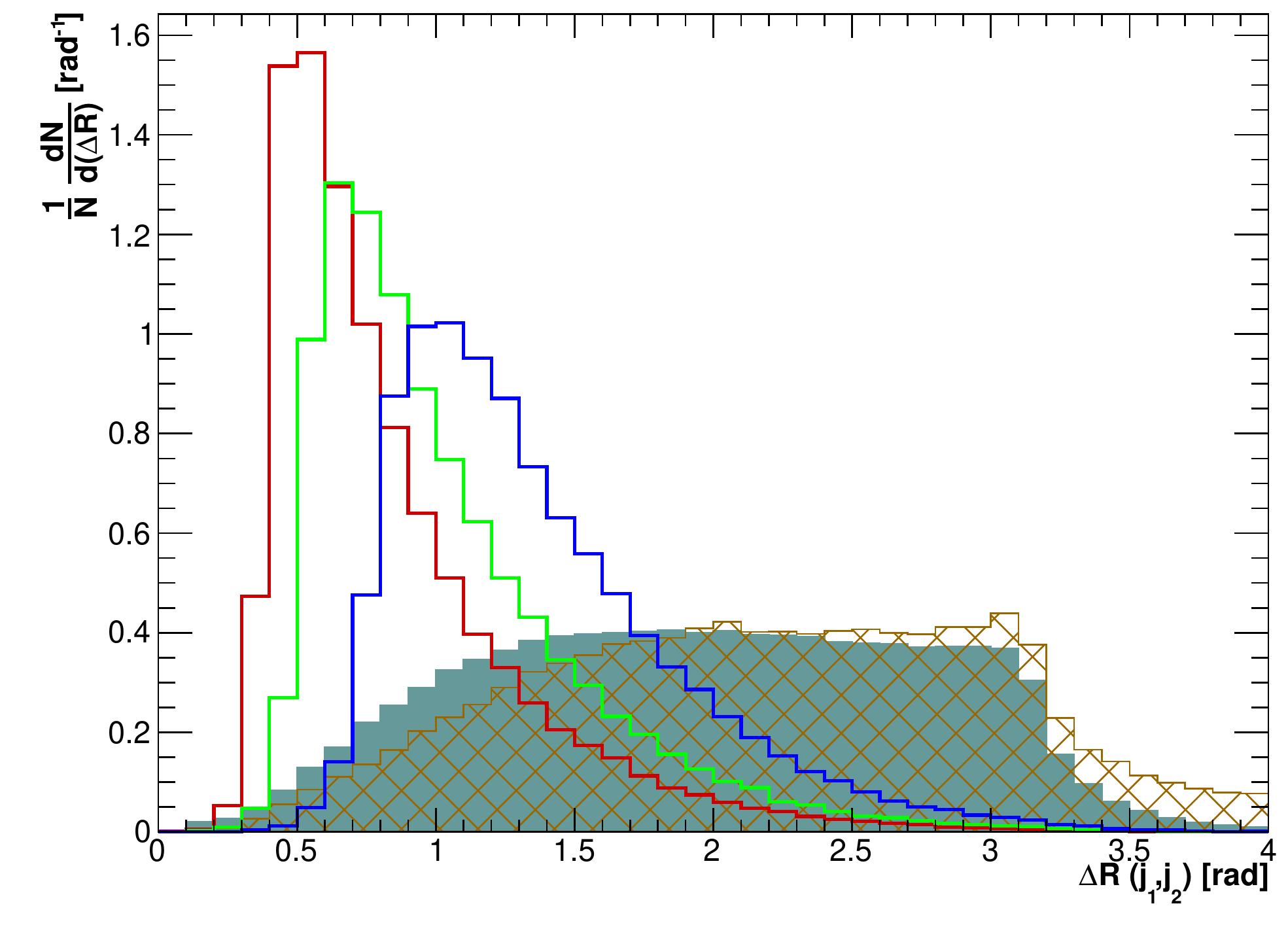}
\caption{Normalized differential distributions for hadronic mono-$Z$ backgrounds (shaded) and signals after baseline selection cuts for the kinematic variables shown in \autoref{table:BPs&Cutsjj} with unpolarized beams. The corresponding distributions for polarized beams are given in \autoref{figure:HistosZjjpol} in the Appendix.}
\label{figure:HistosZjj}
\end{figure*}

\begin{table*}[t]
\caption{\label{table:CutBZjj}%
Cut-flow chart for different operators and BPs for the  mono-$Z$ hadronic signal and backgrounds at $\sqrt{s}=1$ TeV and $\mathcal{L}_{\rm int} = 8\, {\rm ab}^{-1}$. The cut efficiencies are calculated with respect to Eq.~\eqref{eq:BSjj}, with the corresponding event numbers shown in the first row. The background cut efficiencies vary across operator types because of the dynamic nature of the cuts chosen (cf.~\autoref{table:BPs&Cutsjj}).}
\begin{ruledtabular}
\begin{tabular}{l ccc ccc ccc} 
  & \multicolumn{9}{c}{Event Numbers (Cut efficiencies)}\\
  \cline{2-10}
  & \multicolumn{3}{c}{$\nu_{\ell}\overline{\nu_{\ell}}jj $} & \multicolumn{3}{c}{$\ell\nu_{\ell}jj$} & 
  \multicolumn{3}{c}{Signal} \\
  \cline{2-10}
 Selection cuts & BP1 & BP2 & BP3 & BP1 & BP2 & BP3 & BP1 & BP2 & BP3 \\
  \colrule
  \multicolumn{10}{l}{SP type:}\\ 
  \colrule
  BS & \multicolumn{3}{c}{$6.05\times10^6$} & \multicolumn{3}{c}{$4.07\times10^6$} & $2.16\times10^4$ & $1.32\times10^4$ & $2.13\times10^3$ \\
  & \multicolumn{3}{c}{$(100\%)$} & \multicolumn{3}{c}{$(100\%)$} & $(100\%)$ & $(100\%)$ & $(100\%)$ \\
  \multirow{2}{4.5em}{After Final cut} 
   & $3.16\times10^5$ & $6.09\times10^5$ & $5.86\times10^5$ & $1.59\times10^4$ & $4.13\times10^4$ & $4.95\times10^4$ & $8.62\times10^3$ & $ 4.57\times10^3 $ & $ 0.97\times10^3 $ \\
   & $(5.23\%)$ & $(10.08\%)$ & $(9.70\%)$ & $(0.39\%)$ & $(1.01\%)$ & $(1.22\%)$ & $(39.87\%)$ & $(34.60\%)$ & $(45.64\%)$ \\
  \colrule
  \multicolumn{10}{l}{VA type:}\\ 
  \colrule
  BS & \multicolumn{3}{c}{$6.05\times10^6$} & \multicolumn{3}{c}{$4.07\times10^6$} & $6.12\times10^3$ & $4.10\times10^3$ & $1.97\times10^3$ \\
  & \multicolumn{3}{c}{$(100\%)$} & \multicolumn{3}{c}{$(100\%)$} & $(100\%)$ & $(100\%)$ & $(100\%)$ \\
  \multirow{2}{4.5em}{After Final cut} 
   & $ 2.30\times10^6 $ & $ 2.12\times10^6 $ & $ 1.88\times10^6 $ & $ 2.44\times10^5 $ & $ 2.36\times10^5 $ & $ 2.2\times10^5 $ & $ 3.38\times10^3 $ & $ 2.29\times10^3 $ & $ 1.20\times10^3 $ \\
   & $(38.01\%)$ & $(35.04\%)$ & $(31.05\%)$ & $(6.00\%)$ & $(5.81\%)$ & $(5.48\%)$ & $(55.28\%)$ & $(55.93\%)$ & $(61.03\%)$ \\
  \colrule
  \multicolumn{10}{l}{TAT type:}\\ 
  \colrule
  BS & \multicolumn{3}{c}{$6.05\times10^6$} & \multicolumn{3}{c}{$4.07\times10^6$} & $5.30\times10^4$ & $2.67\times10^4$ & $7.46\times10^3$ \\
  & \multicolumn{3}{c}{$(100\%)$} & \multicolumn{3}{c}{$(100\%)$} & $(100\%)$ & $(100\%)$ & $(100\%)$ \\
  \multirow{2}{4.5em}{After Final cut} 
   & $ 3.51\times10^5 $ & $ 3.59\times10^5 $ & $ 5.90\times10^5 $ & $ 1.8\times10^4 $ & $ 1.90\times10^4 $ & $ 4.72\times10^4 $ & $ 2.62\times10^4 $ & $ 1.15\times10^4 $ & $ 3.66\times10^3 $ \\
   & $(5.81\%)$ & $(5.94\%)$ & $(9.76\%)$ & $(0.45\%)$ & $(0.47\%)$ & $(1.16\%)$ & $(49.37\%)$ & $(43.04\%)$ & $(49.12\%)$ \\

\end{tabular}
\end{ruledtabular}
\end{table*}
\begin{table*}[tb]
\caption{\label{table:SigZjj}%
Signal significances of the mono-$Z$ hadronic channel  at $\sqrt{s}=1~{\rm TeV}$ and $\mathcal{L}_{\rm int} = 8~{\rm ab}^{-1}$ after implementing all analysis cuts mentioned in the text. The values in the parentheses denote the significance with a 1\% background systematic uncertainty.}
\begin{ruledtabular}
\begin{tabular}{l ccc ccc ccc} 
   & \multicolumn{9}{c}{Signal significance for ${\cal L}_{\rm int}=8\,{\rm ab}^{-1}$}\\
   \cline{2-10}
   & \multicolumn{3}{c}{Unpolarized beams} & \multicolumn{3}{c}{H20 scenario} & \multicolumn{3}{c}{Optimally polarized beams}\\
   \cline{2-10}
  Operator types & BP1 & BP2 & BP3 & BP1 & BP2 & BP3 & BP1 & BP2 & BP3 \\
\colrule
  SP 
  & $14.8\;(2.7)$ & $5.6\;(0.7)$ & $1.2\;(0.2)$\phantom{0} 
  & \phantom{0}$20.2\;(12.7)$ & $ 8.2\;(4.3)$ & $1.8\;(1.0)$\phantom{0} 
  & \phantom{0}$32.8\;(11.6)$ & $12.9\;(3.4)$ & $2.8\;(0.8)$ \\
  VA 
  & $2.1\;(0.1)$ & $1.5\;(0.1)$ & $0.8\;(0.1)$\phantom{0} 
  & \phantom{0}$7.8\;(2.3)$ & $5.5\;(1.7)$ & $3.1\;(1.0)$\phantom{0}
  & \phantom{0}$12.1\;(2.3)$ & $8.5\;(1.7)$ & $4.8\;(1.0)$ \\
TAT 
  & $41.6\;(7.3)$ & $18.4\;(3.1)$ & $4.6\;(0.6)$\phantom{0} 
  & \phantom{0}$54.7\;(34.7)$ & $26.3\;(16.5)$ & $6.7\;(3.6)$\phantom{0} 
  & \phantom{0}$87.0\;(31.6)$ & $41.4\;(14.2)$ & $10.5\;(2.8)$ \\

 \end{tabular}
\end{ruledtabular}  
\end{table*}

Cut 1: 
The contribution to the $\nu\overline{\nu}jj$ background comes mainly from the process $e^+e^-\to Z+\gamma/Z$, where the dijet comes from the $\gamma/Z$ boson, whereas the $\ell\nu jj$ background comes largely from $e^+e^-\to W^+W^-$, where the $W$ boson pair decays semileptonically. So, by selecting a dijet-invariant mass around the $Z$ mass, i.e., $85\le M_{\rm inv}(jj)\le100\text{ GeV}$, the backgrounds are reduced considerably. In particular, the $\ell\nu jj$ background with a peak around the $W$ boson mass (see \autoref{figure:HistosZjj} top row) is reduced by $90\%$, whereas $42\%$ of the $\nu\overline{\nu}jj$ background after baseline selection is rejected (cf.~\autoref{table:CutBZjj}). 
Cut 2: 
As was the case for the leptonic channel, the missing energy ($\slashed{E}$) distribution for the signal depends on the DM mass. As the mass of the DM increases, the distribution of missing energy ($\slashed{E}$) becomes narrower and shift toward the value of the c.m. energy. So we choose a dynamic cut (with respect to the benchmark points) to efficiently select the signal events and reject as many background events as possible (as shown in \autoref{table:BPs&Cutsjj}). We see that this reduces the $\nu\overline{\nu}jj$ background up to $16\%$ and the $\ell\nu jj$ background up to $1.7\%$ of their respective values after BS, varying with different operators, while keeping more than $50\%$ of the signal events.

Cut 3: 
The cosine of the polar angle of the missing momentum is again very efficient as a variable selection cut for pretty much the same reason as in the leptonic channel, that heavier invisible particles tend to scatter at larger angles with beam direction. 

\begin{figure*}[ht!]
\centering 
\includegraphics[width=0.51\linewidth]{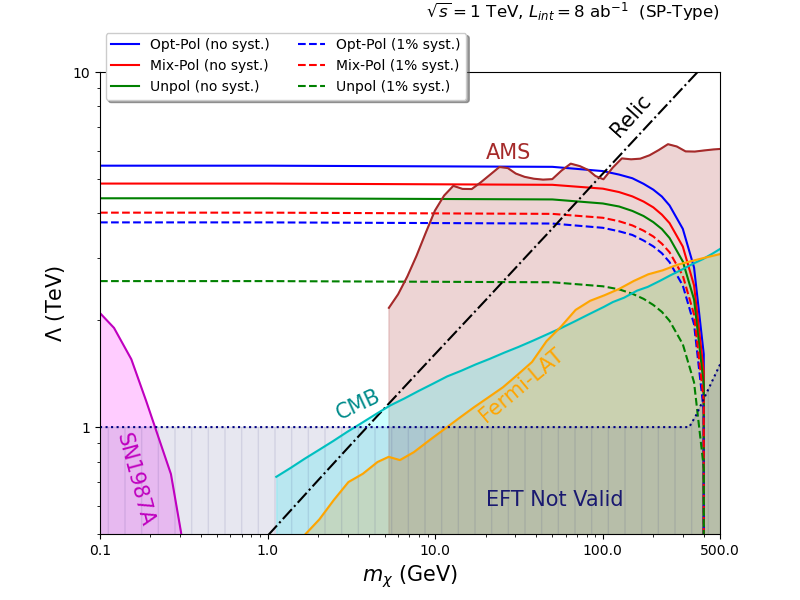}
\hspace{-0.7cm}
\includegraphics[width=0.51\linewidth]{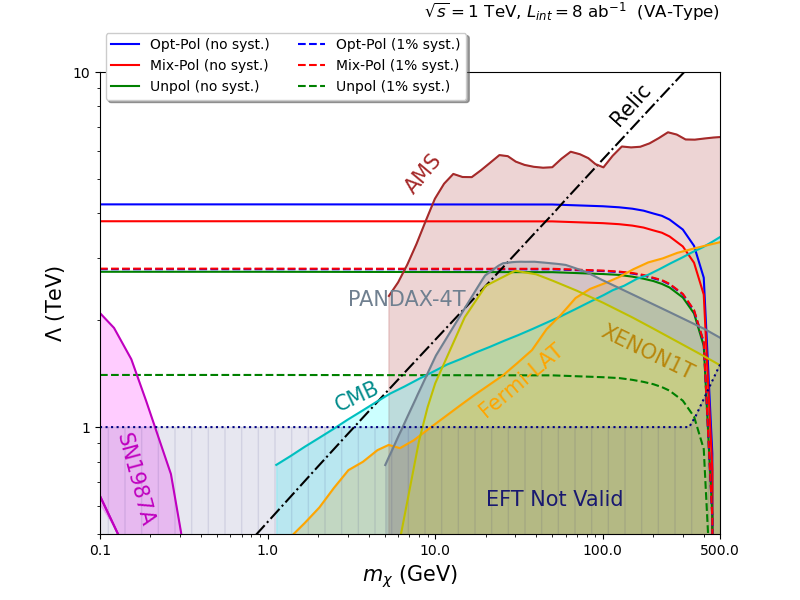} \\
\vspace{-0.1cm}
\includegraphics[width=0.51\linewidth]{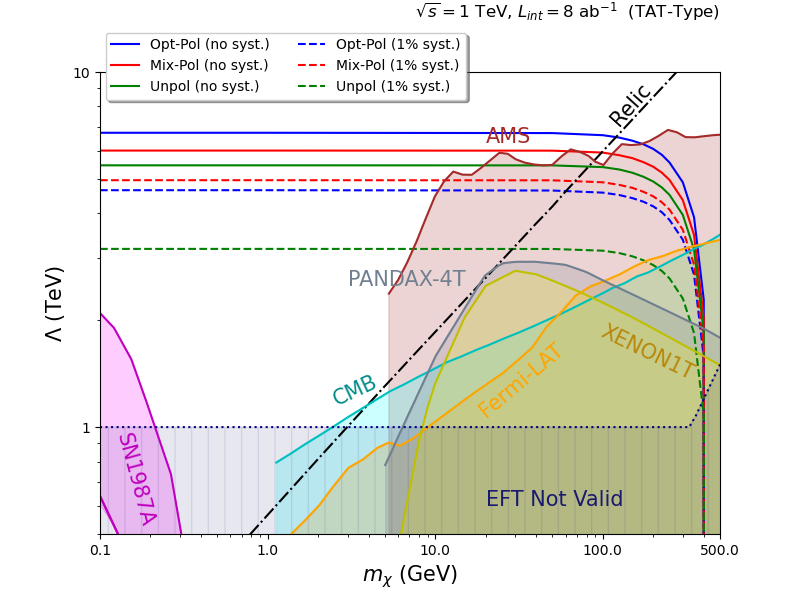}
\caption{$3\sigma$ sensitivity contours in the mono-$Z$ hadronic channel. The labels are the same as in \autoref{figure:ContourZ}. 
}
\label{figure:ContourZjj}
\end{figure*}

Cut 4: 
We put a cut on $\Delta R_{jj}$ to remove a significant amount of events for both the backgrounds that are less collinear than the signals where the jets come from on shell resonant decay of the $Z$ boson, as discussed for the leptonic channel, and as shown in \autoref{figure:ContourZjj} (final row). We avoid using this cut for the VA-type operator because, similar to the leptonic case, here also it appears not to be sensitive (also evident from the distribution) after the previous three cuts. In \autoref{table:CutBZjj}, we reported the number of selected events and the selection efficiencies after the application of all the cuts for unpolarised beams. The corresponding numbers for the polarised beams are in \autoref{table:CutBZjjpol} in the Appendix. 
\begin{figure*}[ht]
\centering
 \includegraphics[width=0.9\linewidth]{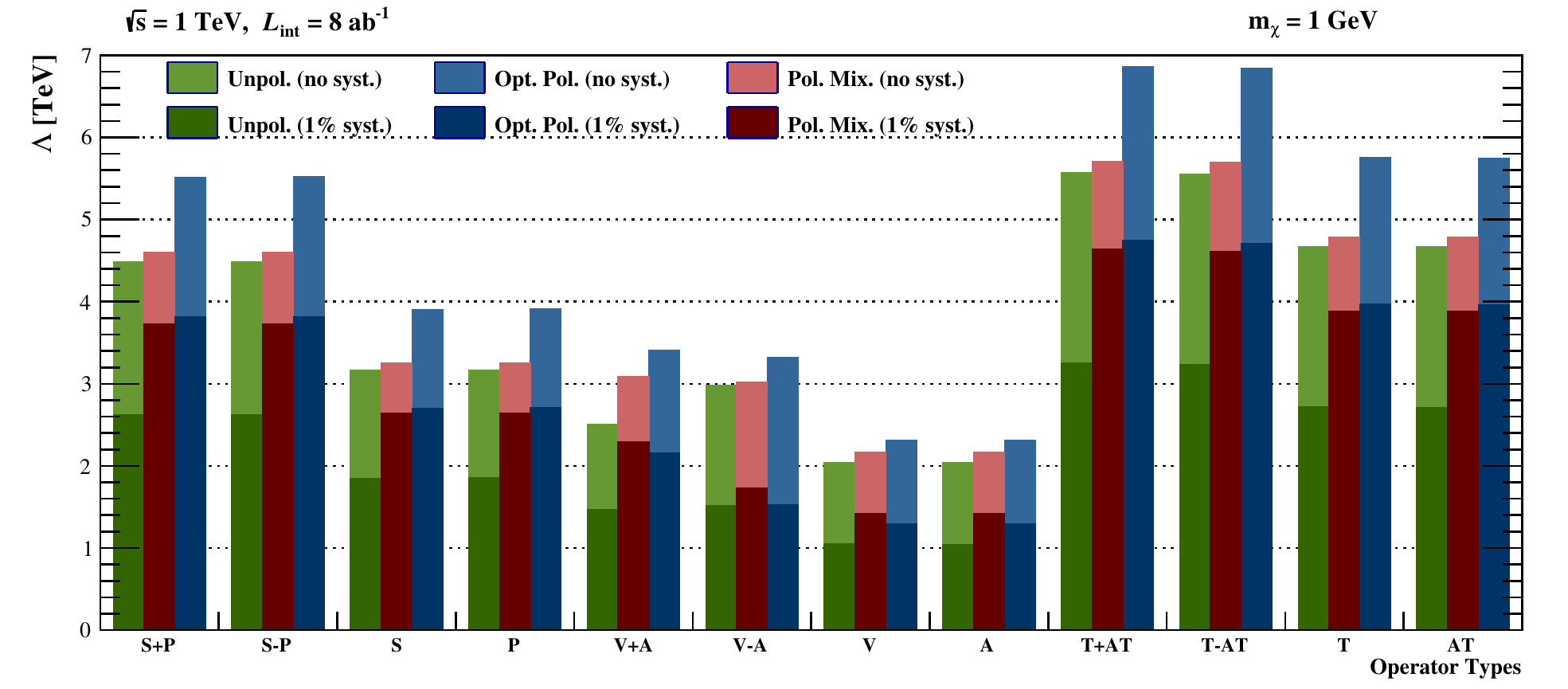}
 \caption{$3\sigma$ sensitivity reach of different operators in the mono-$Z$ hadronic channel with DM mass of $1$ GeV  at $\sqrt{s}=1$ TeV and $\mathcal{L}_{\rm int} = 8~{\rm ab}^{-1}$. The green, red  and blue bars show the sensitivities with unpolarized, mixed and optimally polarized beams respectively and the lighter (darker) shade corresponds to zero (1\%) background systematics. }
 \label{fig:barZjj}
\end{figure*}
\subsubsection{Results} 
The signal significances calculated using Eq.~\eqref{eq:significance} are tabulated in \autoref{table:SigZjj}. We see similar behavior for the different BPs as in the previously discussed channels, i.e., enhanced signal significance with decreasing mass of the DM. The selection cuts are most efficient for SP- and TAT-type operators. For BP$1$ we remove more than $94\%$--$97\%$ of the background events, while keeping more than $43\%$ of the signal events for the two operator types, yielding a large signal significance, especially for the TAT-type operator with polarized beams. \par
Varying the DM mass, we display the $3\sigma$ sensitivity contours for all three operators in  \autoref{figure:ContourZjj}. It is clear that the TAT-type operator has the best sensitivity, which reaches up to $5.5$~TeV with unpolarized beams and $6.7$~TeV with optimally polarized beams. The SP-type operator also has a sensitivity comparable to the monophoton channel and can reach up to $4.4\; (5.5)$~TeV with unpolarized (optimally polarized) beams. The VA-type operator has a modest sensitivity in this channel, only up to $2.7\; (4.2)$~TeV with unpolarized (optimally polarized) beams. The mixed polarization case of the H20 scenario also produces good numbers particularly with systematic uncertainty outperforming the optimal polarization case with the $3\sigma$ reach of $\Lambda$ reaching up to $4.9$ ($4.0$), $3.8$ ($2.8$), and $6.0$ ($5.0$), respectively, for SP-, VA-, and TAT-type operators without (with) uncertainty in background estimation.

We also compare the hadronic mono-$Z$ sensitivities for different operator types in \autoref{fig:barZjj} for a fixed DM mass of 1 GeV. The green, blue and red bars show the $3\sigma$ sensitivity with unpolarized, optimal and mixed polarization case, respectively, and the lighter (darker) shade corresponds to zero (1\%) background (both) systematics. We again find the best sensitivity for T$\pm$AT operators, followed by the S$\pm$P operators, while the vector- and axial-vector-type operators have the weakest sensitivity in this channel, mainly because of its similarity to the background in the event distributions, just as in the leptonic mono-$Z$ case. In general, the hadronic mono-$Z$ channel is more effective than the leptonic one, primarily because of a larger signal cross section due to higher hadronic branching ratio (69.9\%) of the $Z$ boson, as compared to its leptonic branching ratio to electrons and muons (only 6.7\%).  
\begin{table*}[tb]
\caption{\label{tab:summary}%
Summary of our results for the $3\sigma$ sensitivity reach of the cutoff scale $\Lambda$ in the three different channels discussed in the text. Here we have fixed the DM mass at 1 GeV. The numbers in parentheses are with 1\% background systematics. The numbers in bold show the highest $\Lambda$ value that can be probed for a given operator.}
\begin{ruledtabular}
\begin{tabular}{lcccc} 
     &  & \multicolumn{3}{c}{$3\sigma$ sensitivity reach of $\Lambda$ (TeV)}\\
    \cline{3-5}
    Process type & Beam configuration\phantom{0} & SP & VA & TAT \\
    \colrule
    \multirow{3}{5em}{Mono-$\gamma$} 
    & Unpolarized & \phantom{0}5.37 (3.15)\phantom{0} & \phantom{0}5.78 (3.39)\phantom{0} & \phantom{0}5.80 (3.41) \\
    & Optimal pol & {\bf 6.60} (3.38) & {\bf 8.76} (5.95) & {\bf 7.14} (4.73) \\
    & H20 scenario & 5.84 ({\bf 4.60}) & 7.86 ({\bf 5.99}) & 6.32 ({\bf 4.97}) \\
    \colrule
    \multirow{3}{5em}{Mono-$Z$ leptonic} 
    & Unpolarized & 3.47 (2.90) & 1.73 (1.45) & 4.22 (3.53) \\
    & Optimal pol & 4.30 (3.98) & 2.69 (2.56) & 5.24 (4.85) \\
    & H20 scenario & 3.84 (3.75) & 2.41 (2.36) & 4.68 (4.57) \\
    \colrule
    \multirow{3}{5em}{Mono-$Z$ hadronic} 
    & Unpolarized & 4.41 (2.58) & 2.74 (1.40) & 5.46 (3.18) \\
    & Optimal pol & 5.45 (3.77) & 4.24 (2.78) & 6.74 (4.65) \\
    & H20 scenario & 4.85 (4.02) & 3.80 (2.79) & 6.01 (4.96) \\

\end{tabular}
\end{ruledtabular}
\end{table*}

\section{Conclusion} \label{sec:5}
Very little is known about the nature of the DM and its interaction with the SM particles. It is possible to think of a scenario where DM only couples to the SM leptons, but not to quarks at tree level. We have explored the physics potential of the future $e^+e^-$ colliders in probing such {\it leptophilic} DM in a model-independent way. As a case study, we have taken the $\sqrt s=1$ TeV ILC with an integrated luminosity of $8$ ab$^{-1}$ and have analyzed the pair-production of fermionic DM using leptophilic dimension-six operators of all possible bilinear structures, namely, scalar-pseudoscalar, vector-axial vector, and tensor-axial tensor. We have performed a detailed cut-based analysis for each of these operators in three different channels based on the tagged particle, namely, monophoton,  mono-$Z$ leptonic, and hadronic. 

We have taken into account one of the most important and powerful features of lepton colliders, i.e., the possibility of beam polarization with different degrees of polarization and helicity orientations. We find that the ${\rm sign}(P(e^-),P(e^+))=(+,+)$ beam configuration is optimal for the SP- and TAT-type operators, while the $(+,-)$ configuration is better for probing the VA-type operators. The maximum value of the cutoff scale $\Lambda$ that can be probed in each channel at $3\sigma$ is given in \autoref{tab:summary}. We find that the monophoton channel provides the best sensitivity across all operator types both with and without background systematics. The mono-$Z$ hadronic channel also performs well particularly for the TAT-type operator with systematics. A comparison between the sensitivity reaches of different operators can also be seen from Figs.~\ref{fig:barG}, \ref{fig:barZll}, and \ref{fig:barZjj}. 
We also demonstrate the complementarity of our lepton collider study with other existing direct and indirect detection searches for leptophilic DM (cf. Figs.~\ref{figure:ContourG}, \ref{figure:ContourZ}, and \ref{figure:ContourZjj}). In particular, we show that lepton colliders will be able to provide the best-ever sensitivity in the still unexplored light DM regime.

\section*{Acknowledgments}
S.K. would like to thank Kajari Mazumdar for useful discussions. He would also like to thank Tanumoy Mandal and Suman Kumar Kundu for the help with using \texttt{ROOT} software package. 
P.K.D. thanks K.~C.~Kong and Partha Konar for the discussions. B.D. would like to thank Juri Smirnov for a discussion on the feebly interacting massive particle DM possibility. A.G. would like to thank the Department of Physics, IISER Pune and Chungnam National University for their support. The work of P.K.D. is partly supported by the SERB Grant No.~EMR/2016/002651. The work of B.D. is supported in part by the U.S. Department of Energy under Award No.~DE-SC0017987 and by a Fermilab Intensity Frontier Fellowship. This work was partly performed at the Aspen Center for Physics, which is supported by National Science Foundation Grant No. PHY-1607611. Work of A.G. is supported by the National Research Foundation of Korea (NRF-2019R1C1C1005073).\par

\appendix 
\section{KINEMATIC DISTRIBUTIONS AND CUT-FLOW TABLES FOR POLARIZED BEAMS} 
\begin{figure*}[!ht]
\includegraphics[width=0.325\linewidth]{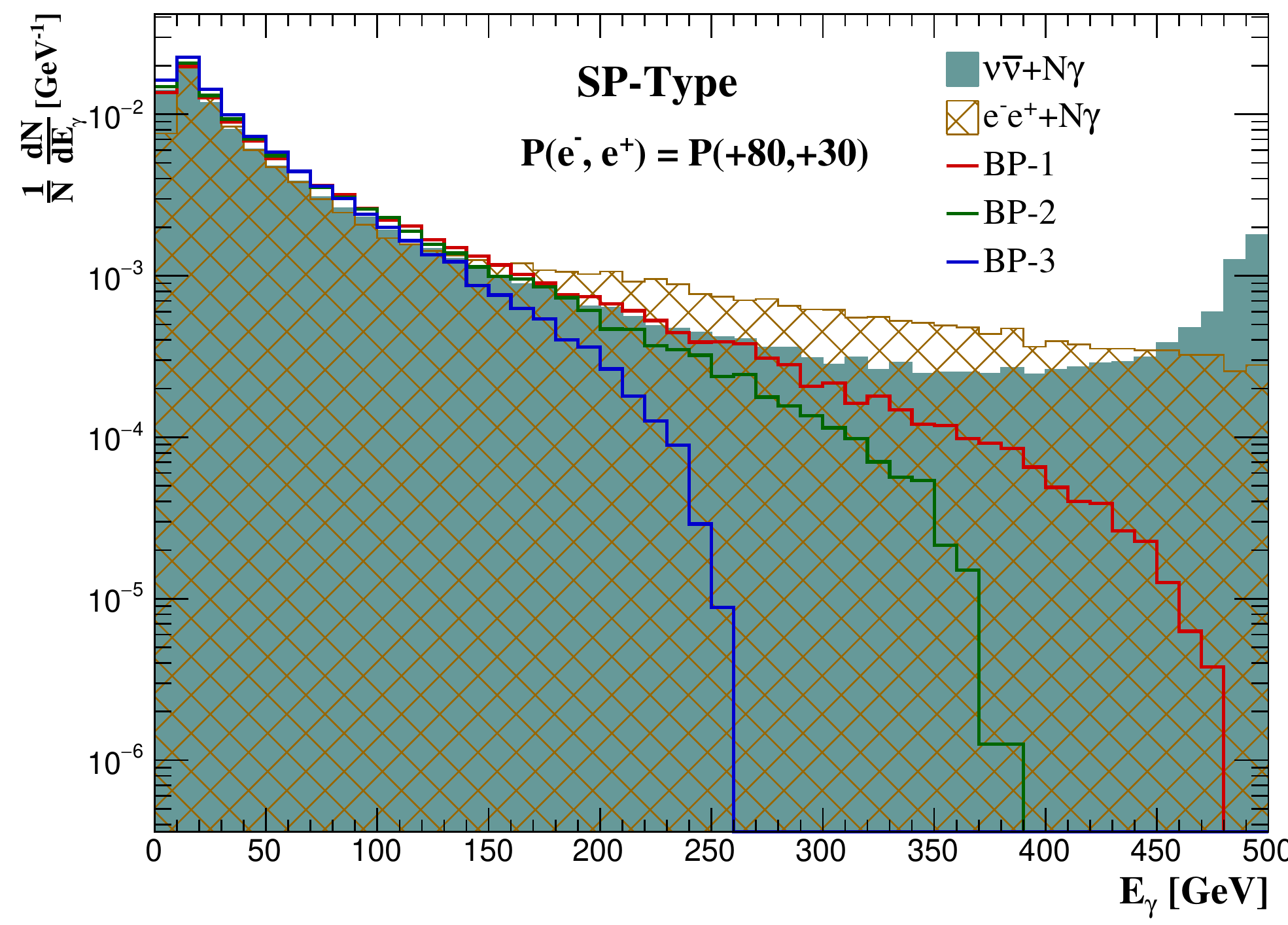}
\includegraphics[width=0.325\linewidth]{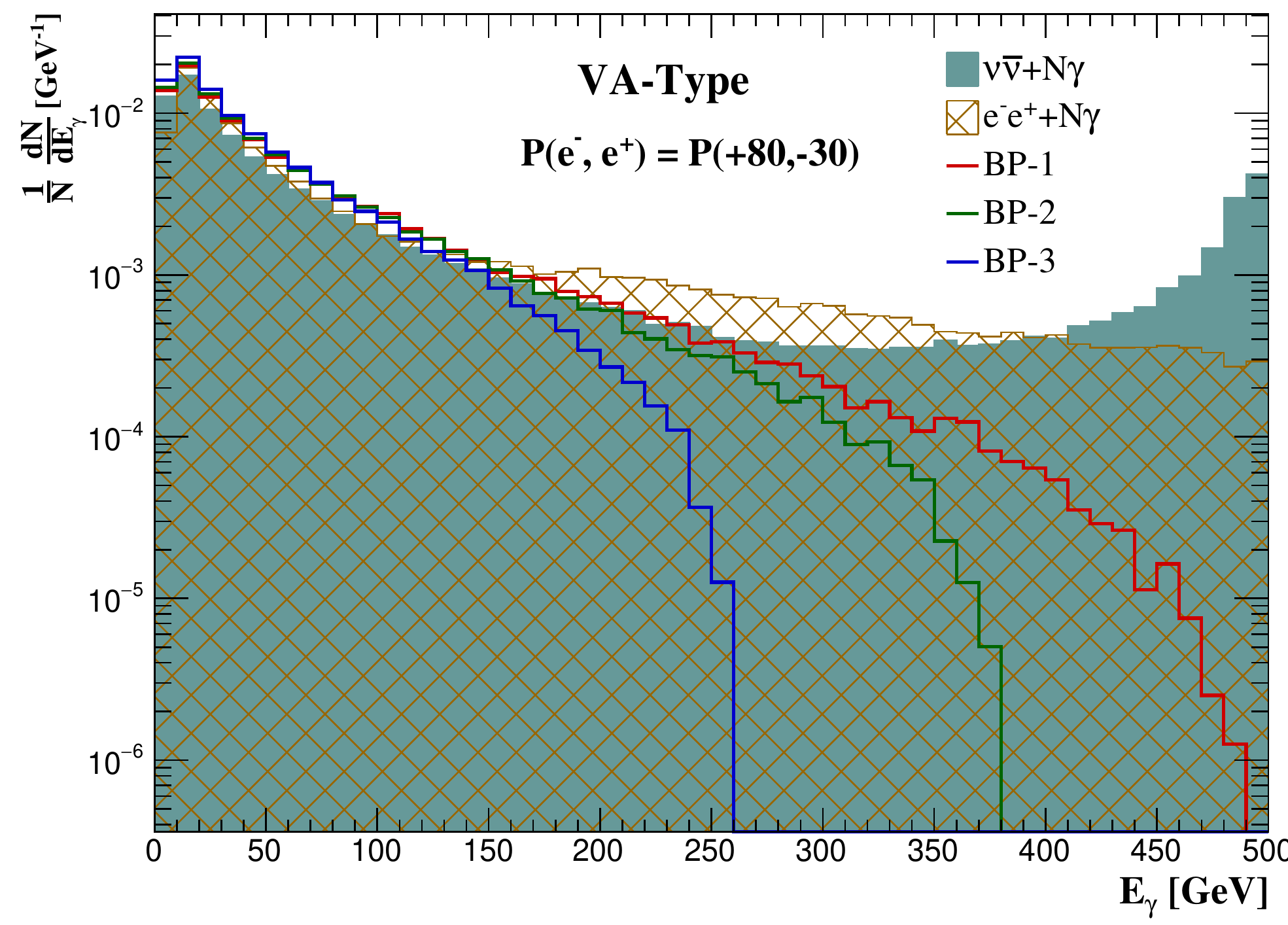}
\includegraphics[width=0.325\linewidth]{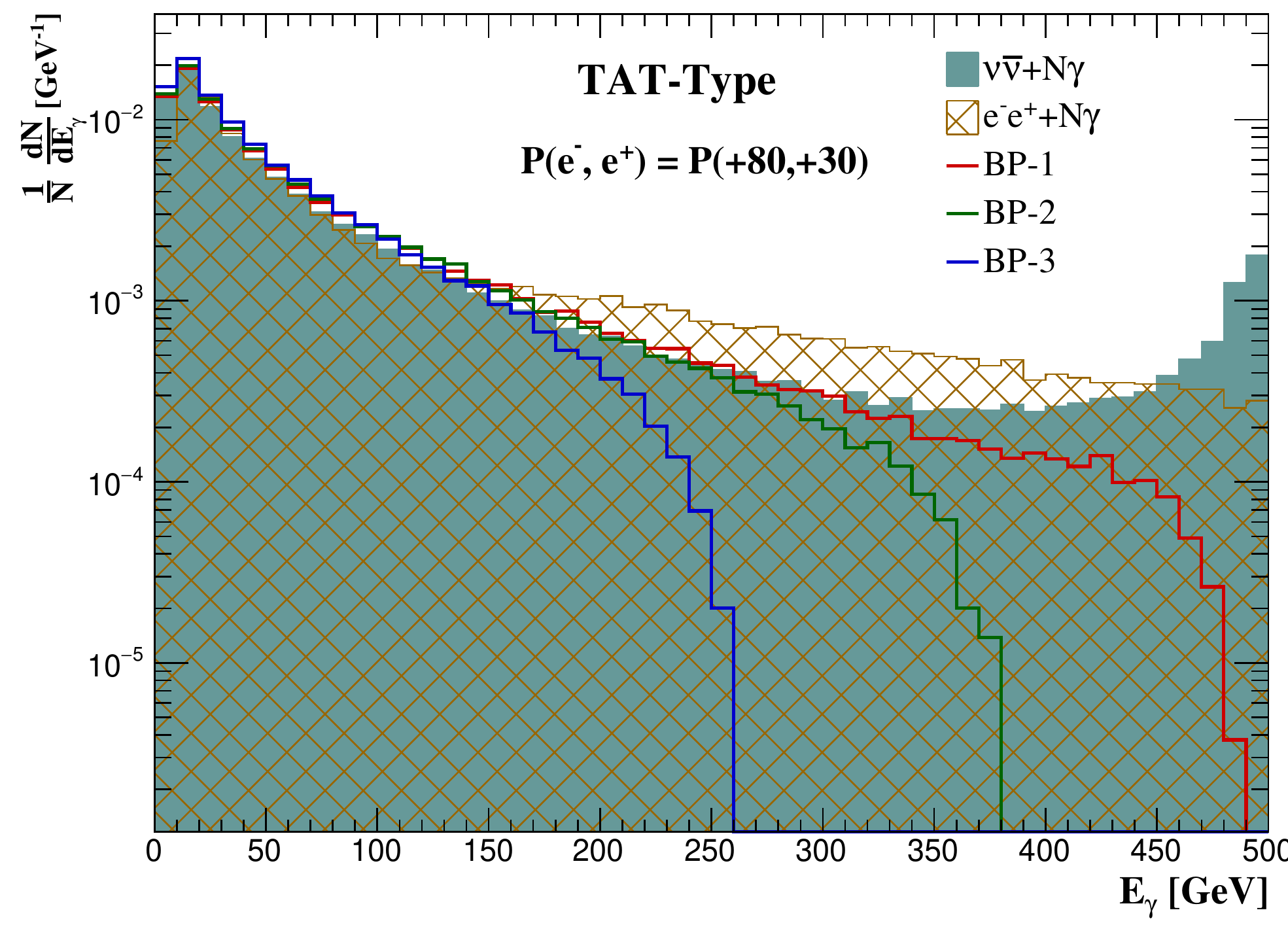}
\includegraphics[width=0.325\linewidth]{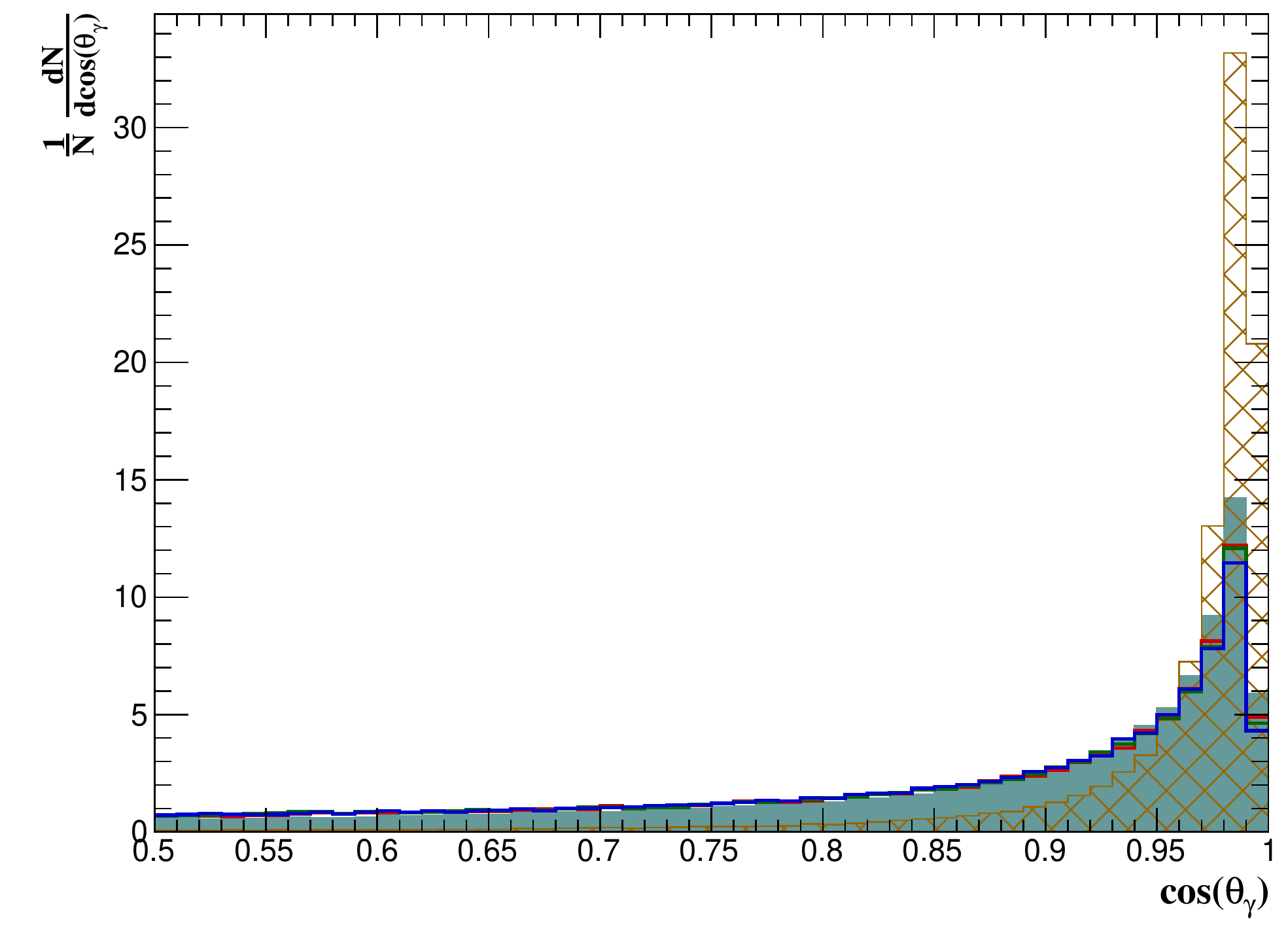}
\includegraphics[width=0.325\linewidth]{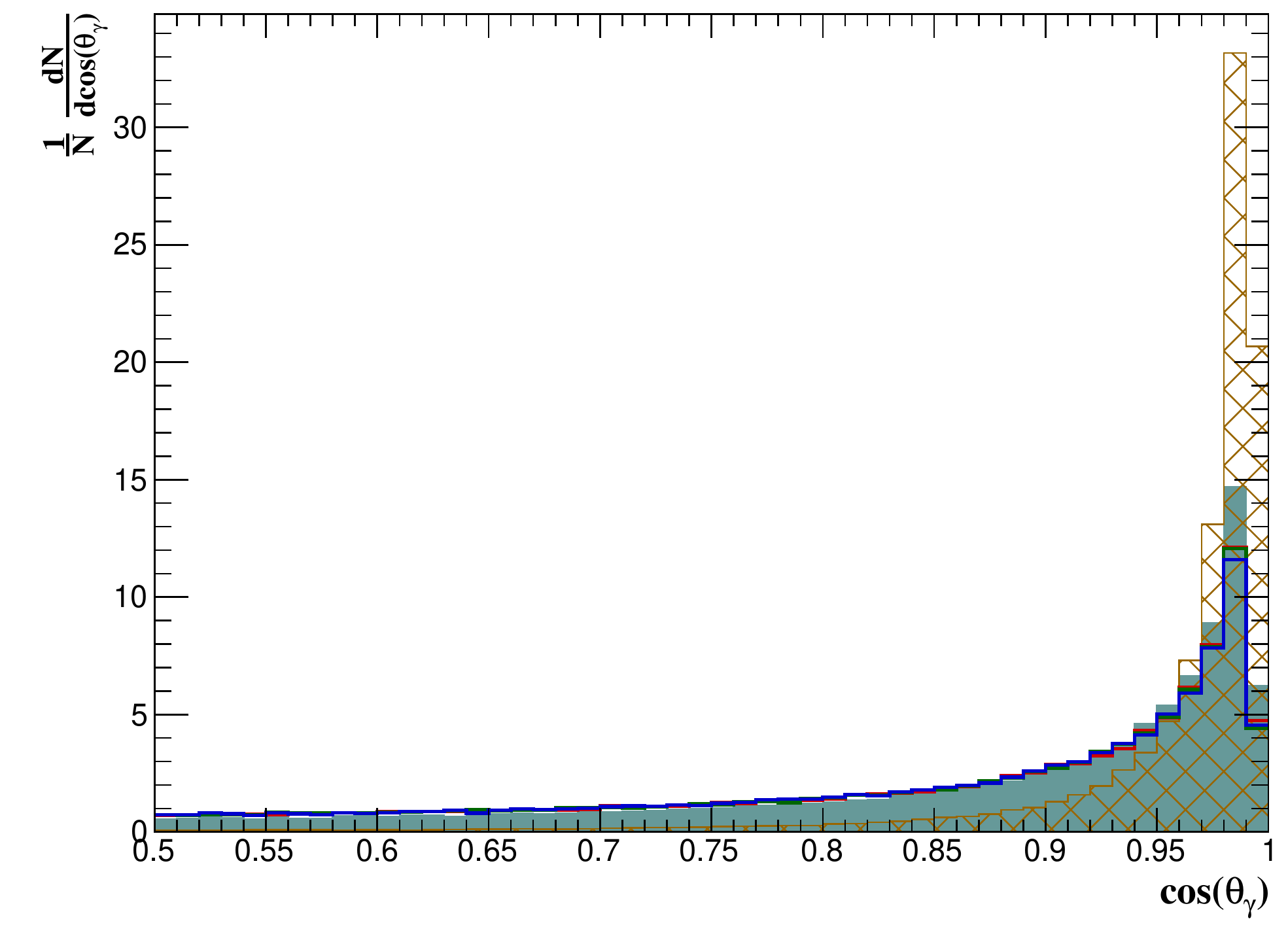}
\includegraphics[width=0.325\linewidth]{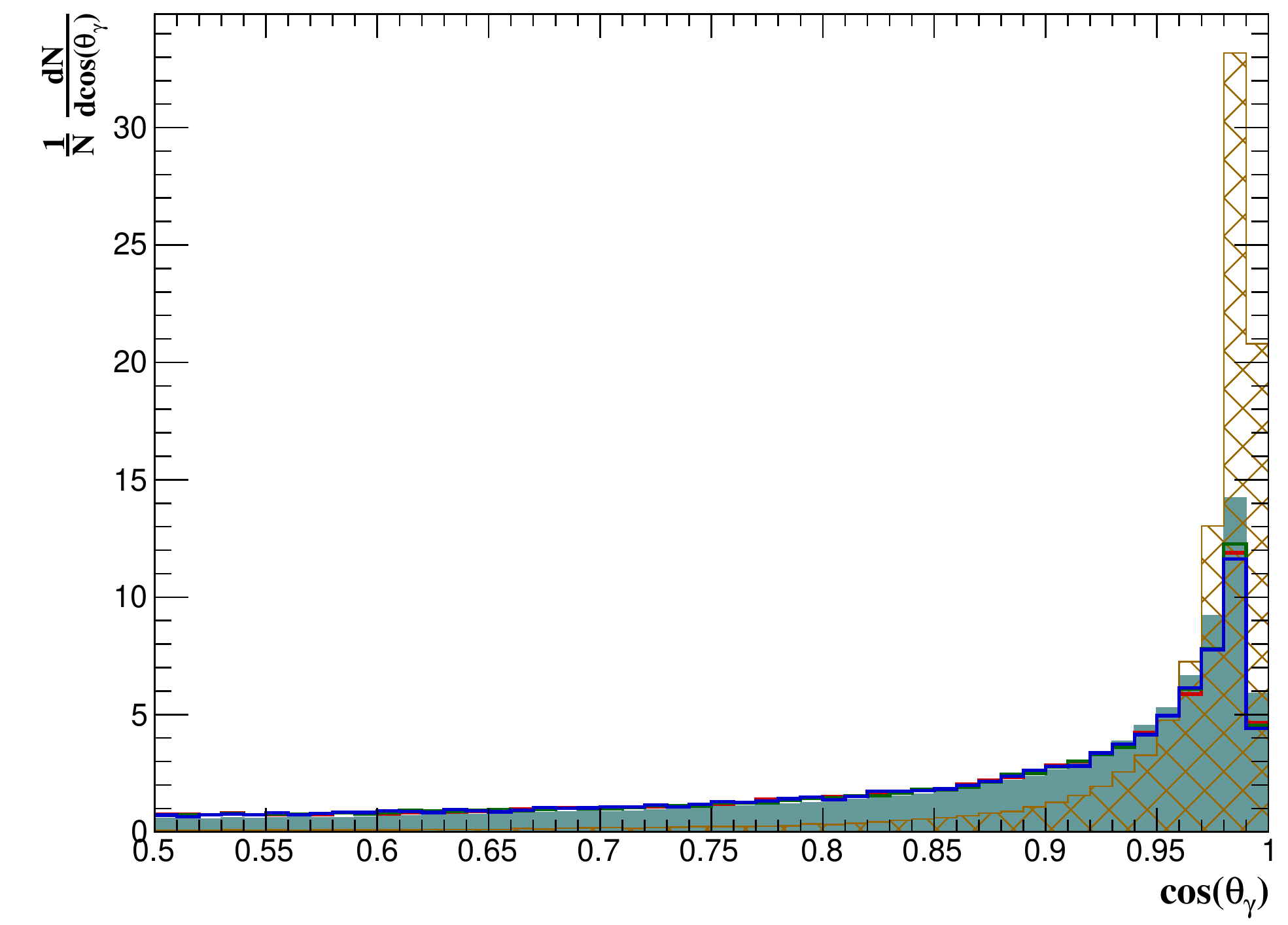}
\caption{Same as in \autoref{figure:HistosPh}, but for optimally polarised beams. }
\label{figure:HistosPhpol}
\end{figure*}

\begin{figure*}[ht]
\includegraphics[width=0.325\linewidth]{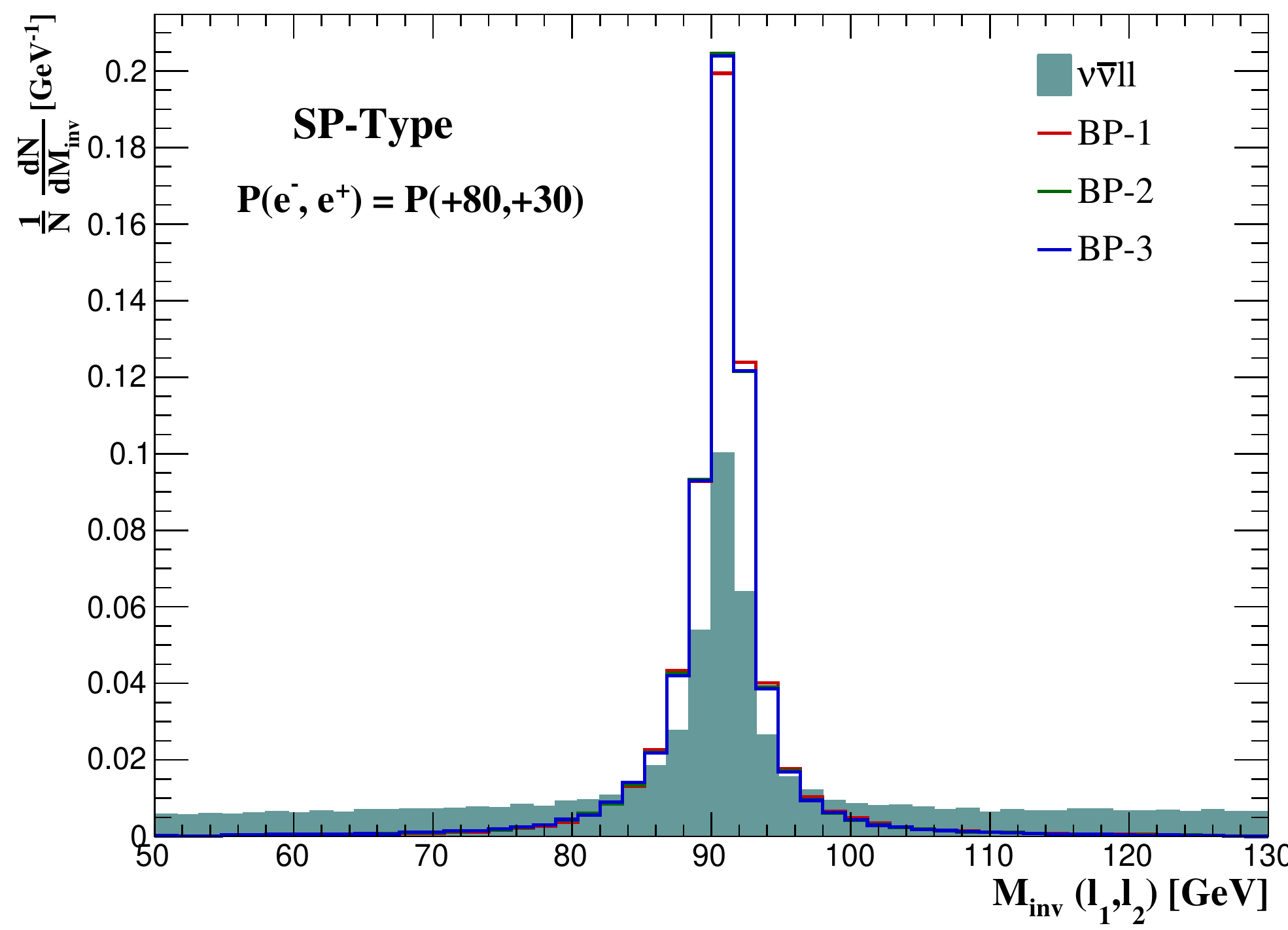}
\includegraphics[width=0.325\linewidth]{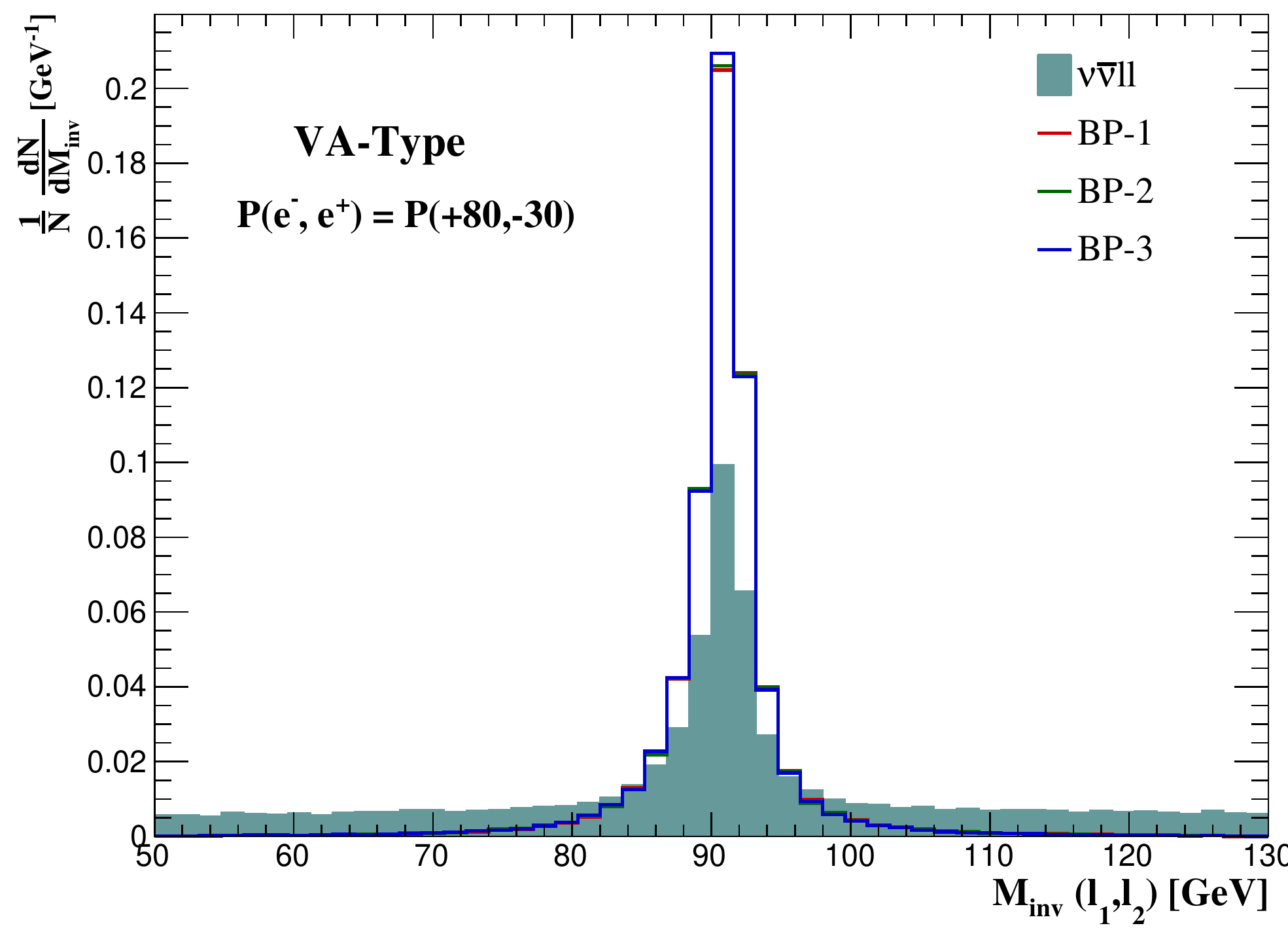}
\includegraphics[width=0.325\linewidth]{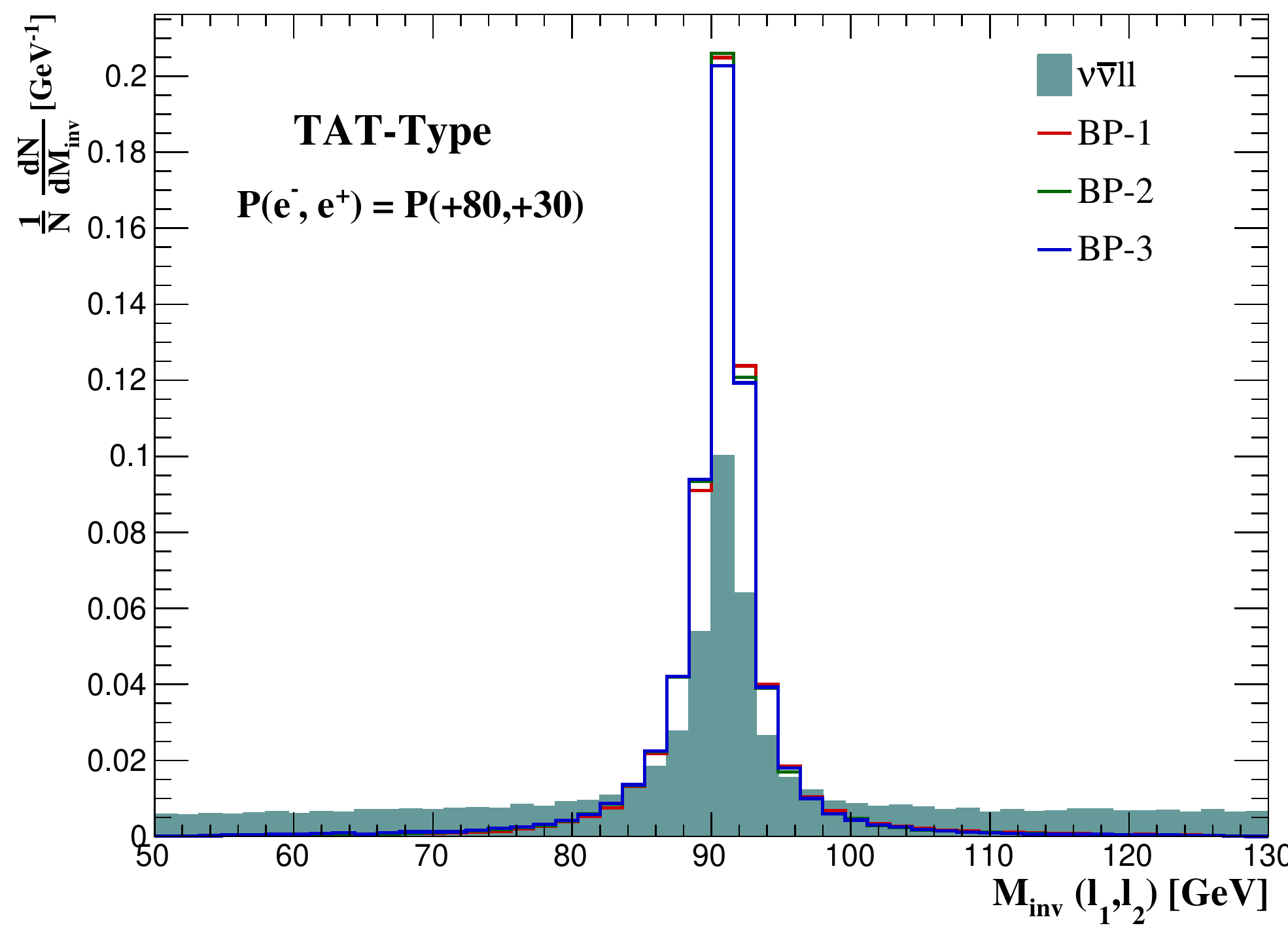}
\includegraphics[width=0.325\linewidth]{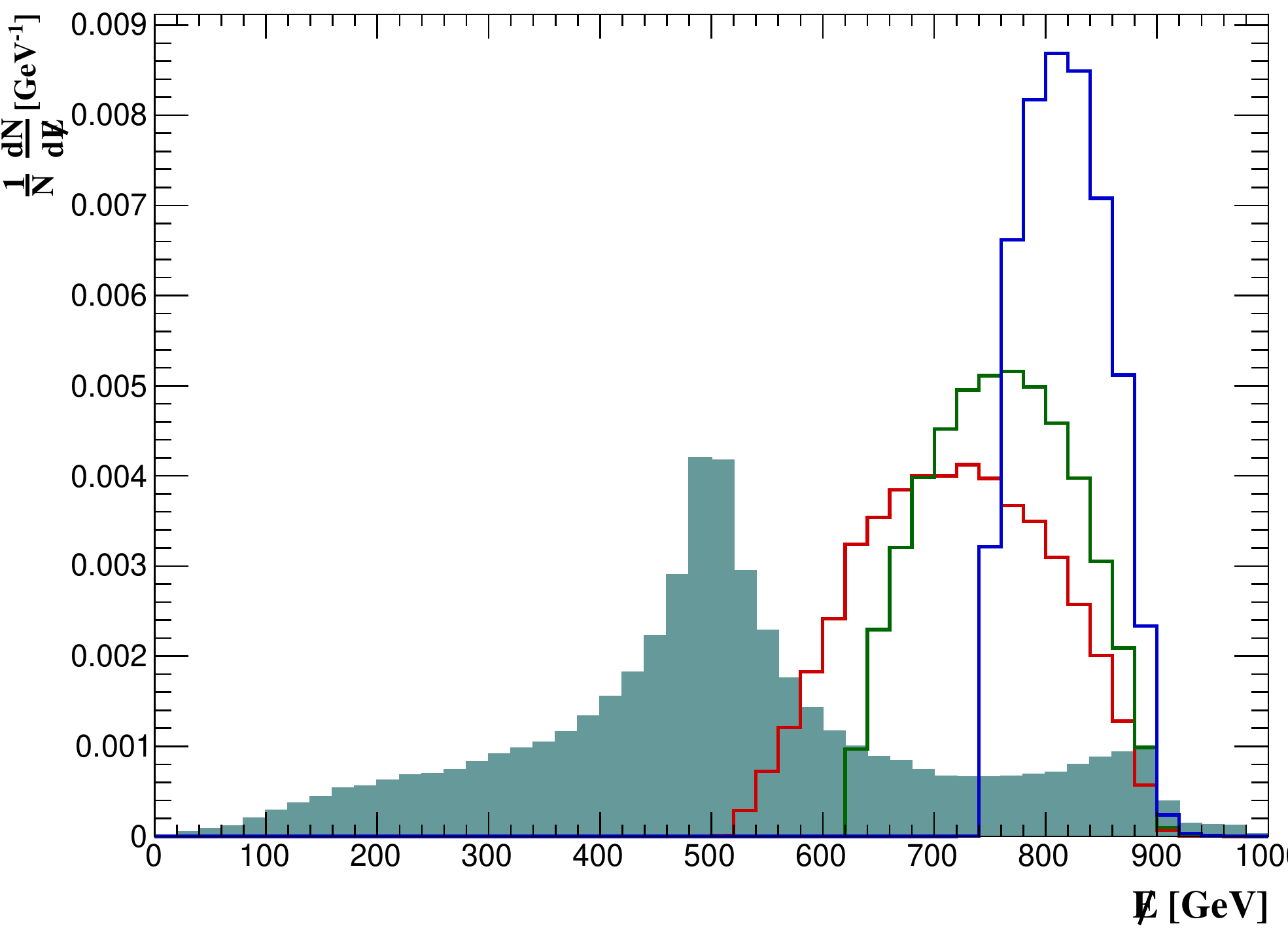}
\includegraphics[width=0.325\linewidth]{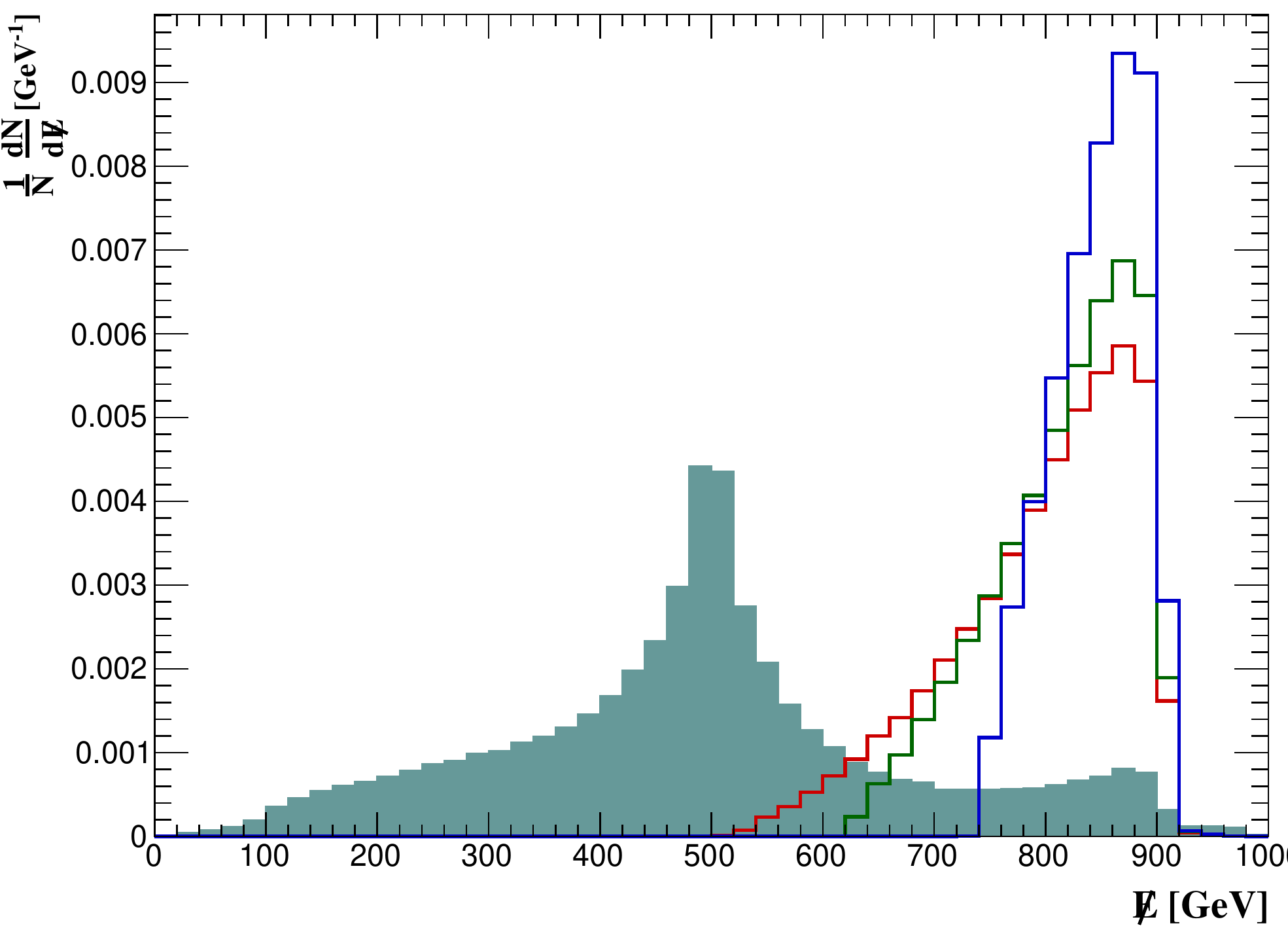}
\includegraphics[width=0.325\linewidth]{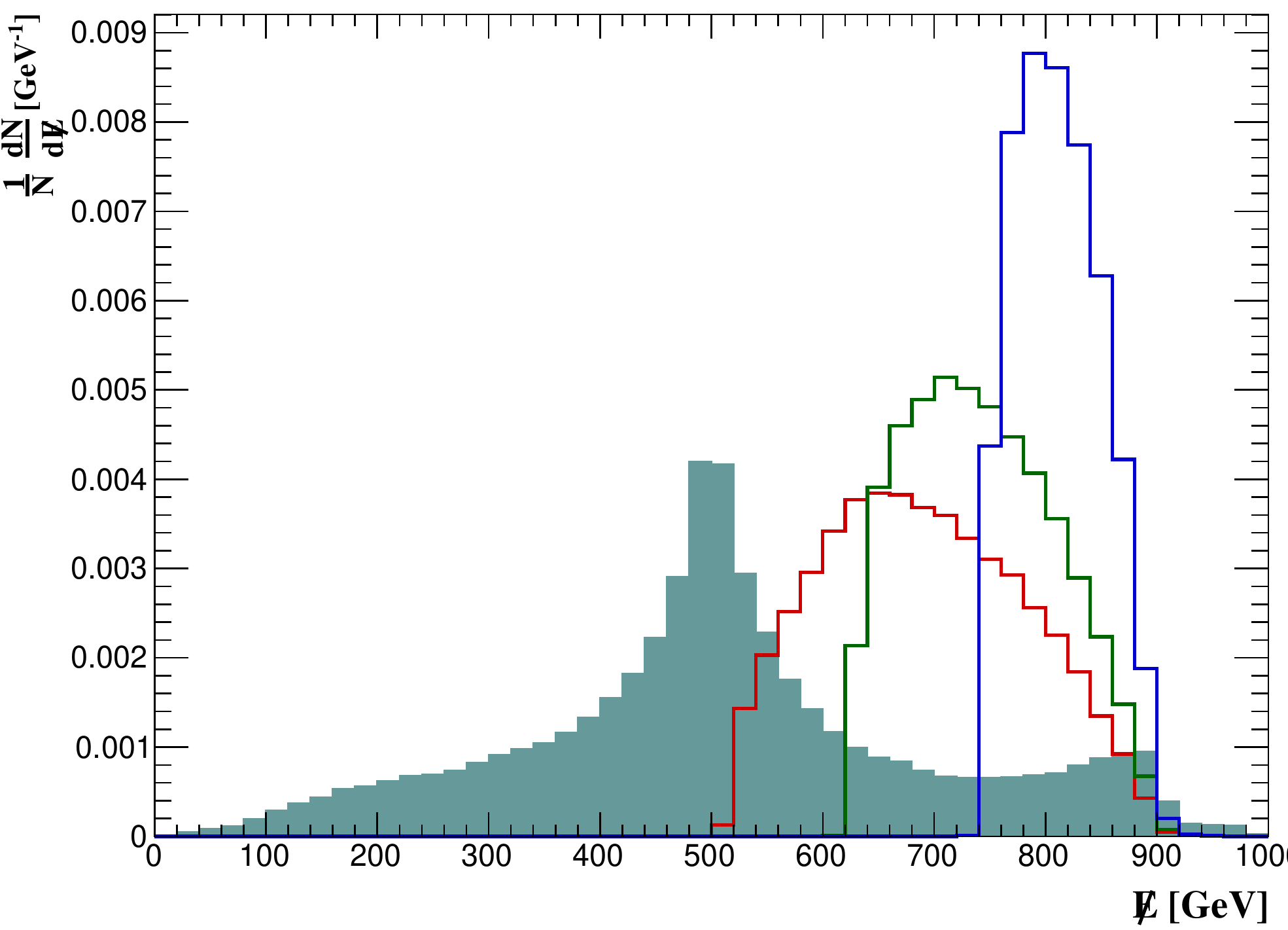}
\includegraphics[width=0.325\linewidth]{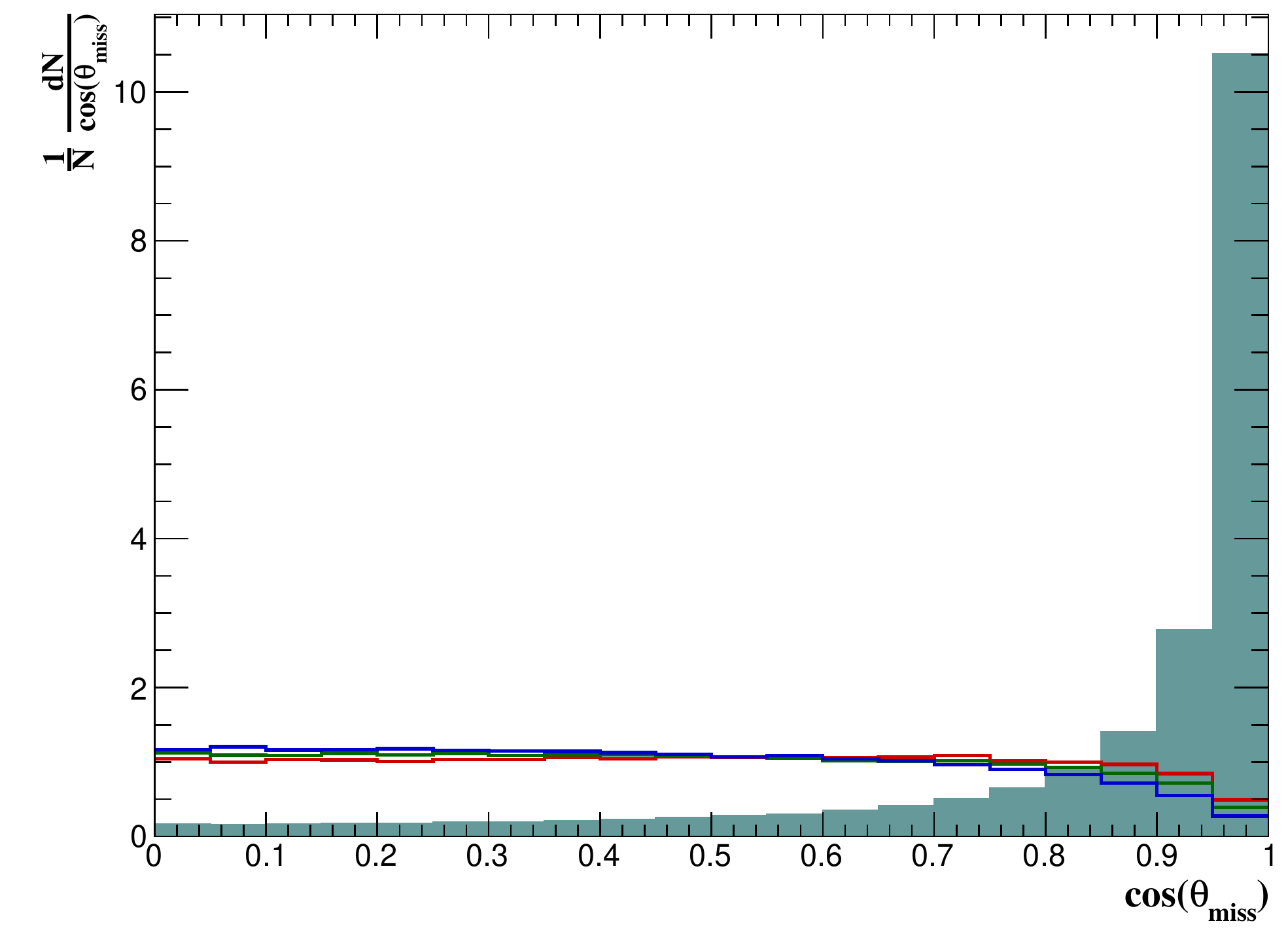}
\includegraphics[width=0.325\linewidth]{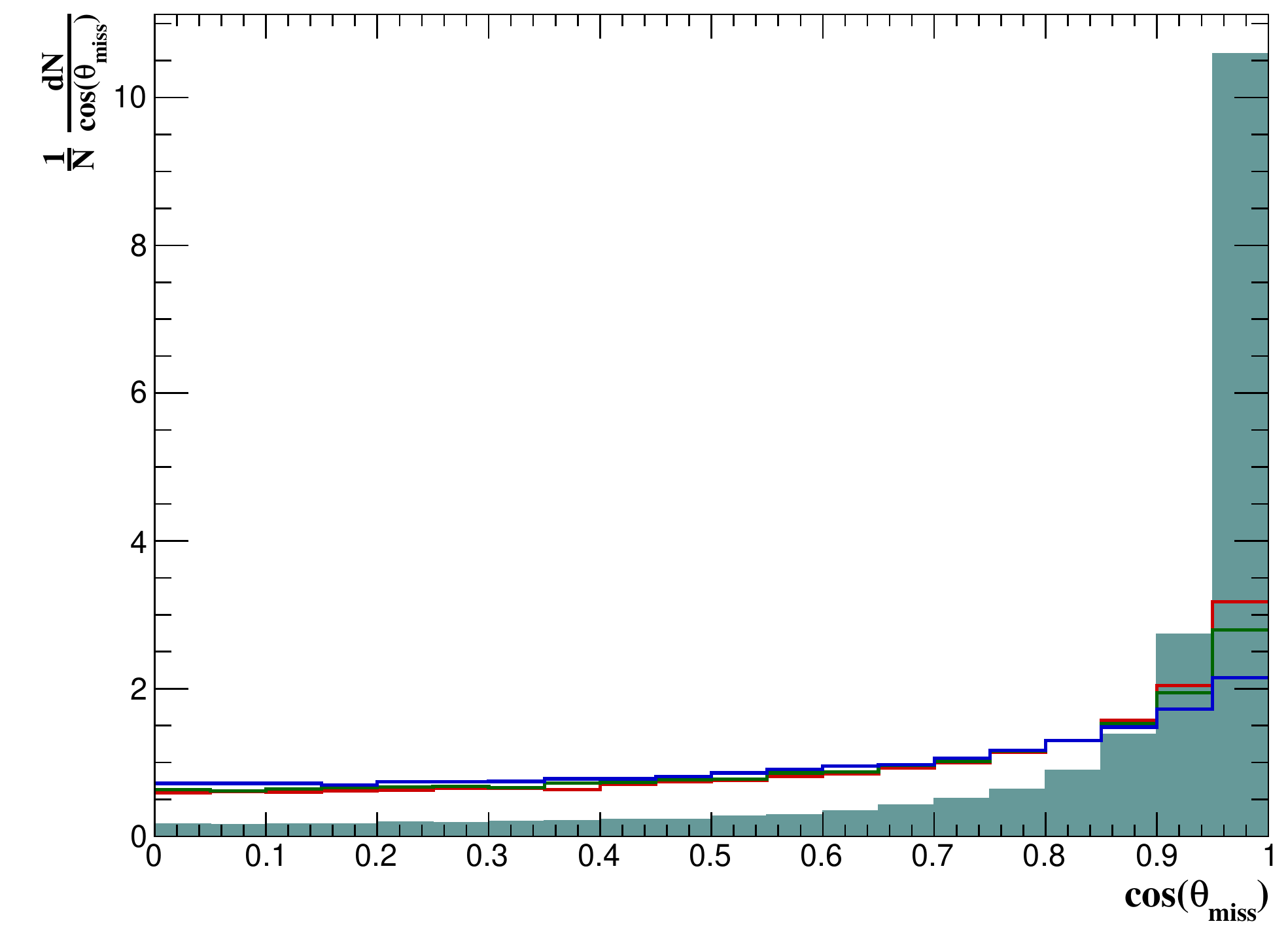}
\includegraphics[width=0.325\linewidth]{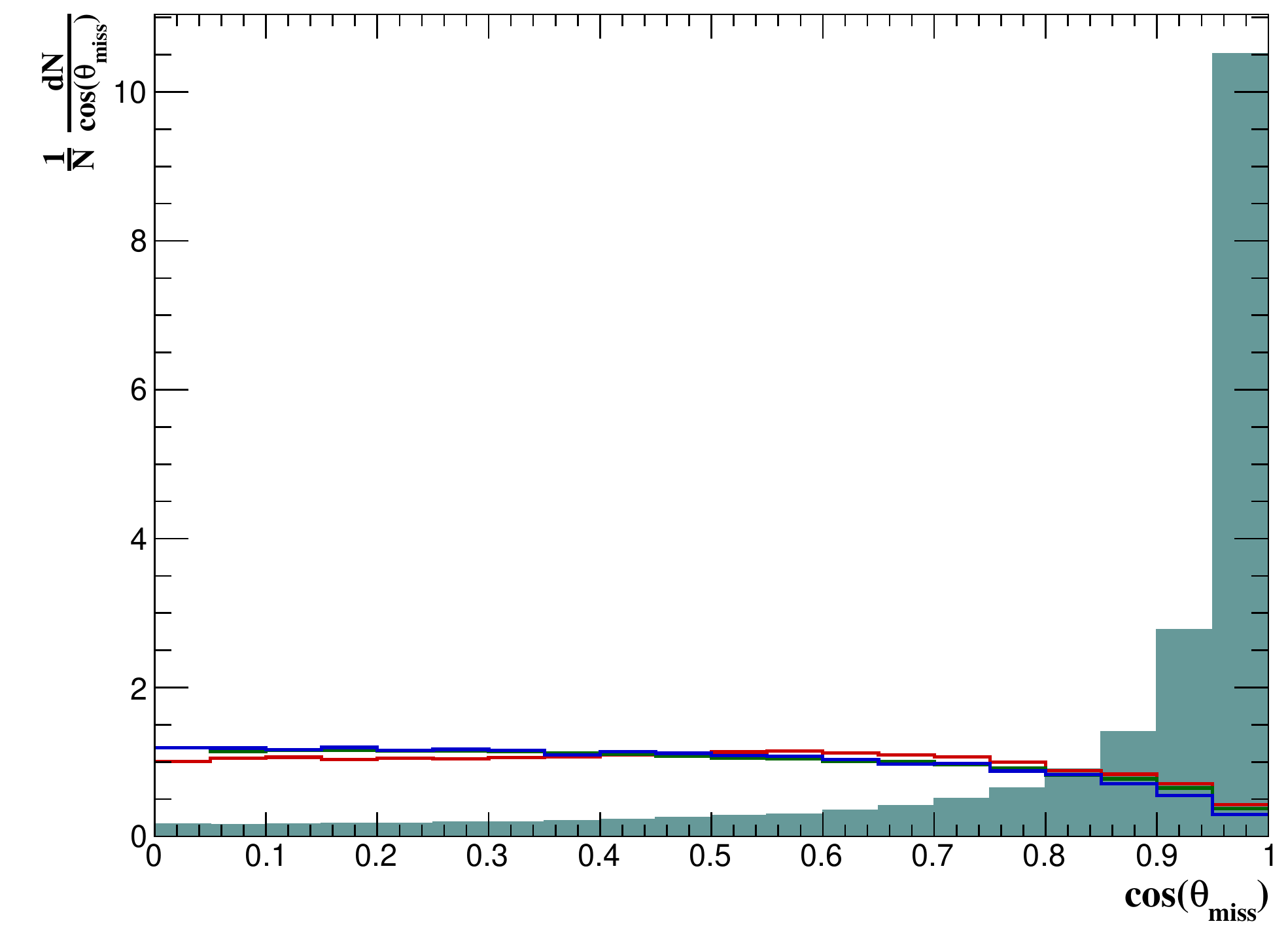}
\includegraphics[width=0.325\linewidth]{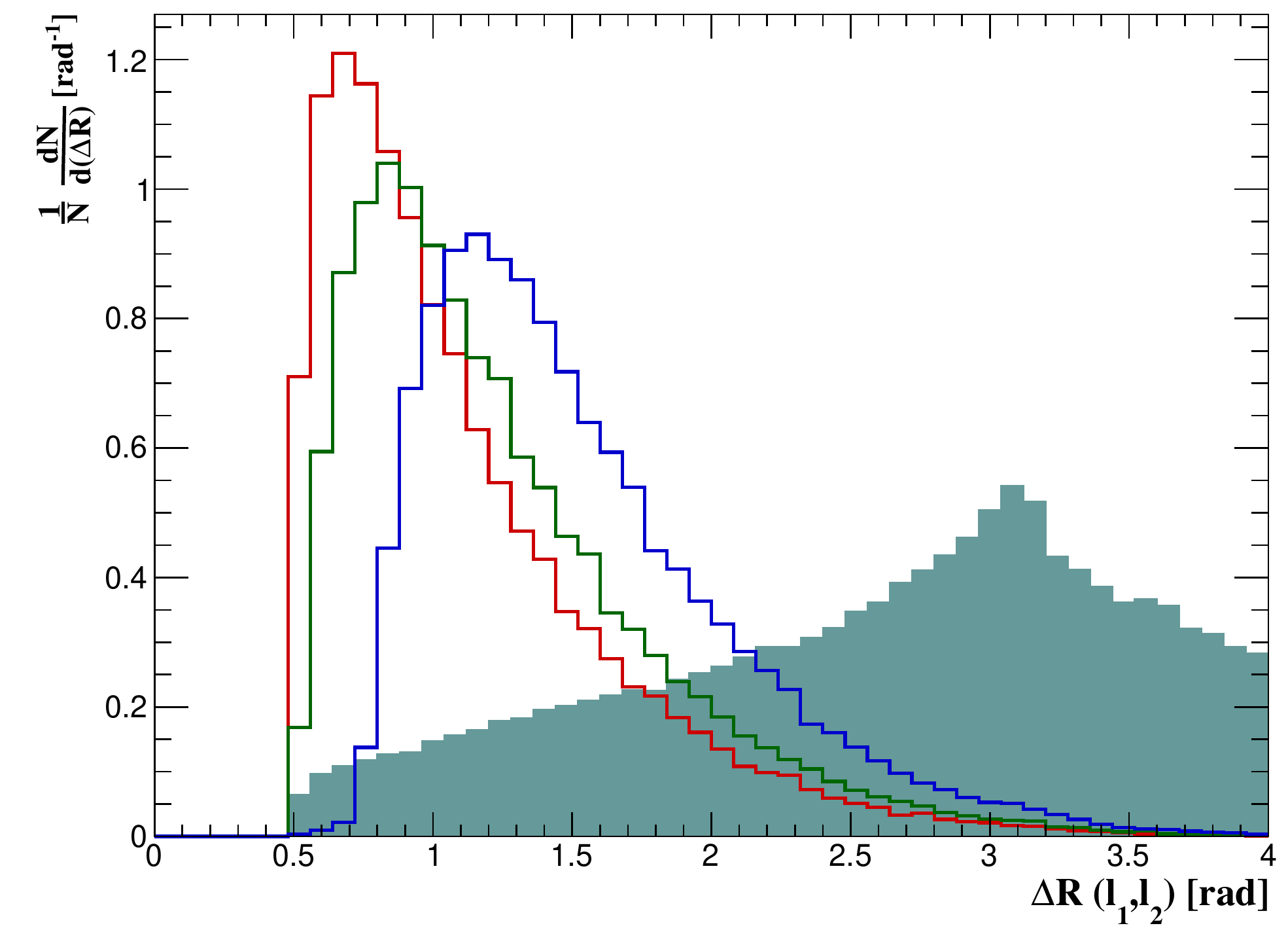}
\includegraphics[width=0.325\linewidth]{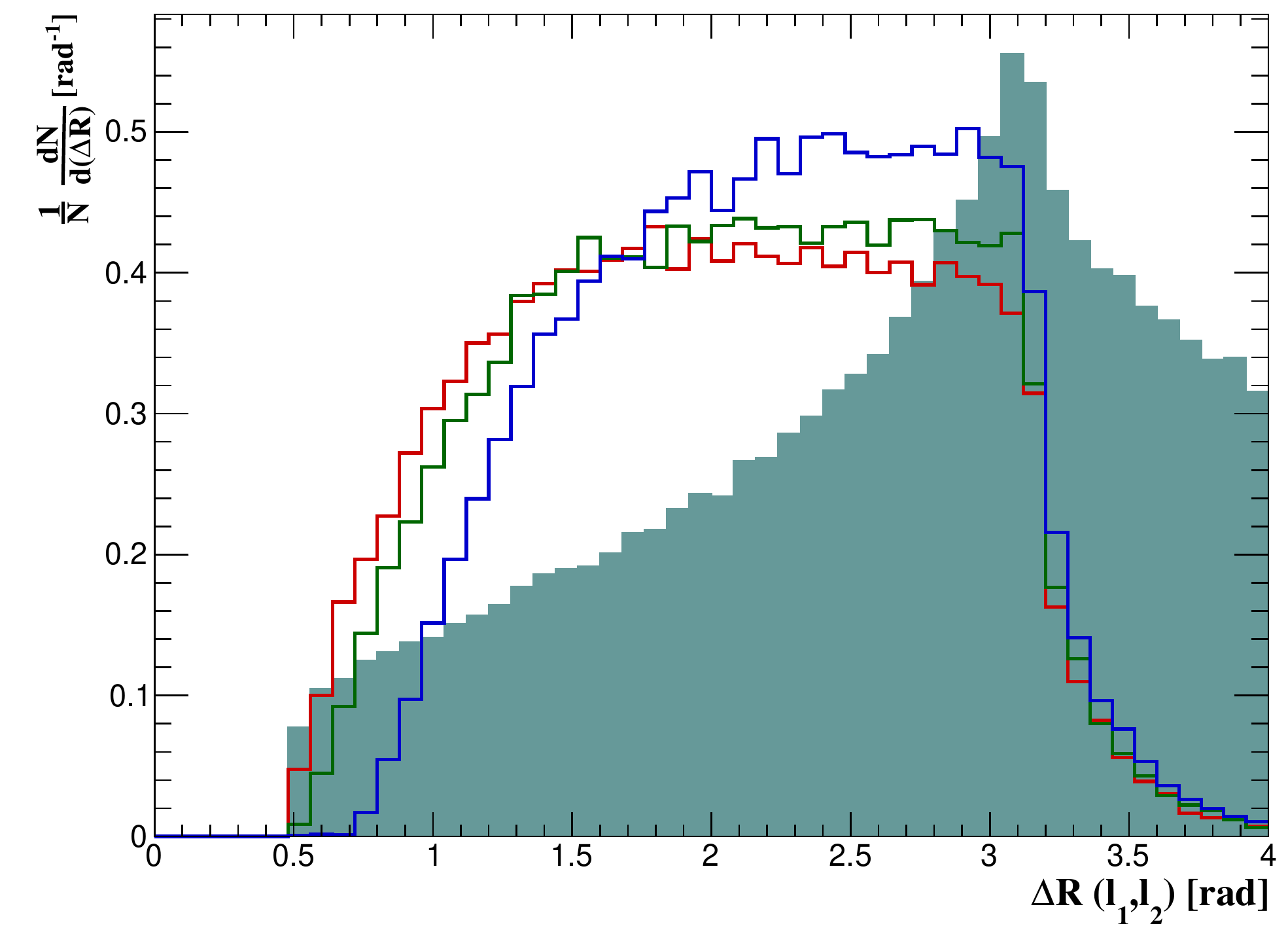}      \includegraphics[width=0.325\linewidth]{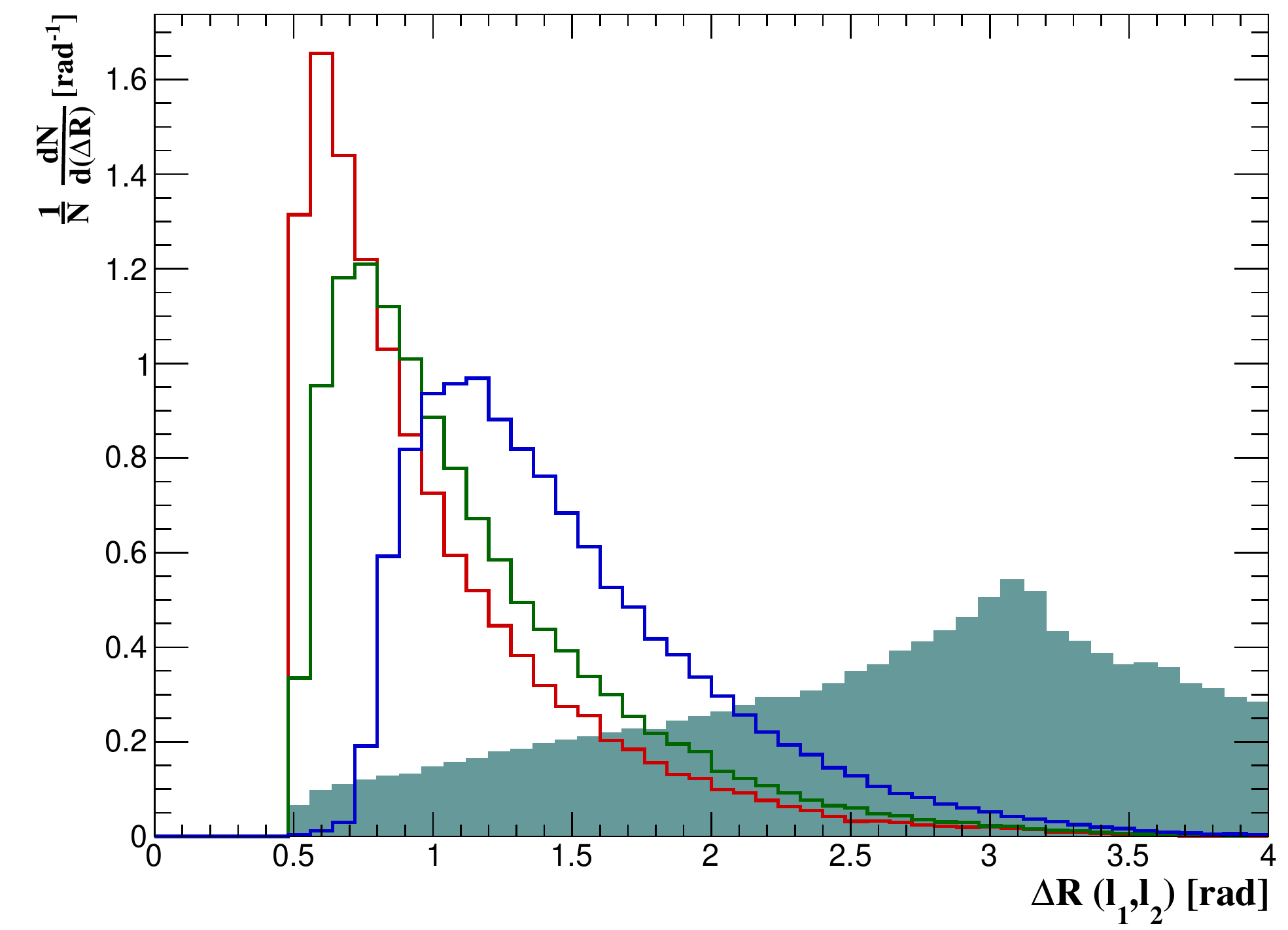}
\caption{Same as in \autoref{figure:HistosZ}, but for optimally polarised beams.}
\label{figure:HistosZpol}
\end{figure*}

\begin{figure*}[htb]
\includegraphics[width=0.325\linewidth]{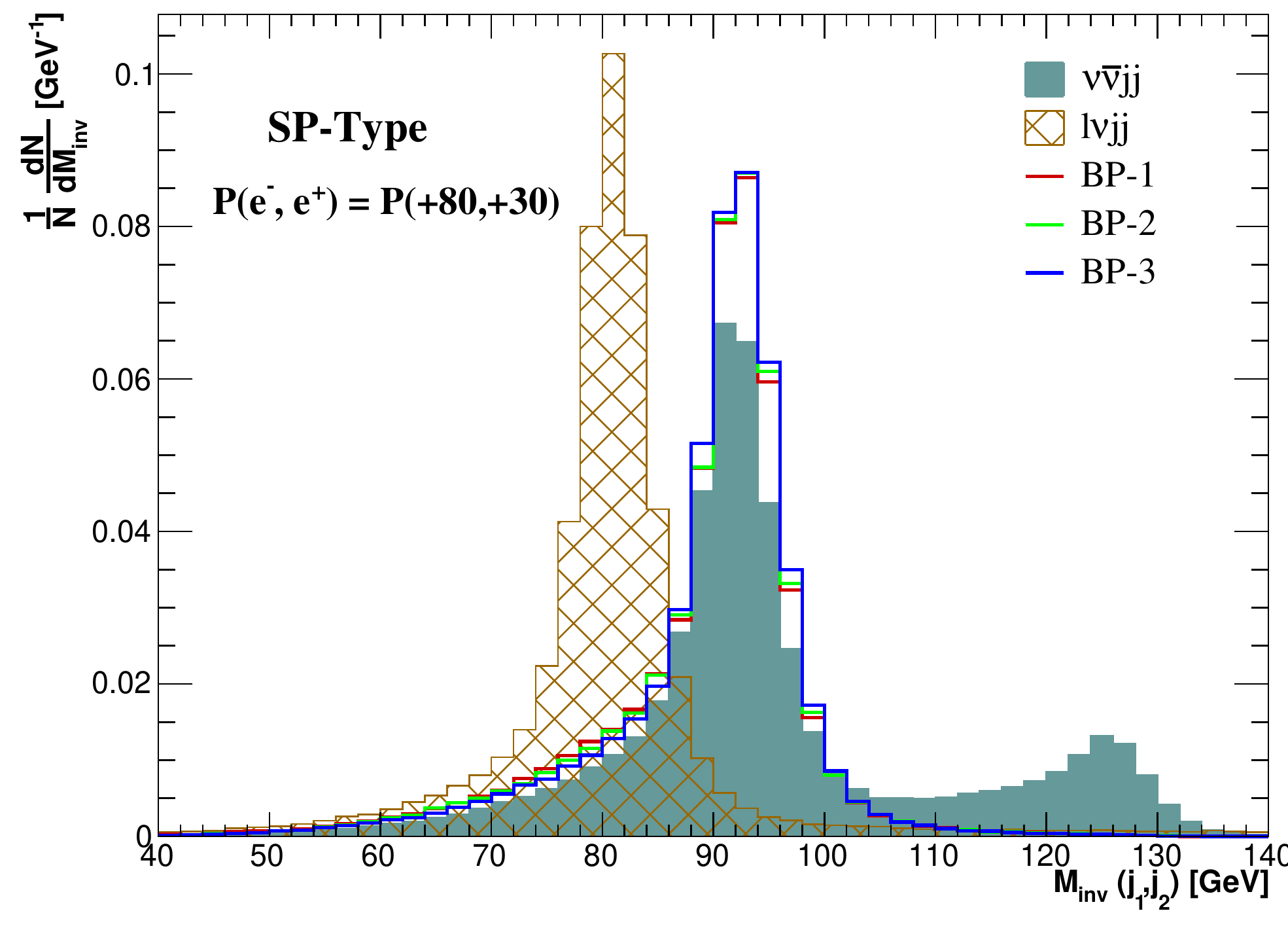}
\includegraphics[width=0.325\linewidth]{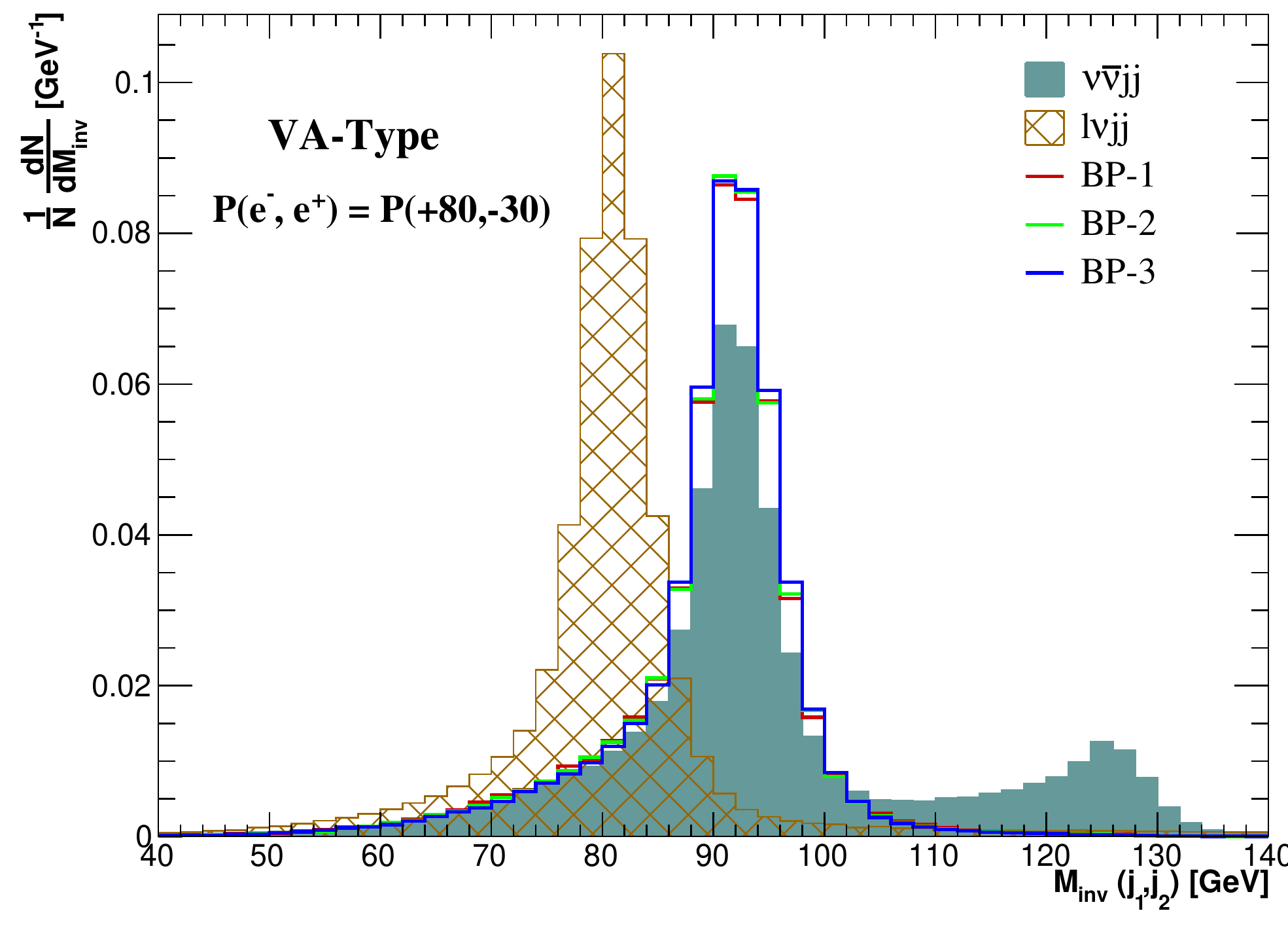}
\includegraphics[width=0.325\linewidth]{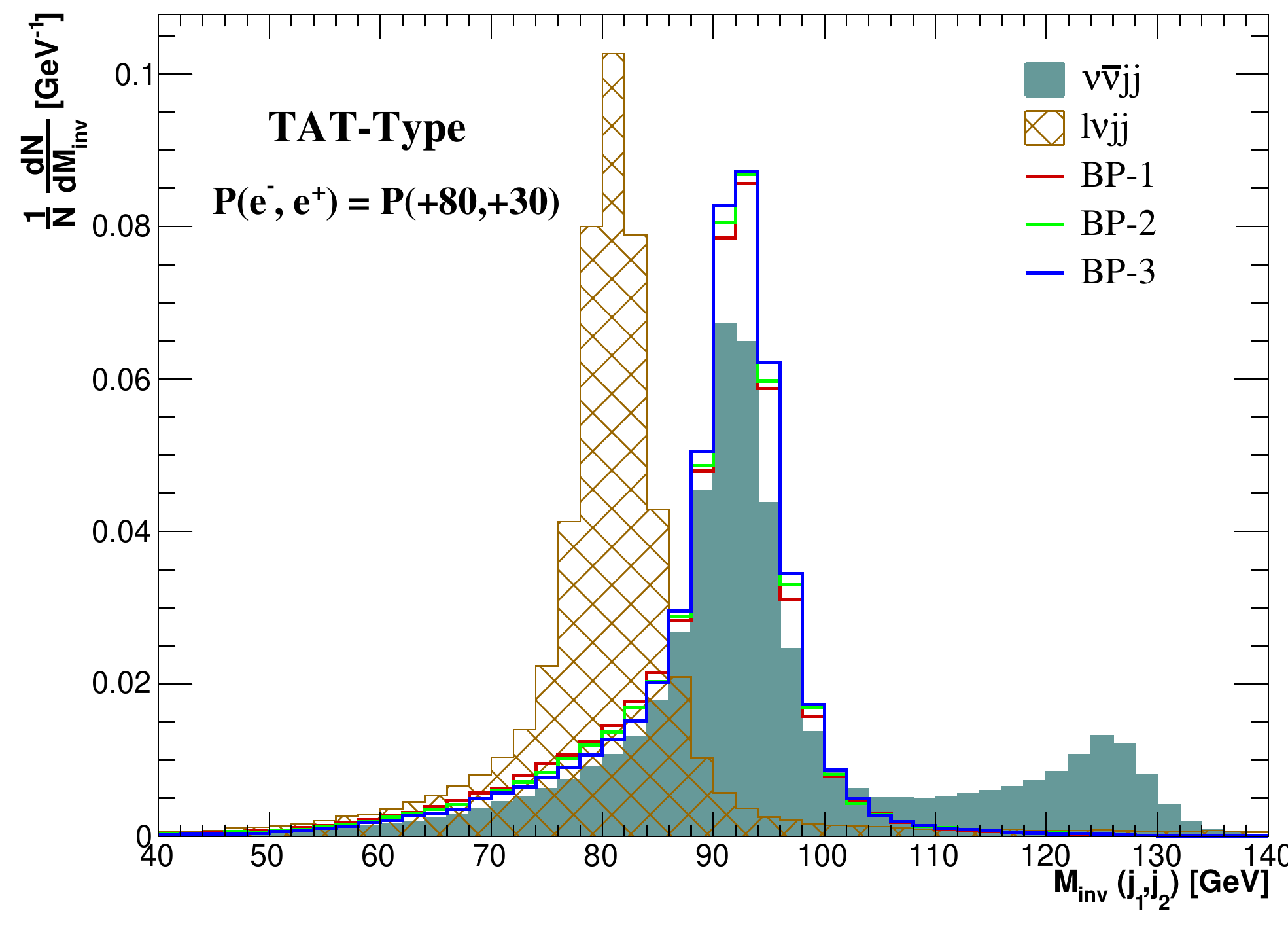}
\includegraphics[width=0.325\linewidth]{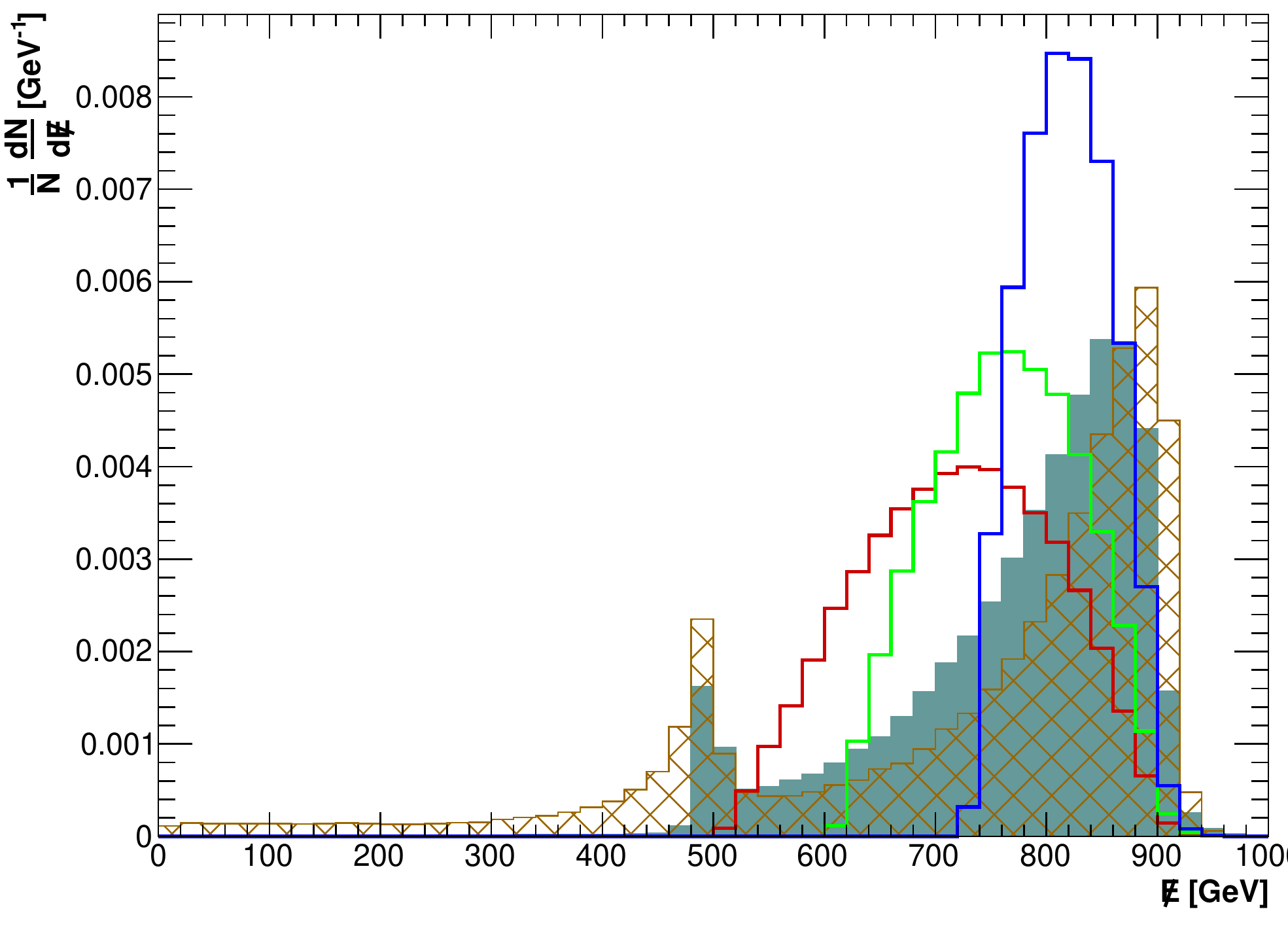}
\includegraphics[width=0.325\linewidth]{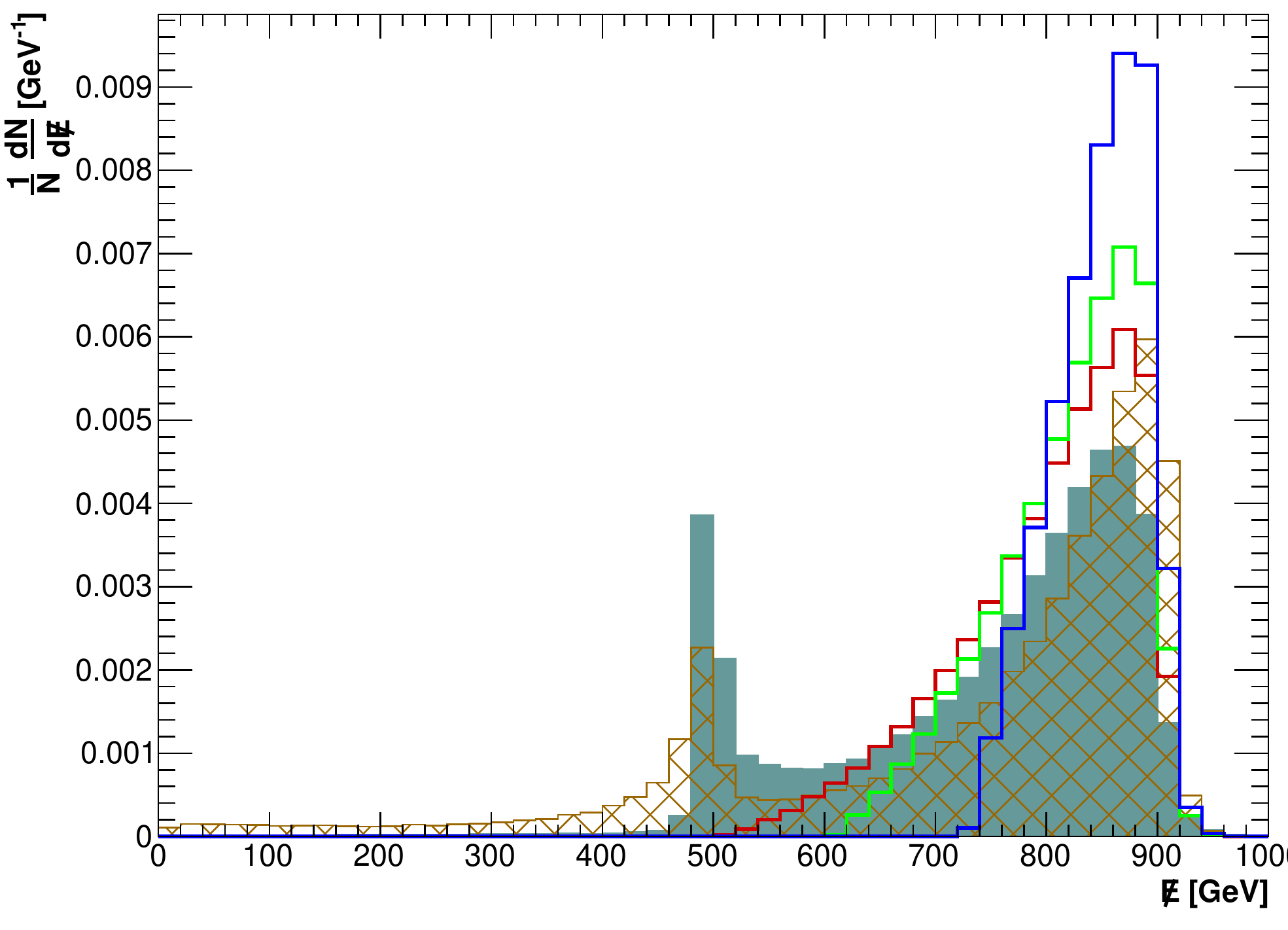}
\includegraphics[width=0.325\linewidth]{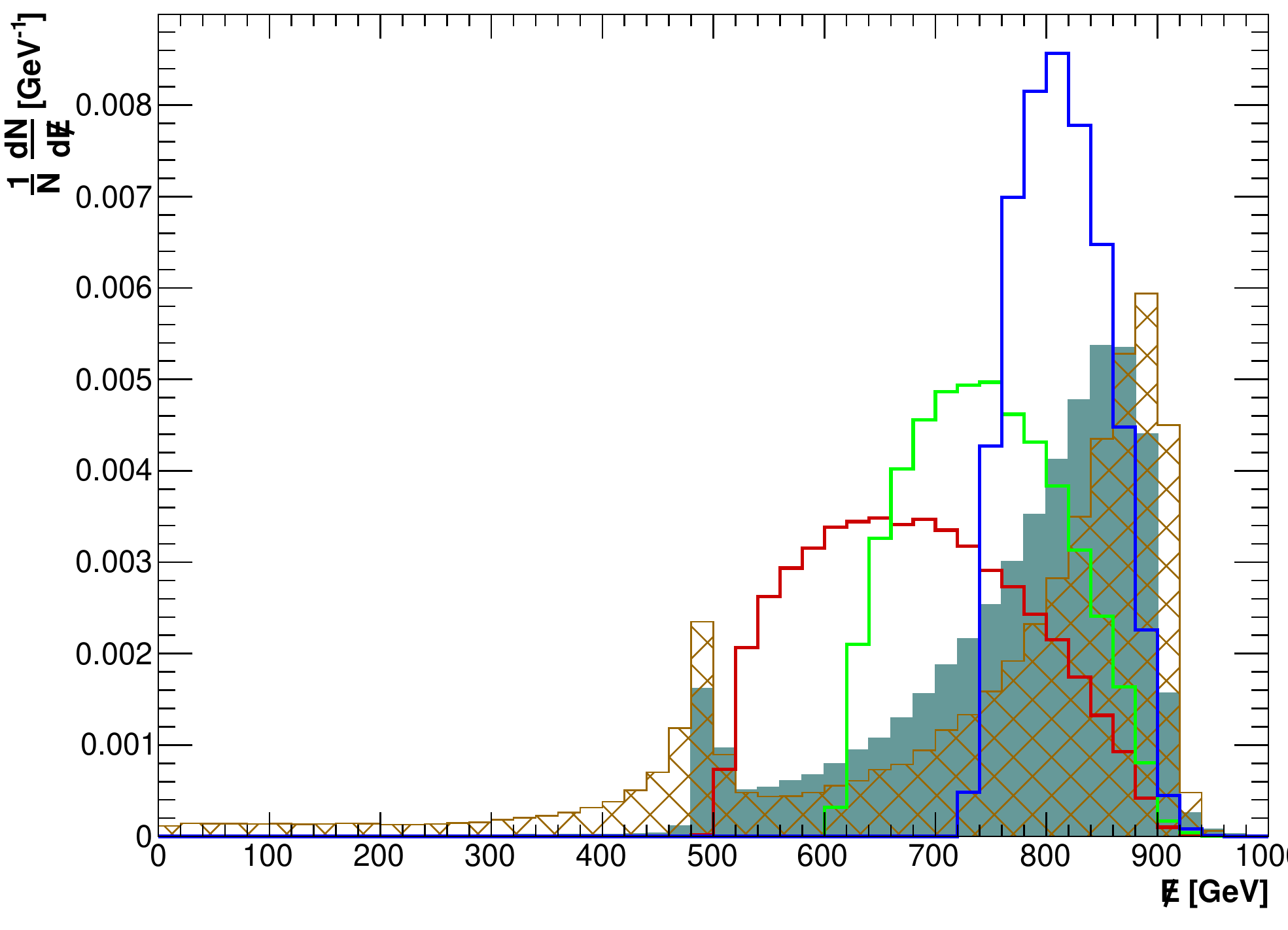}
\includegraphics[width=0.325\linewidth]{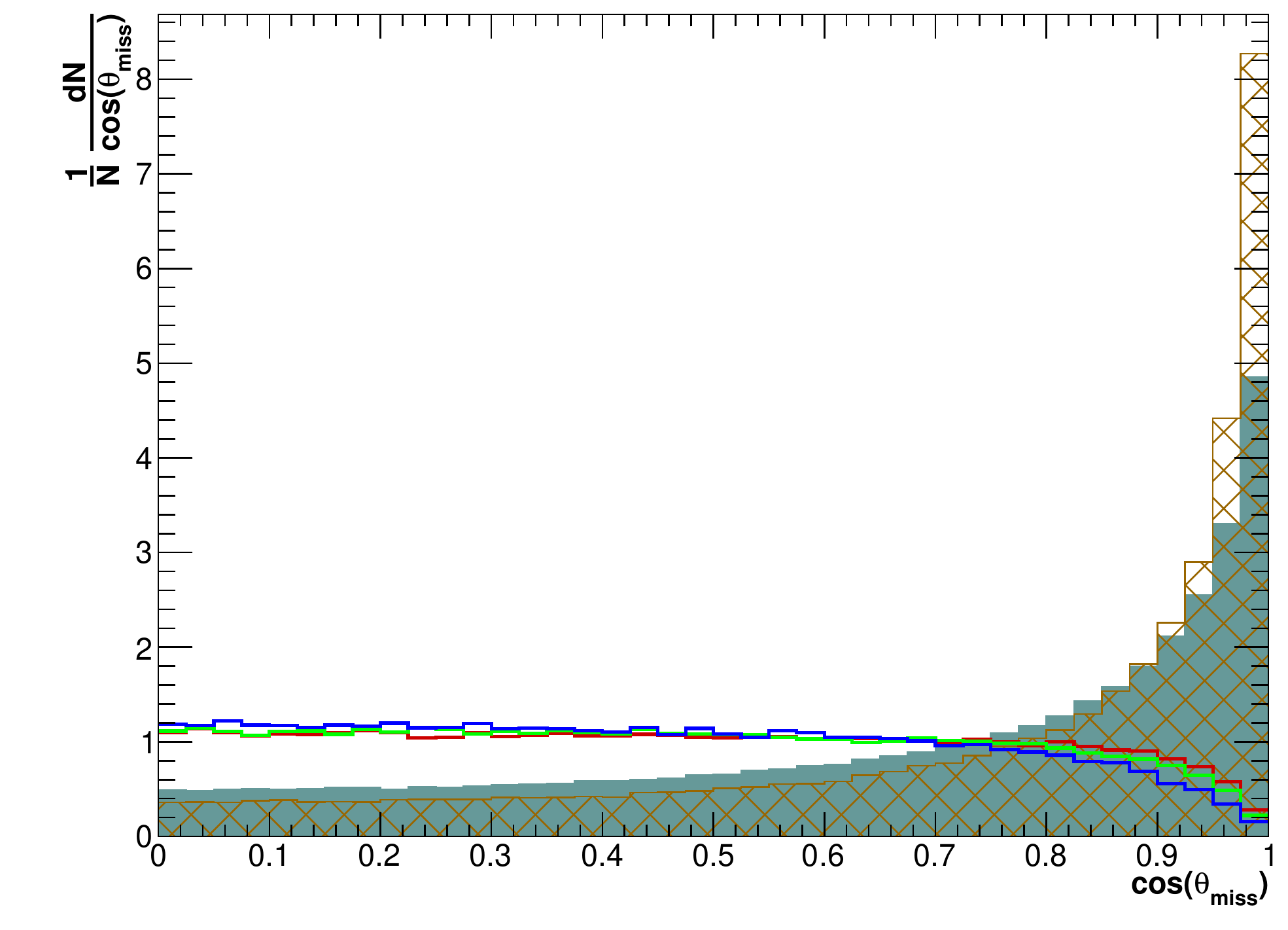}
\includegraphics[width=0.325\linewidth]{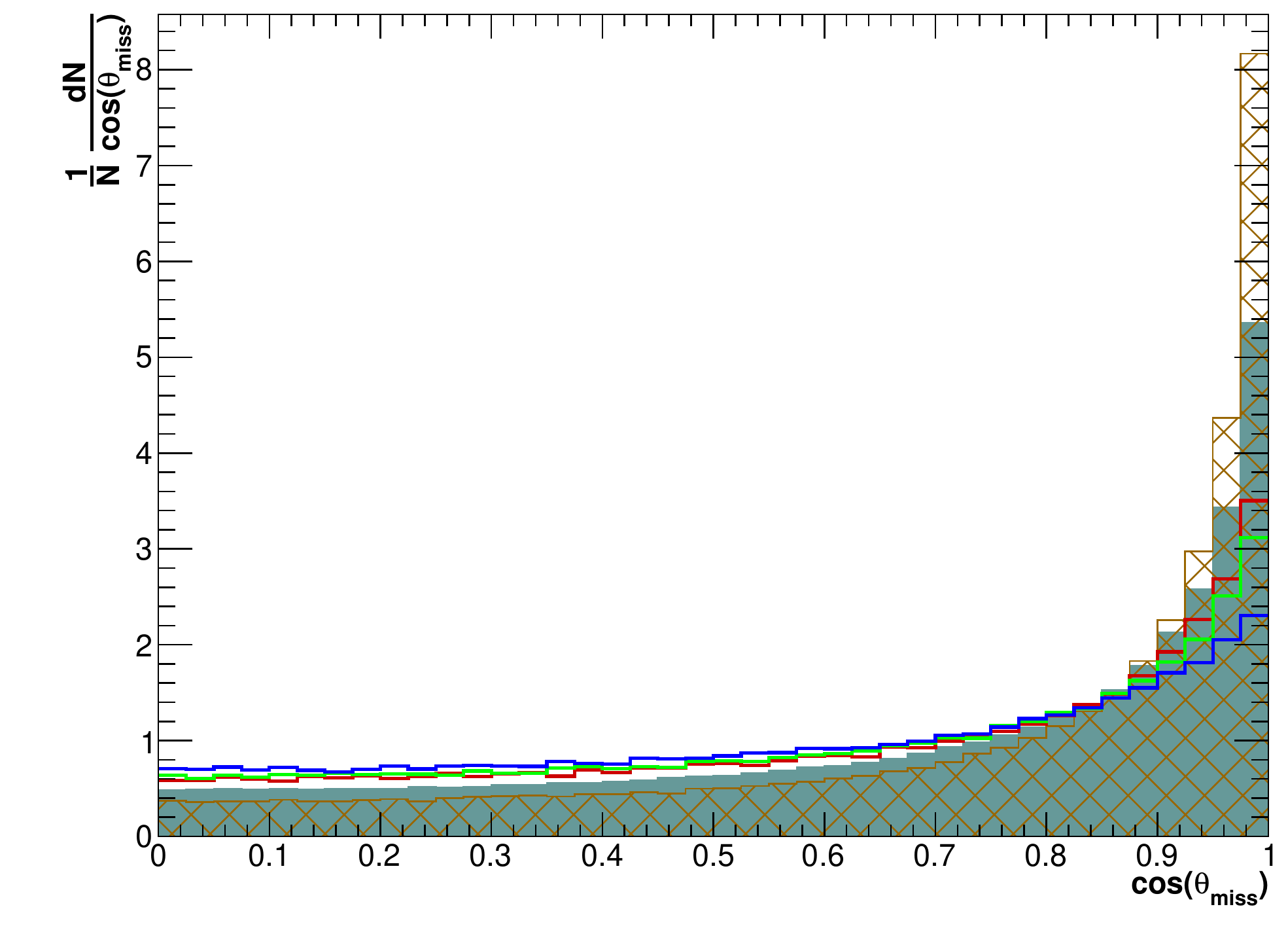}
\includegraphics[width=0.325\linewidth]{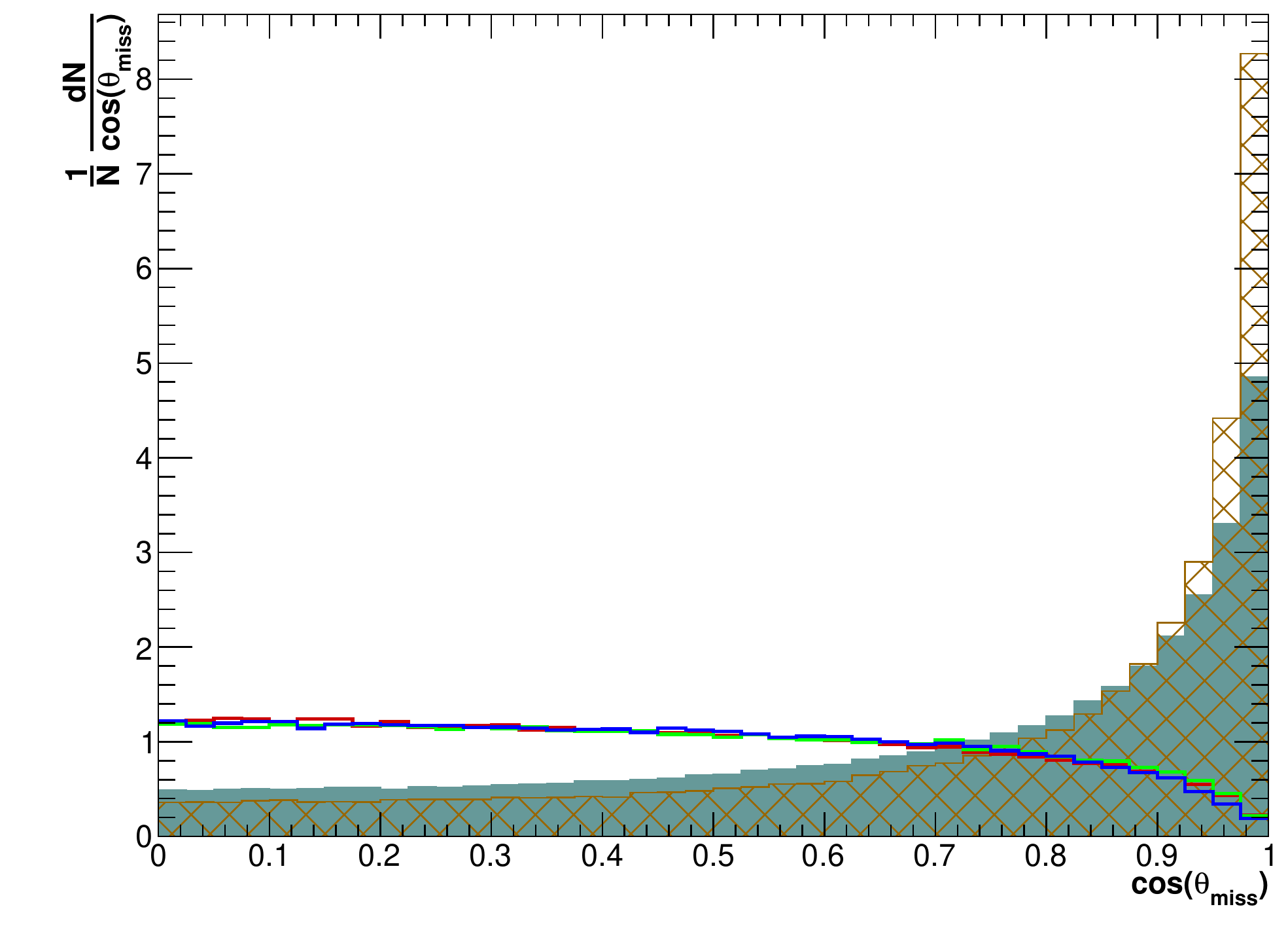}
\includegraphics[width=0.325\linewidth]{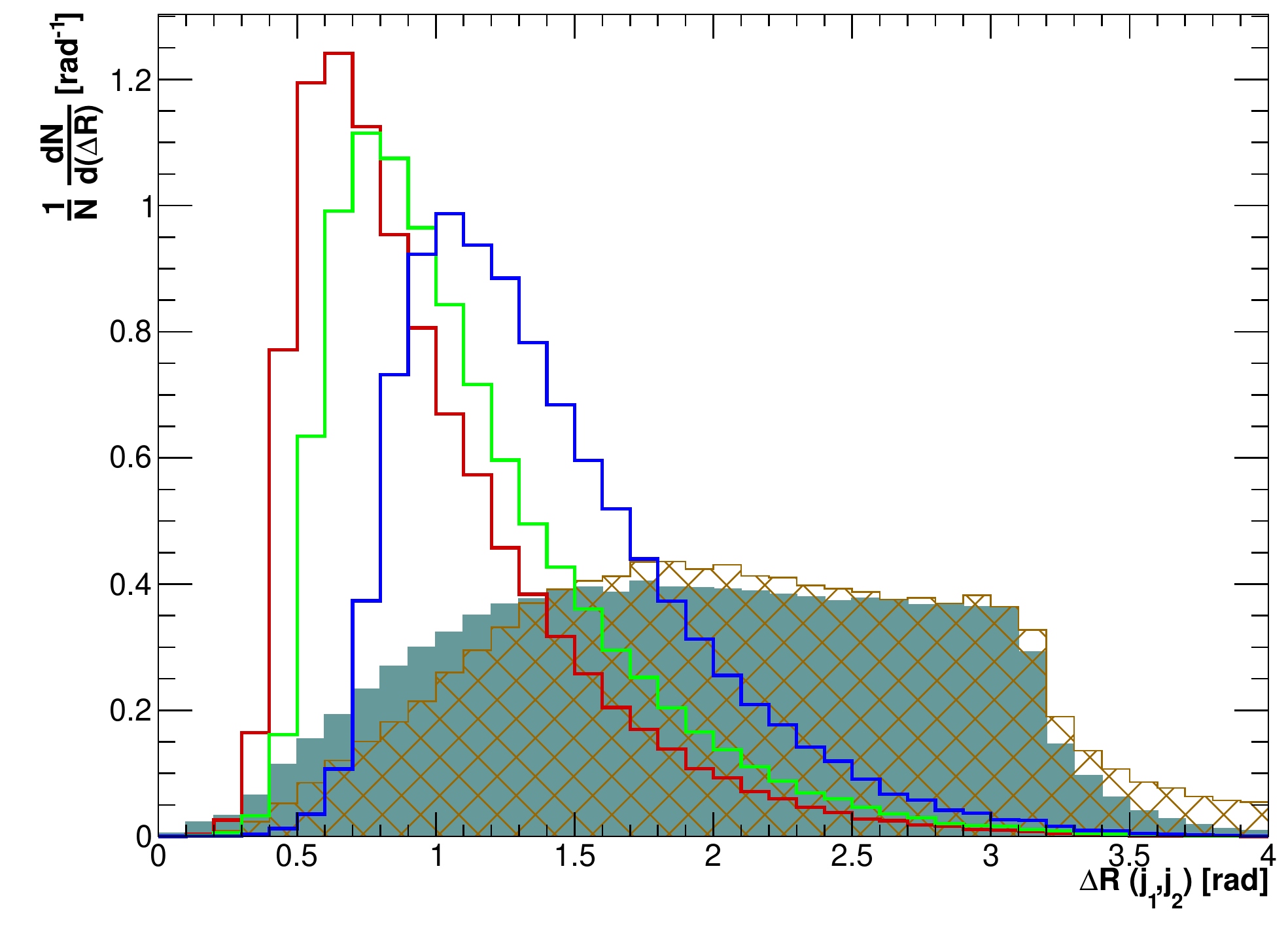}
\includegraphics[width=0.325\linewidth]{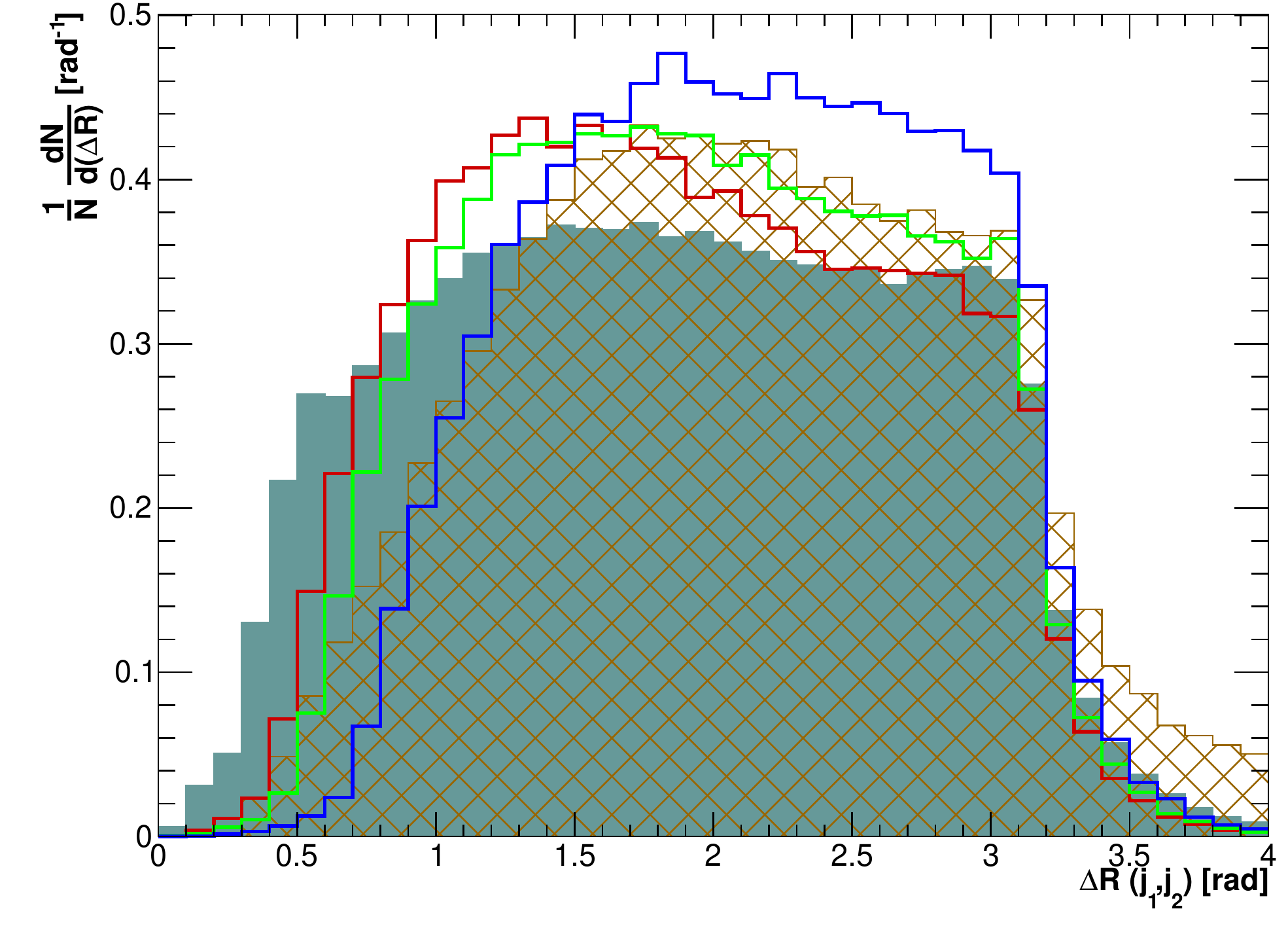}
\includegraphics[width=0.325\linewidth]{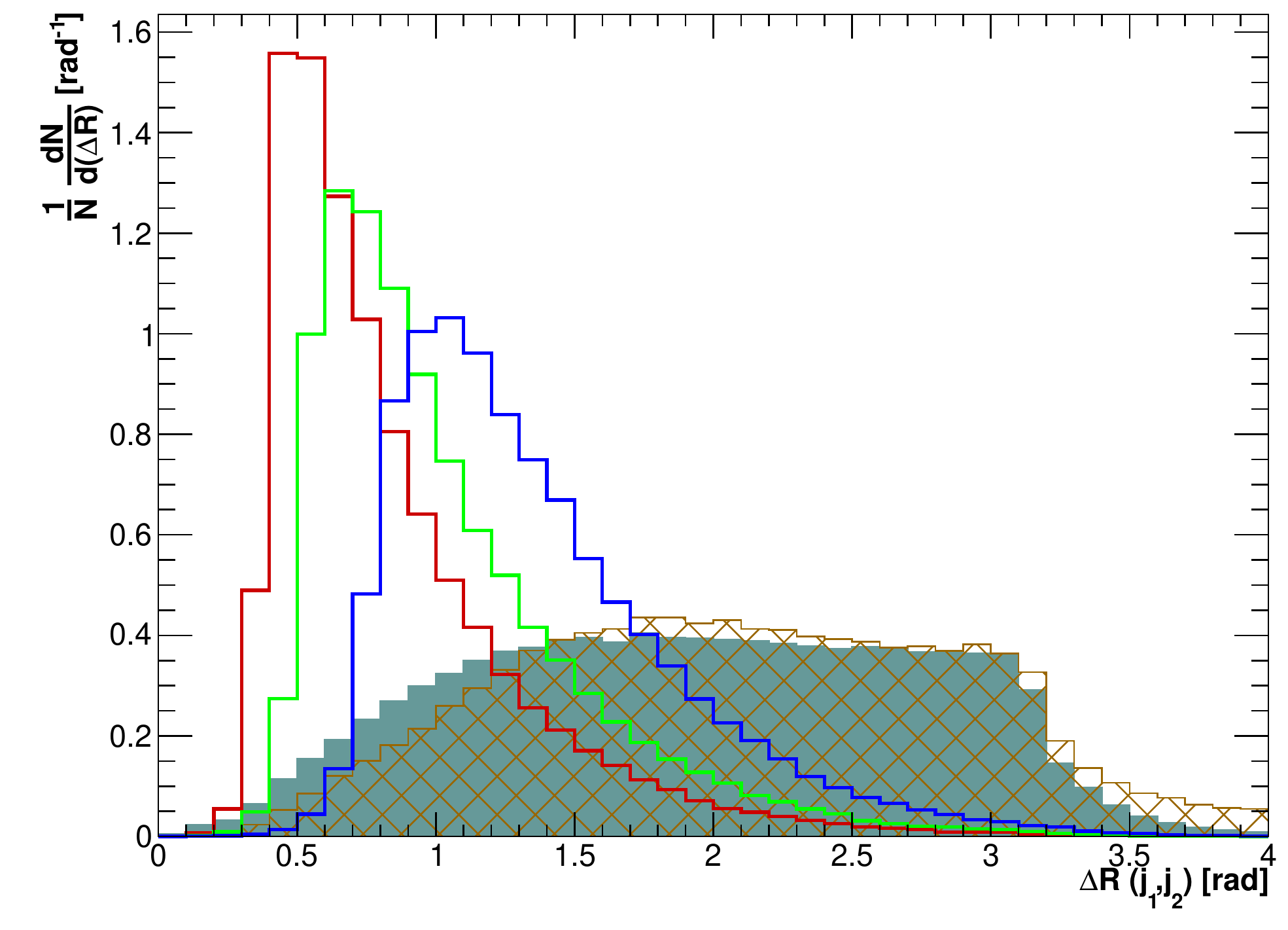}
\caption{Same as in \autoref{figure:HistosZjj}, but for optimally polarised beams.}
\label{figure:HistosZjjpol}
\end{figure*}

For completeness, we present in Figs.~\ref{figure:HistosPhpol}--\ref{figure:HistosZjjpol} the kinematic distributions with optimally polarized beams for the same set of variables as shown in Figs.~\ref{figure:HistosPh}, \ref{figure:HistosZ}, and \ref{figure:HistosZjj} for the monophoton, mono-$Z$ leptonic, and hadronic channels, respectively.  
\begin{table*}[ht]
\caption{\label{table:CutBGGpol}%
Same as \autoref{table:CutBGG}, but for optimally polarised beams.}
\begin{ruledtabular}
\begin{tabular}{l ccc ccc ccc} 
   & \multicolumn{9}{c}{Event Numbers (Cut efficiencies)}\\
  \cline{2-10}
  & \multicolumn{3}{c}{Neutrino pair} & \multicolumn{3}{c}{Radiative Bhabha} & 
  \multicolumn{3}{c}{Signal} \\
  \cline{2-10}
  Selection cuts & BP1 & BP2 & BP3 & BP1 & BP2 & BP3 & BP1 & BP2 & BP3 \\
  \colrule 
  \multicolumn{10}{l}{SP type:}\\ 
  \colrule 
  \multirow{2}{4.5em}{BS} & \multicolumn{3}{c}{$8.55\times10^6$} & \multicolumn{3}{c}{$6.87\times10^7$} & $2.49\times10^5$ & $1.79\times10^5$ & $1.09\times10^5$ \\
  & \multicolumn{3}{c}{$(100\%)$} & \multicolumn{3}{c}{$(100\%)$} & $(100\%)$ & $(100\%)$ & $(100\%)$ \\
  \multirow{2}{4.5em}{After Final cut} 
   & $4.70\times10^6$ & $4.61\times10^6$ & $4.50\times10^6$ & $7.19\times10^5$ & $7.19\times10^5$ & $7.19\times10^5$ & $1.52\times10^5$ & $ 1.11\times10^5 $ & $ 6.84\times10^4 $ \\
   & $(54.97\%)$ & $(53.88\%)$ & $(52.60\%)$\phantom{0} & \phantom{0}$(1.05\%)$ & $(1.05\%)$ & $(1.05\%)$\phantom{0} & \phantom{0}$(61.14\%)$ & $(62.10\%)$ & $(62.79\%)$ \\
  \colrule 
  \multicolumn{10}{l}{VA type:}\\ 
  \colrule 
  \multirow{2}{4.5em}{BS} & \multicolumn{3}{c}{$5.41\times10^6$} & \multicolumn{3}{c}{$6.85\times10^7$} & $6.32\times10^5$ & $4.92\times10^5$ & $3.30\times10^5$ \\
  & \multicolumn{3}{c}{$(100\%)$} & \multicolumn{3}{c}{$(100\%)$} & $(100\%)$ & $(100\%)$ & $(100\%)$ \\
  \multirow{2}{4.5em}{After Final cut} 
   & $ 2.74\times10^6 $ & $ 2.63\times10^6 $ & $ 2.55\times10^6 $ & $ 6.42\times10^5 $ & $ 6.42\times10^5 $ & $ 6.42\times10^5 $ & $ 3.87\times10^5 $ & $ 3.05\times10^5 $ & $ 2.07\times10^5 $ \\
   & $(50.72\%)$ & $(48.66\%)$ & $(47.18\%)$ & $(0.94\%)$ & $(0.94\%)$ & $(0.94\%)$ & $(61.22\%)$ & $(61.93\%)$ & $(62.78\%)$ \\
  \colrule 
  \multicolumn{10}{l}{TAT type:}\\ 
  \colrule 
  \multirow{2}{4.5em}{BS} & \multicolumn{3}{c}{$8.55\times10^6$} & \multicolumn{3}{c}{$6.87\times10^7$} & $3.56\times10^5$ & $3.30\times10^5$ & $2.65\times10^5$ \\
  & \multicolumn{3}{c}{$(100\%)$} & \multicolumn{3}{c}{$(100\%)$} & $(100\%)$ & $(100\%)$ & $(100\%)$ \\
  \multirow{2}{4.5em}{After Final cut} 
   & $ 4.70\times10^6 $ & $ 4.61\times10^6 $ & $ 4.50\times10^6 $ & $7.19\times10^5$ & $7.19\times10^5$ & $7.19\times10^5$ & $ 2.20\times10^5 $ & $ 2.04\times10^5 $ & $ 1.66\times10^5 $ \\
   & $(54.97\%)$ & $(53.88\%)$ & $(52.60\%)$ & $(1.05\%)$ & $(1.05\%)$ & $(1.05\%)$ & $(61.73\%)$ & $(61.77\%)$ & $(62.52\%)$ \\

\end{tabular}
\end{ruledtabular}
\end{table*}

\begin{table*}[ht]
\caption{\label{table:CutBGZpol}%
Same as \autoref{table:CutBGZ}, but for optimally polarised beams. }
\begin{ruledtabular}
\begin{tabular}{l ccc ccc} 
   & \multicolumn{6}{c}{Event Numbers (Cut efficiencies)}\\
  \cline{2-7}
  & \multicolumn{3}{c}{$ \nu\bar{\nu}\ell^{+}\ell^{-} $} & \multicolumn{3}{c}{Signal}\\
  \cline{2-7}
  Selection cuts & BP1 & BP2 & BP3 & BP1 & BP2 & BP3  \\
  \colrule
  \multicolumn{7}{l}{SP:}\\ 
  \colrule
  \multirow{2}{3.em}{Baseline selection} & \multicolumn{3}{c}{$7.08\times10^5$} & $1.99\times10^4$ & $8.84\times10^2$ & $2.13\times10^2$\\
  & \multicolumn{3}{c}{$(100\%)$} & $(100\%)$ & $(100\%)$ & $(100\%)$ \\
  \multirow{2}{4.0em}{After Final cut} 
  & $7.41\times10^3$ & $1.16\times10^4$ & $1.24\times10^4$ & $9.70\times10^2$ & $5.00\times10^2$ & $1.21\times10^2$ \\
  & $(1.04\%)$ & $(1.64\%)$ & $(1.75\%)$\phantom{00} & \phantom{00}$(48.87\%)$ & $(56.56\%)$ & $(56.81\%)$  \\
  \colrule
  \multicolumn{7}{l}{VA:}\\ \hline 
  \multirow{2}{3.em}{Baseline selection} & \multicolumn{3}{c}{$4.63\times10^5$} & $1.20\times10^3$ & $8.05\times10^2$ & $3.85\times10^2$\\
  & \multicolumn{3}{c}{$(100\%)$} & $(100\%)$ & $(100\%)$ & $(100\%)$ \\
  \multirow{2}{4.0em}{After Final cut} 
  & $3.97\times10^4$ & $3.79\times10^4$ & $3.08\times10^4$ & $9.34\times10^2$ & $6.49\times10^2$ & $3.24\times10^2$ \\
  & $(8.59\%)$ & $(8.20\%)$ & $(6.66\%)$ & $(77.77\%)$ & $(80.62\%)$ & $(84.16\%)$  \\
  \colrule
  \multicolumn{7}{l}{TAT:}\\ 
  \colrule
  \multirow{2}{3.em}{Baseline selection} & \multicolumn{3}{c}{$7.08\times10^5$} & $4.40\times10^3$ & $2.57\times10^3$ & $7.44\times10^2$\\
  & \multicolumn{3}{c}{$(100\%)$} & $(100\%)$ & $(100\%)$ & $(100\%)$ \\
  \multirow{2}{4.0em}{After Final cut} 
  & $1.12\times10^4$ & $8.17\times10^3$ & $1.06\times10^4$ & $2.86\times10^3$ & $1.36\times10^3$ & $4.08\times10^2$ \\
  & $(1.58\%)$ & $(1.15\%)$ & $(1.49\%)$ & $(65.03\%)$ & $(52.98\%)$ & $(54.84\%)$  \\

\end{tabular}
\end{ruledtabular}
\end{table*}

\begin{table*}[ht]
\caption{\label{table:CutBZjjpol}%
Same as \autoref{table:CutBZjj}, but for optimally polarised beams.}
\begin{ruledtabular}
\begin{tabular}{l ccc ccc ccc} 
   & \multicolumn{9}{c}{Event Numbers (Cut efficiencies)}\\
  \cline{2-10}
  & \multicolumn{3}{c}{$\nu_{\ell}\overline{\nu_{\ell}}jj $} & \multicolumn{3}{c}{$\ell\nu_{\ell}jj$} & 
  \multicolumn{3}{c}{Signal} \\
  \cline{2-10}
  Selection cuts & BP1 & BP2 & BP3 & BP1 & BP2 & BP3 & BP1 & BP2 & BP3 \\
  \colrule 
  \multicolumn{10}{l}{SP type:}\\ 
  \colrule 
  \multirow{2}{4.5em}{BS} & \multicolumn{3}{c}{$1.65\times10^6$} & \multicolumn{3}{c}{$2.07\times10^6$} & $2.68\times10^4$ & $1.13\times10^4$ & $2.65\times10^3$ \\
  & \multicolumn{3}{c}{$(100\%)$} & \multicolumn{3}{c}{$(100\%)$} & $(100\%)$ & $(100\%)$ & $(100\%)$ \\
  \multirow{2}{4.5em}{After Final cut} 
   & $8.54\times10^4$ & $1.59\times10^5$ & $1.53\times10^5$ & $9.80\times10^3$ & $2.54\times10^4$ & $2.94\times10^4$ & $1.07\times10^4$ & $ 5.63\times10^3 $ & $ 1.19\times10^3 $ \\
   & $(5.18\%)$ & $(9.65\%)$ & $(9.27\%)$\phantom{0} & \phantom{0}$(0.47\%)$ & $(1.23\%)$ & $(1.42\%)$\phantom{0} & \phantom{0}$(39.86\%)$ & $(49.90\%)$ & $(44.96\%)$ \\
  \colrule 
  \multicolumn{10}{l}{VA type:}\\ 
  \colrule 
  \multirow{2}{4.5em}{BS} & \multicolumn{3}{c}{$1.01\times10^6$} & \multicolumn{3}{c}{$1.22\times10^6$} & $1.43\times10^4$ & $9.58\times10^3$ & $4.60\times10^3$ \\
  & \multicolumn{3}{c}{$(100\%)$} & \multicolumn{3}{c}{$(100\%)$} & $(100\%)$ & $(100\%)$ & $(100\%)$ \\
  \multirow{2}{4.5em}{After Final cut} 
   & $ 3.29\times10^5 $ & $ 3.01\times10^5 $ & $ 2.66\times10^5 $ & $ 8.60\times10^4 $ & $ 8.32\times10^4 $ & $ 7.81\times10^4 $ & $ 7.88\times10^3 $ & $ 5.33\times10^3 $ & $ 2.81\times10^3$ \\
   & $(32.44\%)$ & $(29.69\%)$ & $(26.23\%)$ & $(7.05\%)$ & $(6.82\%)$ & $(6.40\%)$ & $(55.10\%)$ & $(55.64\%)$ & $(61.17\%)$ \\
  \colrule 
  \multicolumn{10}{l}{TAT:}\\ 
  \colrule 
  \multirow{2}{4.5em}{BS} & \multicolumn{3}{c}{$1.65\times10^6$} & \multicolumn{3}{c}{$2.07\times10^6$} & $6.57\times10^4$ & $3.31\times10^4$ & $9.27\times10^3$ \\
  & \multicolumn{3}{c}{$(100\%)$} & \multicolumn{3}{c}{$(100\%)$} & $(100\%)$ & $(100\%)$ & $(100\%)$ \\
  \multirow{2}{4.5em}{After Final cut} 
   & $ 9.45\times10^4 $ & $ 9.44\times10^4 $ & $ 1.53\times10^5 $ & $ 1.12\times10^4 $ & $ 1.15\times10^4 $ & $ 2.86\times10^4 $ & $ 3.23\times10^4 $ & $ 1.44\times10^4 $ & $ 4.55\times10^3 $ \\
   & $(5.73\%)$ & $(5.72\%)$ & $(9.30\%)$ & $(0.54\%)$ & $(0.55\%)$ & $(1.38\%)$ & $(49.21\%)$ & $(43.43\%)$ & $(48.99\%)$ \\
  
 \end{tabular}
\end{ruledtabular}  
\end{table*}

\bibliographystyle{apsrev4-2}
\bibliography{Reference.bib}

\end{document}